\documentclass[a4paper,11pt]{article}
\usepackage{jheppub} 

\usepackage[english]{babel}

\usepackage{amsmath}
\usepackage{graphicx}

\usepackage{amssymb}
\usepackage{amsthm}
\usepackage{braket}
\usepackage{tikz
, pgfplots}
\usetikzlibrary{decorations, positioning, snakes}
\usetikzlibrary{shapes.geometric}

\usepackage{setspace}

\newtheorem{prop}{Proposition}

\pgfdeclaredecoration{complete sines}{initial}
{
    \state{initial}[
        width=+0pt,
        next state=sine,
        persistent precomputation={\pgfmathsetmacro\matchinglength{
            \pgfdecoratedinputsegmentlength / int(\pgfdecoratedinputsegmentlength/\pgfdecorationsegmentlength)}
            \setlength{\pgfdecorationsegmentlength}{\matchinglength pt}
        }] {}
    \state{sine}[width=\pgfdecorationsegmentlength]{
        \pgfpathsine{\pgfpoint{0.25\pgfdecorationsegmentlength}{0.5\pgfdecorationsegmentamplitude}}
        \pgfpathcosine{\pgfpoint{0.25\pgfdecorationsegmentlength}{-0.5\pgfdecorationsegmentamplitude}}
        \pgfpathsine{\pgfpoint{0.25\pgfdecorationsegmentlength}{-0.5\pgfdecorationsegmentamplitude}}
        \pgfpathcosine{\pgfpoint{0.25\pgfdecorationsegmentlength}{0.5\pgfdecorationsegmentamplitude}}
}
    \state{final}{}
}

\title{Gravity MHV amplitudes via Berends-Giele currents}
\date{June 2025}
\author{Chanon Hasuwannakit}
\author[1]{and Kirill Krasnov\note{Corresponding author.}}
\affiliation{School of Mathematical Sciences, University of Nottingham, NG7 2RD, UK}
\emailAdd{kirill.krasnov@nottingham.ac.uk}

\abstract{Berends and Giele derived the Parke–Taylor formula for Yang–Mills MHV amplitudes by computing Berends–Giele currents involving gluons of all-plus and all-but-one-plus helicities. Remarkably, the all-plus current already encodes much of the Parke–Taylor formula structure. The all-but-one-plus current satisfies a more intricate recursion relation than the all-plus case, but one that can still be solved explicitly. This current turns out to be proportional to the all-plus current, which explains why the essential features of the MHV formula are already present at the all-plus level.

In this paper, we carry out an analogous program for gravity. The all-plus graviton Berends–Giele current satisfies a recursion relation that is more involved than in the Yang–Mills case, but whose explicit solution is known: a sum over spanning trees of the complete graph on 
$n$ vertices. We derive and solve the recursion relation for the all-but-one-plus graviton current. The solution is again given by a sum over spanning trees, where each tree contributes a term proportional to the corresponding all-plus current, multiplied by a factor given by a sum over subtrees. Only a small subset of these terms contributes to the MHV amplitude, which we recover explicitly. This provides a direct derivation of the gravity MHV formula from the gravitational Feynman rules—achieving what Berends, Giele, and Kuijf in their 1987 paper regarded as "hard to obtain directly from quantum gravity."}

\pgfplotsset{compat=1.18}
\begin{document}

\maketitle

\flushbottom

\section{Introduction}

The remarkable Parke-Taylor formula \cite{Parke:1986gb} for MHV gluon scattering amplitudes has been proven in \cite{Berends:1987me} using the technology of what became known as Berends-Giele (BG) currents. The idea is to consider an amplitude with $n$ legs on-shell, and one off-shell leg. These one off-shell leg amplitudes are known as currents. One can then sew currents for smaller numbers of particles into bigger currents, which produces a recursive relation for these objects. For the current with all on-shell legs of the same helecity gluons, the recursive relation for the BG currents is easy to solve, producing an expression that already carries all the essential features of the MHV Parke-Taylor formula. However, this all-plus BG current by itself does not produce any non-trivial scattering amplitude, because it gets multiplied by zero in the process of its conversion into an amplitude (more specifically, in the process of amputating the final propagator). 

The explanation for why the all-plus BG current contains all the essential features of the Parke-Taylor formula, or rather why Parke-Taylor formula is very closely related to the all-plus current comes by considering the BG current for a single negative helicity gluon and a general number of positive helicity gluons. This current also satisfies a BG-type recursion, in the sense that the current for $n$ positive helicity particles is determined by all the currents for a smaller number of gluons. Remarkably, even thought more complicated, the all but one positive helicity gluon current recursion relation can still be explicitly solved. The solution is more involved than in the all-plus case, but it turns out to be proportional to the all-plus current. The proportionality coefficient is a sum of many terms. However, in the process of conversion of the current into the amplitude only one of these terms survives. This explains why most of the non-triviality of the Parke-Taylor formula is captured by the all-plus current. The sketched derivation of the Parke-Taylor formula is one from \cite{Berends:1987me}. We also review it in the Appendix for completeness and for a comparison with the gravity case. 

The situation with gravity MHV amplitudes is more complicated. First, a formula for gravity MHV amplitudes, in the form of sum of squares of gluon MHV amplitudes, was proposed in \cite{Berends:1988zp}. It is much more complicated than in the gluon case. Similar formulas were later obtained, using the BCFW recursion relations \cite{Britto:2005fq}, in \cite{Bedford:2005yy}, \cite{Cachazo:2005ca}, \cite{Benincasa:2007qj}, see also 
\cite{Elvang:2007sg}. A twistor proof of the Berends-Giele-Kuijf (BGK) formula was given in \cite{Mason:2009afn}. 

A completely different type of graviton MHV formula was proposed in Section 6 of \cite{Bern:1998sv}, via so-called half-soft functions. These were shown to satisfy Berends-Giele-type recursion relations, see formula (B.9) of this paper. Essentially the same formula was  rediscovered in \cite{Nguyen:2009jk}, where it was motivated by the inverse-soft reconstruction procedure, and proven using a variant of the BCFW recursion. The paper \cite{Nguyen:2009jk} explicitly described the MHV amplitude as a sum over trees. In \cite{Hodges:2012ym} the \cite{Nguyen:2009jk} formula was interpreted in terms of a reduced matrix determinant. More recently, the gravity MHV formula, in its sum over trees form, was proven using the perturbiner approach in \cite{Miller:2024oza}. 

The purpose of this paper is to revisit the derivation of the MHV gravity formula, in its \cite{Bern:1998sv} and \cite{Nguyen:2009jk} form. We follow the original approach of Berends-Giele \cite{Berends:1987me}, and use currents as well as the corresponding recursion relations. The recursion relation for the all-plus gravity BG current was already derived in \cite{Bern:1998sv}. This all-plus recursion, together with its solution, also appeared in \cite{Krasnov:2013wsa}, where it was derived using the so-called pure connection formalism for gravity, using the Feynman rules explained in \cite{Delfino:2014xea}. This last reference also derived a recursion relation for the all-but-one-plus BG currents. A first step towards the solution of this more complicated recursion was taken in \cite{Delfino:2014xea} in that it was shown how to decouple the part of the BG current that depends on the reference spinor of the negative helicity graviton. But no complete solution was given. In this paper we solve the BG recursion relation for the all-but-one-plus BG currents. This is then used to extract the MHV formula. Thus, the present paper describes a derivation of the MHV gravity formula that follows exactly the same logic as that of the original Berends-Giele paper. This is the first derivation of the gravity MHV formula of this type in the literature. It is interesting to note that the paper \cite{Berends:1988zp} concludes with "these string based relations make it possible to obtain a general $n$-graviton amplitude, which would be hard to obtain directly from quantum gravity". What is achieved in this paper is precisely a derivation of the type deemed too hard by the authors of \cite{Berends:1988zp}.

The benefits of our method are in its elementary nature. We do not need to resort to a somewhat more involved (but intimately related) logic of the perturbiner approach as in \cite{Miller:2024oza}. Neither do we need to resort to twistor arguments. There is some combinatorial complexity to our derivation, but the problem is not simple, so there must be some price to pay. 

We first show that it is not difficult to obtain the recursion for the all-but-one-plus BG currents from gravitational Feynman rules. This recursion already apepared in \cite{Delfino:2014xea}, but we spell out a new derivation that, unlike in this reference, does not need any tricks of using the pure connection formalism, and then taking the flat limit. Our starting point is the chiral Einstein-Cartan Lagrangian (a variant of Plebanski Lagrangian for GR), and we obtain the recursion by a relatively simple analysis of the arising perturbation theory. In particular, we will see that the derivation of the all-but-one-plus BG recursion relation is not more complicated than the derivation of the all-plus one. 

The most technical part of our derivation is in solving the all-but-one-plus BG recursion relation. However, this is a purely combinatorial problem, with an interesting solution. Thus, apart from a new derivation of the gravity MHV formula, this paper describes an explicit solution for the all-but-one-plus BG current. This is a new result. 

We also revisit the derivation of the all-but-one-plus BG current of gluons, providing a different argument from that in \cite{Berends:1987me}. Our solution for this current is also slightly more general than the one in \cite{Berends:1987me} because we do not make an assumption about the negative gluon reference spinor, unlike in this reference. This produces an additional term, not present in the formula of \cite{Berends:1987me}. We need this slightly more general treatment in order to motivate considerations of the gravitational case. 

For convenience of the reader, we briefly describe the main results already in the Introduction. We first describe the facts relevant for Yang-Mills theory. 

The YM BG currents have a very specific tensorial structure, at least for processes involving all plus gluons, as well as all-but-one-plus. The current is then a product of certain prefactor that is best described in spinor notations, see the main text, times a scalar factor that depends on just the momenta (and of course helicities) of the on-shell particles. Let us denote the scalar part of the BG current in the all-plus case by $J(1\dots n)$. Then these scalar currents satisfy the following BG recursion
\begin{align}\label{YM recursion-intr}
    J(1\dots n) &= \frac{1}{\Box}\bigg(\sum_{m=1}^{n-1}(q|1\dots m| m+1\dots n|q) J(1\dots m)J(m+1 \dots n)\bigg).
\end{align}
Here $\Box$ stands for the square of the final momentum, so $\Box=(1+\ldots + n)^2$. As is usual in the scattering amplitudes literature, we refer to the momentum of a particle with number $i=1,\ldots ,n$ as simply $i$. The object $q$ is the reference spinor of the positive helicity gluons, see the main text for more information. The object $(q|1\dots m| m+1\dots n|q)$ can be described in words as the Clifford multiplication of the momenta $(1+\dots+ m)$ and $(m+1+\dots+ n)$ with $q$, with then spinor inner product with another copy of $q$ taken. This recursion appears already in \cite{Berends:1987me}, with differences in notations. The solution of this recursion is given by
\begin{align}
    J(1\dots n) = \frac{1}{(q1)(12)\dots(n-1~n)(nq)}.
\end{align}
As we already explained, many of the essential features of the Parke-Taylor formula are already visible here. We recall that the amplitudes under discussion are color-ordered, and this is why gluons are ordered in a specific way in the formula. 

To prove the Parke-Taylor formula, one needs to compute a different current, which is one with a single negative helicity gluon (which we number as gluon 1), as well as a general number of positive helicity gluons, which we number as $2\ldots n$. This current has the same tensorial structure as the all-plus one (after chosing the reference spinor $q$ to coincide with the momentum spinor of the negative helicity gluon). Thus, it is also completely described by its scalar part, which can be shown to satisfy the following recursion
\begin{align}\label{YM-recursion-minus-intr}
    J(1|2\dots n) = \frac{1}{\Box}\bigg(\frac{(q|2\dots n|p]}{[1p]}J(2\dots n) + \sum_{m=2}^{n-1} (q|1\dots m|m+1\dots n|q) J(1|2\dots m)J(m+1\dots n)\bigg).
\end{align}
Here $p$ is the reference spinor of the negative helicity gluon. The solution to this recursion is
\begin{align}\label{YMSol-intr}
    J(1|2\dots n) = J(2\dots n) \bigg(\frac{[2p]}{[12][1p]} + \sum_{m=2}^{n-1} \frac{(q|1\dots m | 1\dots m+1|q)}{(1\dots m)^2(1\dots  m+1)^2}\bigg).
\end{align}
This solution also appears in \cite{Berends:1987me}, with the only difference being that $p$ was chosen to be $2$ in this reference, so that the first term does not arise. While this is a possible choice in the color-ordered case, there is no analog of this choice in the unordered gravity case, and so we kept it in the YM formula as well. Thus, \eqref{YMSol-intr} is a slight generalization of what appears in \cite{Berends:1987me}. 

The solution \eqref{YMSol-intr} can be alternatively described as follows. We can write
\begin{align}
    J(1|2\ldots n)/J(2\ldots n) = \phi(2) + \phi(23) + \ldots \phi(2\ldots n),
\end{align}
where the functions $\phi(2),\phi(23),\ldots,\phi(2\ldots n)$ depend solely on the momenta of the particles that appear as numbers in their arguments. We can also interpret this sum as the sum over subgraphs (necessarily containing the first node $2$) of the simple linear graph on nodes $2\ldots n$. Only the last term in this sum contains a factor of $1/\Box$, where $\Box=(1+2+\ldots n)^2$, and so only the last term survives when one uses the all-but-one-plus current to extract the MHV gluon amplitude. This explains why the Parke-Taylor formula contains the all-plus current $J(2\ldots n)$ as its most essential building block.

The story for gravity is exactly parallel. Again, the all-plus BG currents have a specific tensor structure, and are completely described by their scalar part. This scalar part satisfies the gravitational BG recursion that reads
\begin{align}\label{gravity recursion-intr}
    J(\mathcal{K}) =\frac{1}{\Box}\left(\sum_{|\mathcal{I}| < |\mathcal{J}| , \mathcal{I}\cup \mathcal{J} = \mathcal{K}} (q|\mathcal{I}|\mathcal{J}|q)^2J(\mathcal{I})J(\mathcal{J}) \right).
\end{align}
This recursion is more complicated than that in the YM case, where the color-ordering implies that there is just a single graph on vertices $2\ldots n$ ordered in a specific way that is relevant. This means that in the YM case there is only $n-1$ ways to split the set of momenta $1\ldots n$ into two groups. In the gravity case one must instead consider all possible splittings of the set of momenta $\mathcal{K}=\{1\ldots n\}$ into subsets $\mathcal{I,J}$. This makes the sum in \eqref{gravity recursion-intr} more complicated. It is also interesting to remark that the gravity all-plus recursion is a "square" of the YM all-plus recursion \eqref{YM recursion-intr} in the sense that the factor $(q|\mathcal{I}|\mathcal{J}|q)$ appears with its first power in the YM case and squared in the gravity case. 
As far as we are aware, this recursion relation has first appeared in \cite{Bern:1998sv}, in the context of half-soft functions. 

The solution to the recursion \eqref{gravity recursion-intr} is given by
\begin{align}\label{tree decomposition-intr}
    J(\mathcal{K}) = \sum_{\Gamma_\mathcal{K}} J(\Gamma_\mathcal{K}), \quad 
        J(\Gamma_\mathcal{K}) = \prod_{i\in \mathcal{K}} (qi)^{2\alpha_i - 4} \prod_{\langle jk\rangle \in \mathcal{K}} \frac{[jk]}{(jk)} .
\end{align}
Here the sum is taken over all spanning trees $\Gamma_\mathcal{K}$ of the complete graph on vertices $1\ldots n$. Each tree contributes a weight given by the product of factors $(qi)^{2\alpha_i-4}$ for each vertex $i\in \Gamma_\mathcal{K}$. Here $\alpha_i$ is the multiplicity of the vertex, that is the number of edges that connect to it. The other building block of $J(\Gamma_\mathcal{K})$ is the product over all edges of the factors $[jk]/(jk)$. As is explained in \cite{Krasnov:2013wsa}, this all-plus current can be represented as a reduced determinant, using the matrix-tree theorem. We note that the all-plus current contains most of the essential features of the gravity MHV formula \cite{Nguyen:2009jk}, in particular in the fact that it is given by a sum over trees. The fact that \eqref{tree decomposition-intr} solves \eqref{gravity recursion-intr} is, somewhat implicitly, contained in \cite{Bern:1998sv}, and an explicit proof is given in \cite{Krasnov:2013wsa}. 

To determine the gravity MHV amplitudes one needs to compute the all-but-one-plus BG current. As was shown in \cite{Delfino:2014xea}, and as we show in a different way in the main text, the scalar part of this current satisfies the following recursion relation
\begin{align}\label{gravity-recursion-intr}
    J(1|\mathcal{K}) = \frac{1}{\Box}\bigg(\frac{(q|2\ldots n|p]^2}{[1p]^2}J(\mathcal{K})+\sum_{|\mathcal{I}| < |\mathcal{J}| , \mathcal{I}\cup \mathcal{J} = \mathcal{K}} (q|\mathcal{I}|\mathcal{J}|q)^2 J(1|\mathcal{I})J(\mathcal{J})  \bigg).
\end{align}
Again this recursion relation is very much the "square" of the YM one \eqref{YM-recursion-minus-intr}.

The new result of this paper is a solution to this all-but-one-plus gravity recursion relation. We show that the solution is the sum over all spanning trees $\Gamma_\mathcal{K}$ on the set of momenta $\mathcal{K}=\{2\ldots n\}$. Each spanning tree contributes the weight given by the product of the all-plus scalar current $J(\Gamma_\mathcal{K})$ for this graph, explicitly given by \eqref{tree decomposition-intr} times a factor $\Phi(\Gamma_\mathcal{K})$ that depends on the graph
\begin{align}\label{solution-ansatz-intr}
    J(1|\mathcal{K}) = \sum_{\Gamma_\mathcal{K}} \Phi(\Gamma_\mathcal{K})J(\Gamma_\mathcal{K}).
\end{align}
Analogously to the YM case, each $\Phi(\Gamma_\mathcal{K})$ factor is given by the sum over all subgraphs, so that we can write
\begin{align}
    J(1|\mathcal{K}) = \sum_{\Gamma_\mathcal{K} }J(1|\Gamma_\mathcal{K}), \qquad 
    J(1|\Gamma_\mathcal{K})/J(\Gamma_\mathcal{K}) \equiv \Phi(\Gamma_\mathcal{K})= \sum_{\Gamma_\mathcal{I} \subseteq \Gamma_\mathcal{K},i = |\mathcal{I|}} {\phi}_i(\Gamma_\mathcal{I}). 
\end{align}
Each ${\phi}_i(\Gamma_\mathcal{I})$ factor is in general given by a sum of several terms, depending on the complexity of $\Gamma_\mathcal{I}$. Some lowest order explicit formulas are
\begin{align}
    {\phi}_1(i) &= \frac{(qi)[ip]^2}{[1i][1p]^2}, \\ \nonumber
    {\phi}_2(ij) &= \frac{(q|i|j|q)^2}{(1+i)^2(1+j)^2(1+i+j)^2}, \\ \nonumber
   {\phi}_3(ijk) &=  \frac{(q|i|j|q)(q|ij|k|q)}{(1+j)^2(1+i+j)^2 (1+i+j+k)^2}+ 
\frac{(q|j|k|q)(q|i|jk|q)}{(1+j)^2(1+j+k)^2(1+i+j+k)^2}.
\end{align}
In general, for a graph with $E$ ends $\phi(\Gamma_\mathcal{I})$ contains $E!$ terms. An explicit general formula is given in the main text. Of this, only the very last term in the arising sum, namely $\phi(\Gamma_\mathcal{K})$ contains a factor of $1/\Box$ and thus survives the process of extracting the gravity MHV amplitude. This is analogous to the YM case, and is in that case, provides an explanation of why most of the complexity of the gravity MHV formula is already contained in the all-plus current $J(\mathcal{K})$.

The organisation of the paper is as follows. We start in Section \ref{sec:YM} by reviewing how the YM Berends-Giele recursion relations are derived from the relevant perturbation theory, and how the YM MHV formula is proven. Some of more technical aspects of this calculation are located in the Appendix. We continue in Section \ref{sec:GR} with the similar story for gravity. We start by establishing the Feynman rules in the amount necessary for our computation. We derive the recurrence relation for the all-plus BG current here, describe its solution, and give a combinatorial proof of this solution. We continue in Section \ref{sec:one-minus-current} to do the same for the all-but-one-plus BG current. We first obtain a recurrence relation for these currents, and then proceed to calculate these currents for simple graphs. Having considered a sufficiently large collection of examples we are led to a conjectural solution for this current, described in Section \ref{sec:solution}. We use this solution to obtain the gravity MHV formula in Section \ref{GravityAmplitude}. 
We then proof that the conjectural solution satisfies the relevant recursive relation in Section \ref{sec:proof}. We conclude with a discussion.

\section{Yang-Mills}
\label{sec:YM}

\subsection{Action and Feynman rules}

We will use the Chalmers-Siegel \cite{Chalmers:1996rq} chiral description of Yang-Mills theory, with the action given by
\begin{align}
    S[A,\varphi] = \int \varphi^{ia} \Sigma^i \wedge ( dA^a + \frac{1}{2} f^{abc} A^b A^c) + g_{YM}^2 (\varphi^{ai})^2. 
\end{align}
Here $\varphi^{ia}$ is the auxiliary field that is needed in any first-order formalism, and $\Sigma^i$ are self-dual 2-forms for the flat metric (often referred to as 't Hooft symbols). The indices $a,b,\ldots$ are the Lie algebra ones, and $i=1,2,3$ enumerates the basis vectors in the space of self-dual 2-forms. The objects $f^{abc}$ are the Lie algebra structure constants. Integrating out the auxiliary field $\varphi^{ia}$ produces the Lagrangian $((F^a)_{SD})^2$, where $(F^a)_{SD}$ is the self-dual part of the field strength $F^a$. This Lagrangian coincides with the usual YM Lagrangian $(F^a)^2$ modulo a surface term. 

For our purposes it will be most convenient to rewrite the above action in spinor notations. We use the gravitational literature spinor conventions and in particular denote the spinor indices by $A,B,\ldots$ and $A',B',\ldots$. The action takes the following form
\begin{align}
    S[A,\varphi] = \int \varphi^{AB a} ( \partial_{A}{}^{A'} A_{BA'}^a + \frac{1}{2} f^{abc} A^b_A{}^{A'} A_{BA'}^c) + g_{YM}^2 (\varphi^{AB a})^2. 
\end{align}
There is only a simple cubic interaction in this formulation of YM theory, given by the second term. A simple rescaling of the fields places the YM coupling constant $g_{YM}$ in front of this term. Indeed, we absorb $g_{YM}$ into $\varphi^{AB a}$ and then $1/g_{YM}$ into $A^a_{AA'}$. This changes the action into 
\begin{align}\label{YM-action}
    S[A,\varphi] = \int \varphi^{AB a} ( \partial_{A}{}^{A'} A_{BA'}^a + \frac{g_{YM}}{2} f^{abc} A^a_A{}^{A'} A_{BA'}^c) +  (\varphi^{AB a})^2. 
\end{align}
The auxiliary field $\varphi^{AB a}$ is symmetric in $AB$. This makes the "kinetic" term $\varphi^{AB a}  \partial_{A}{}^{A'} A_{BA'}^a$ degenerate, which reflects the gauge invariance of the theory. There is a simple gauge-fixing procedure, explained in more details in e.g. \cite{Krasnov:2016emc}, which consists in adding an $AB$ anti-symmetric component to $\varphi^{AB a}$, which imposes the Lorentz (or, if this is also added to the $(\varphi^{AB a})^2$ term, Feynman) gauge. With this gauge-fixing the operator in $\varphi^{AB a}  \partial_{A}{}^{A'} A_{BA'}^a$ becomes invertible. The arising operator is, in fact, a version of the Dirac operator, and thus its inverse is given by the Dirac operator divided by the $\Box$. 

The term $(\varphi^{AB a})^2$ is also part of the quadratic part of the action, and so also contributes to the propagators. It is not difficult to see that there are two types of propagators in this theory, namely $\langle A^a_{AA'} \varphi^{CD b}\rangle$ and $\langle A^a_{AA'} A^b_{BB'}\rangle$. Only the first propagator is relevant for our problem, as we will explicitly see below, so we don't even need to state the second propagator. The propagator of interest (in momentum space) is
\begin{align}
    \langle A^a_{AA'} \varphi^{CD b}\rangle = \delta^{ab} \frac{1}{k^2} \epsilon_A{}^C  k_{A'}{}^{D}.
\end{align}

\subsection{Polarisation vectors}

We now resort to the spinor helicity formalism, in which polarisation vectors for the on-shell gluons are parametrised by the momentum spinors of the particles being scattered. We assume familiarity of the readers with this formalism. The polarisation vectors for positive and negative helicity gluons take the following standard form
\begin{align}\label{polar}
    \varepsilon^+_{MM'}(k) = \frac{q_M k_{M'}}{(qk)}, ~~~~~\varepsilon^-_{MM'}(k) = \frac{k_M p_{M'}}{[pk]}.
\end{align}
Here $q$ and $p$ are two reference spinors, which are arbitrary, this arbitrariness reflecting the gauge freedom of the theory. The spinors $k_M, k_{M'}$ are the momentum spinors, which arise for any null 4-vector $k_{MM'}=k_M k_{M'}$. 

In free theory, the on-shell value of the auxiliary field $\varphi$ is given by 
\begin{align}
    \varphi_{AB}^a = \partial_{(A}{}^{A'} A^a_{B)A'},
\end{align}
where round brackets denote symmetrisation. Evaluating for the two on-shell gluons with polarisations \eqref{polar} shows that 
\begin{align}
    \varphi^{+ a}_{AB} = 0, \qquad \varphi^{+ a}_{AB} = k_A k_B.
\end{align}
This means that for processes scattering positive helicity gluons the $\varphi$ leg of the cubic vertex of the theory cannot be an on-shell leg. It can only be the internal leg, to which a propagator $\langle \varphi A\rangle$ is connected. This will have important implications in analysing the types of diagrams that contribute to the processes of interest. 

\subsection{The all-plus recursion derived}

Let us now derive the recursion for the all-plus Berends-Giele currents. By definition, the all-plus BG current is the sum of all tree-level Feynman diagrams with $n$ on-shell positive helicity gluons and one off-shell leg. In YM theory, the color-ordered amplitudes are individually gauge-invariant, which makes them simpler objects to concentrate on. So, from now on we concentrate on processes with a number of gluons that are ordered. This, in particular, means that there are no colour factors in the color ordered amplitudes. 

The polarisation vector of a single positive gluon can be rewritten in the following form
\begin{align}
    \varepsilon^+_{MM'}(k) = q_M (q|k|_{M'} J(k), \qquad 
  J(k) =  \frac{1}{(qk)^2},
\end{align}
where we introduced a convenient notation $(q|k|_{M'} = q^A k_{AM'}= q^A k_A k_{M'} = (qk) k_{M'}$. The last two relations only hold for a null momentum. So, $(q|k|$ is basically the result of Clifford multiplication of the 4-vector $k$ with the auxiliary spinor $q$. We have rewritten the polarisation vector in the form that, as we will see, is common to all all-plus BG currents. 

Thus, anticipating the result, we assume that the all-plus BG currents for on-shell gluons with momenta $1\ldots n$ are given by
\begin{align}\label{current-ansatz}
    J_{MM'}(1\ldots n) = q_M (q| 1\ldots n|_{M'} J(1\ldots n).
\end{align}
The notation $1\ldots n$ stands for the sum of momenta $1+\ldots + n$. We also use the convention in which the momentum of $i$-th particle is referred to as simply $i$, that is $k_i \equiv i$. This declutters the formulas. 

We now assume that \eqref{current-ansatz} holds for all currents with the number of gluons up to $n-1$, and derive the $n$-th current. This will show that the $n$-th current is also of the same form, as well as give us a recursion relation for the scalar part $J(1\ldots n)$. 

It is easiest to motivate the arising recursion relation by considering the simplest process in which two positive helicity gluons combine via the cubic vertex of the theory and produce an off-shell state. The convention for BG currents is that the final propagator is part of the BG current. This produces the following current
\begin{align}\label{J12-calc}
    J(12)_{MM'} = \frac{1}{(1+2)^2} (1+2)_{M'}{}^{N} \frac{q_M 1^{A'}}{(q1)} \frac{q_N 2_{A'}}{(q2)}.
\end{align}
The first factor $(1+2)^{-2} (1+2)_{M'}{}^N$ comes from the propagator converting the object $\varphi_{MN}$ into a connection field $A_{MM'}$. The object $\varphi_{MN}$ is the product of two last factors. As per the cubic vertex of the theory, this is simply the product $\epsilon^+_{M}{}^{A'}(1) \epsilon^+_{NA'}(2)$ of two polarisation vectors, contracted in their primed indices. We work in conventions in which overall factors such as $1/2$ or the minus sign are ignored. We also ignore the factor of the YM coupling constant. All this can be easily reconstructed if needed. 

Massaging the $J(12)$ current we get
\begin{align}
    J(12)_{MM'} = q_M (q|1+2|_{M'} J(12), \qquad J(12) = \frac{[12]}{(1+2)^2 (q1) (2q)} = \frac{1}{(q1)(12)(2q)}. 
\end{align}
We have used $(1+2)^2 = (12)[12]$, which is true in our conventions. We note that we needed a minus sign to convert $(1+2)_{M'}{}^{N} q_N$ into $(q|1+2|_{M'}= q^N (1+2)_{NM'}$, and this minus sign was provided by flipping $(q2)$ into $(2q)$. Quite satisfyingly, this made the formula for $J(12)$ look more canonical in the sense that the denominator contains momentum spinors always next to each other in the product of round brackets. 

It is now not difficult to obtain the all-plus BG recursion. We see that the current $J(12)$ has the structure \eqref{current-ansatz}. As we have already said, we now assume that this holds for all currents with up to $n-1$ particles, and derive the $n$-th current. The current $J(1\ldots n)$ is obtained by the sum of all Feynman diagrams that preserve the ordering. The combinatorics of this sum is such that its terms can be grouped in such a way that every group of terms can be identified with some $m$-th current connected through the cubic vertex to the $(n-m)$-th current. Considering the process in which two such currents connect we have, in exact parallel with \eqref{J12-calc}
\begin{align}
    \frac{1}{\Box} (1\ldots  n)_{M'}{}^N q_M (q|1\ldots  m|^{A'} q_N (q|m+1\ldots  n|_{A'} J(1\ldots m) J(m+1\ldots n) = \\ \nonumber
    q_M (q| 1\ldots  n|_{M'} \frac{1}{\Box} (q|1\ldots m|m+1\ldots n|q) J(1\ldots m) J(m+1\ldots n).
\end{align}
This shows that $J(1\ldots n)_{MM'}$ is still of the same form \eqref{current-ansatz}, because it is given by a sum of terms that are all of the same form. This also gives us the recursion relation for the scalar part of the current which reads
\begin{align}\label{YM recursion}
    J(1\dots n) &= \frac{1}{\Box}\bigg(\sum(q|1\dots m| m+1\dots n|q) J(1\dots m)J(m+1 \dots n)\bigg)
\end{align}

\subsection{Solution for the all-plus current}

The solution to \eqref{YM recursion} is
\begin{align}\label{YM solution}
    J(1\dots n) = \frac{1}{(q1)(12)\dots(n-1~n)(nq)}.
\end{align}
Because of this explicit form of the solution we have
\begin{align}
    J(1\dots m)J(m+1 \dots n) = \frac{(m~m+1)}{(qm)(m+1~q)} J(1\dots \dots n).
\end{align}
Putting this back into \eqref{YM recursion}, we obtain a very important identity
\begin{align}\label{YM Identity}
    \sum_{m=1}^{n-1}(q|1\dots m| m+1\dots n|q)\frac{(m~m+1)}{(qm)(m+1~q)} = (1+2+\dots+n)^2.
\end{align}
Of course one can use this argument in the reverse order and see that it is this identity that shows that \eqref{YM solution} is the solution to the recursion. On the other hand, the identity \eqref{YM Identity} can be verified independently. It is a simple consequence of the Schouten identity $(ab)(cd) = (ac)(bd) - (ad)(bc)$ that holds for the product of two spinor contractions. 

It will be useful for the later to rewrite this identity graphically. We consider the ordered n-vertex graph 

\begin{center}
\scalebox{0.5}{
    \begin{tikzpicture}[
Vertex/.style={circle, draw=black, fill=white, very thick, minimum size=7mm},
Equation/.style={rectangle, draw=white, fill=white, very thick, minimum size=40},]
\node[Vertex] (11) {1};
\node[Vertex] (12) [right=of 11]{2};
\node[Vertex] (13) [right=of 12]{3};
\node[Vertex] (14) [right=of 13]{4};
\node[Equation] (dot) [right=of 14]{\Large$\dots$};
\node[Vertex] (15) [right=of dot]{n-1};
\node[Vertex] (16) [right=of 15]{n};

\draw[-,very thick](11.east) to (12.west);
\draw[-,very thick](12.east) to (13.west);
\draw[-,very thick](13.east) to (14.west);
\draw[-,very thick](14.east) to (dot.west);
\draw[-,very thick](dot.east) to (15.west);
\draw[-,very thick](15.east) to (16.west);
\end{tikzpicture}
}
\end{center}

Then, we replace one of the edges with a wiggly line as 

\begin{center}
\scalebox{0.5}{
    \begin{tikzpicture}[
Vertex/.style={circle, draw=black, fill=white, very thick, minimum size=7mm},
Equation/.style={rectangle, draw=white, fill=white, very thick, minimum size=40},]
\node[Vertex] (11) {1};
\node[Vertex] (12) [right=of 11]{2};
\node[Vertex] (13) [right=of 12]{3};
\node[Vertex] (14) [right=of 13]{4};
\node[Equation] (dot) [right=of 14]{\Large$\dots$};
\node[Vertex] (15) [right=of dot]{i};
\node[Vertex] (16) [right=of 15]{i+1};
\node[Equation] (dot1) [right=of 16]{\Large$\dots$};
\node[Vertex] (17) [right=of dot1]{n-1};
\node[Vertex] (18) [right=of 17]{n};

\draw[-,very thick](11.east) to (12.west);
\draw[-,very thick](12.east) to (13.west);
\draw[-,very thick](13.east) to (14.west);
\draw[-,very thick](14.east) to (dot.west);
\draw[-,very thick](dot.east) to (15.west);
\draw[snake,very thick](15.east) to (16.west);
\draw[-,very thick](16.east) to (dot1.west);
\draw[-,very thick](dot1.east) to (17.west);
\draw[-,very thick](17.east) to (18.west);
\end{tikzpicture}
}
\end{center}
where the wiggly line splits the graph into two sides. The term corresponding to the splitting is defined as
\begin{align}\label{Wiggly}
    (q|1\dots i|i+1 \dots n|q) \frac{(i ~ i+1)}{(q~i)(i+1~q)}
\end{align}

It is then clear that if we take the sum of all graphs with one of edges replaced by a wiggly line, we arrive at the left-hand side of \eqref{YM Identity}. This means we can write, graphically

\begin{center}
\scalebox{0.5}{
    \begin{tikzpicture}[
Vertex/.style={circle, draw=black, fill=white, very thick, minimum size=7mm},
Equation/.style={rectangle, draw=white, fill=white, very thick, minimum size=10},]
\node[Vertex] (11) {1};
\node[Vertex] (12) [right=of 11]{2};
\node[Vertex] (13) [right=of 12]{3};
\node[Vertex] (14) [right=of 13]{4};
\node[Equation] (dot) [right=of 14]{\Large$\dots$};
\node[Vertex] (15) [right=of dot]{n-1};
\node[Vertex] (16) [right=of 15]{n};

\node[Vertex] (21) [below=of 11]{1};
\node[Vertex] (22) [right=of 21]{2};
\node[Vertex] (23) [right=of 22]{3};
\node[Vertex] (24) [right=of 23]{4};
\node[Equation] (dot2) [right=of 24]{\Large$\dots$};
\node[Vertex] (25) [right=of dot2]{n-1};
\node[Vertex] (26) [right=of 25]{n};

\node[Equation] (dotn) [below=of 24]{\Large$\dots$};

\node[Vertex] (34) [below=of dotn]{4};
\node[Vertex] (33) [left=of 34]{3};
\node[Vertex] (32) [left=of 33]{2};
\node[Vertex] (31) [left=of 32]{1};

\node[Equation] (dot3) [right=of 34]{\Large$\dots$};
\node[Vertex] (35) [right=of dot3]{n-1};
\node[Vertex] (36) [right=of 35]{n};

\node[Equation] (eq) [below right=of 26]{\huge $~~~~= ~~~(1+2+\dots+n)^2$};

\draw[snake,very thick](11.east) to (12.west);
\draw[-,very thick](12.east) to (13.west);
\draw[-,very thick](13.east) to (14.west);
\draw[-,very thick](14.east) to (dot.west);
\draw[-,very thick](dot.east) to (15.west);
\draw[-,very thick](15.east) to (16.west);

\draw[-,very thick](21.east) to (22.west);
\draw[snake,very thick](22.east) to (23.west);
\draw[-,very thick](23.east) to (24.west);
\draw[-,very thick](24.east) to (dot2.west);
\draw[-,very thick](dot2.east) to (25.west);
\draw[-,very thick](25.east) to (26.west);

\draw[-,very thick](31.east) to (32.west);
\draw[-,very thick](32.east) to (33.west);
\draw[-,very thick](33.east) to (34.west);
\draw[-,very thick](34.east) to (dot3.west);
\draw[-,very thick](dot3.east) to (35.west);
\draw[snake,very thick](35.east) to (36.west);
\end{tikzpicture}
}
\end{center}

Below we shall see that the same identity is true for any graph, not necessarily just a linear graph of the type of relevance for color ordered YM amplitudes. 

\subsection{All-but-one-plus current}

We now derive the all-but-one-plus BG current recursion relation. We take the negative polarisation gluon to be $1$. The complication comes from the fact that the polarisation vector of the negative gluon is {\it not} of the general form \eqref{current-ansatz}. However, we can achieve more similarity by making the choice $q=1$. In other words, we now take the reference spinor of all the poositive helicity gluons $2\ldots n$ to be equal to the momentum spinor of the gluon $1$. With this choice the polarisation vector becomes
\begin{align}
    \epsilon^-(1)_{MM'} = \frac{ q_M p_{M'}}{[1p]}. 
\end{align}
While this is still not of the type \eqref{current-ansatz}, we can attempt to start connecting this polarisation vector with the all-plus current $J(2\ldots n)$ and see what happens. This gives
\begin{align}
    \frac{1}{\Box} (1+\ldots n)_{M'}{}^N \frac{ q_M p^{A'}}{[1p]} q_N (q|2\ldots n|_{A'} J(2\ldots n) = \\ \nonumber
     q_M (q|1\ldots n|_{M'}  \frac{1}{\Box} \frac{(q|2\ldots n|p]}{[1p]} J(2\ldots n). 
\end{align}
This calculation shows that, because we have chosen the reference spinor $q$ to be the momentum spinor $1$, the negative helicity gluon connected to the all-plus current of gluons $2\ldots n$ is still of the form \eqref{current-ansatz}. And we know from previous considerations that two currents of the form \eqref{current-ansatz} connected through the cubic vertex are again of the same form. This shows that the all-but-one-plus current is of the same general form \eqref{current-ansatz}
\begin{align}
    J(1|2\ldots n)_{MM'} = q_M (q|1\ldots n|_{M'} J(1|2\ldots n), 
\end{align}
and the scalar part of the current satisfies the recursion
\begin{align}
    J(1|2\dots n) = \frac{1}{\Box}\bigg(\frac{(q|2\dots n|p]}{[1p]}J(2\dots n) + \sum_{m=2}^{n-1} (q|1\dots m|m+1\dots n|q) J(1|2\dots m)J(m+1\dots n)\bigg)
\end{align}

The solution of this recursion is 
\begin{align}
    J(1|2\dots n) = J(2\dots n) \bigg(\frac{[2p]}{[12][1p]} + \sum_{m=2}^{n-1} \frac{(q|1\dots +m | 1\dots m+1|q)}{(1+\dots m)^2(1+\dots  +m+1)^2}\bigg).
\end{align}
It is significantly harder to obtain than the all-plus solution. We present the proof in the Appendix \ref{Appendix1}. The proof is also contained in the original Berends-Giele paper \cite{Berends:1987me}, but our derivation is different from the one in this reference, and is adapted to parallel what happens in the gravitational case. 

We can rewrite the all-but-one-plus solution in an alternative way as follows
\begin{align}\label{YM-single-minus-explicitly}
    \frac{J(1|2\ldots n)}{J(2\ldots n)} = \frac{[2p]}{[12][1p]} + \frac{ (q|1+2|1+2+3|q)}{(1+2)^2 (1+2+3)^2} + \ldots + \frac{(q|1\ldots n-1|1\ldots n|q)}{(1\ldots n-1)^2 (1\ldots n)^2},
\end{align}
or 
\begin{align}
    \frac{J(1|2\ldots n)}{J(2\ldots n)} = \sum_{j=2}^n \phi_j, \quad \phi_2 = \frac{[2p]}{[12][1p]}, \quad \phi_j = \frac{(q|1\ldots j-1|1\ldots j|q)}{(1\ldots j-1)^2 (1\ldots j)^2}.
\end{align}
This makes it clear that the all plus single minus current divided by the all plus current is the sum of contributions from lowest order up to the order $n$. Below we will find that the gravitational all-but-one-plus current exhibits the same structure. 

\subsection{MHV amplitude}

We can now extract the MHV amplitude from the all-but-one-plus current. This is done by contracting the current $J(1|2\ldots n)_{MM'}$ with the relevant polarisation vector, and multiplying the result by the squared momentum of the final particle. 

Let us label the momentum of the particle attached to the off-shell leg by $0$. The amplitude is given by
\begin{align}
    A(0^\pm 1^-|2\dots n) = \varepsilon_{\pm}^{MM'}(0) \Box J(1|2\dots n)_{MM'}.
\end{align}
If we are to take the $\varepsilon_{+}^{MM'}(0)$ here, the amplitude vanishes because there is a contraction of two copies of the auxiliary spinor $q$. For the contraction with $\varepsilon_{-}^{MM'}(0)$, the result is non-zero and reproduces the Park-Taylor formula for the MHV amplitude. 

We have
\begin{align}
    A(0^-1^-|2\dots n) &= (01)  \frac{(q|1+2+\dots+n|p]}{[0 p]} (1+2+\dots + n)^2 J(1|2\ldots n). 
\end{align}
As we see from \eqref{YM-single-minus-explicitly}, there is only one term in the current $J(1|2\ldots n)$ that has the pole that can cancel the zero that arises when the momentum $0=1+2+\dots + n$ goes on-shell. Note that zero on the left-hand side of this formula is the momentum of the final particle. Canceling the $(1+2+\dots + n)^2$ numerator-denominator factors we get
\begin{align}
    A(0^-1^-|2\dots n) &= (01)  \frac{(q|1+2+\dots+n|p]}{[0 p]} \frac{(q|2\dots n-1|n|q)J(2\dots n)}{(1+2+\dots + n-1)^2}
    \\
    &=(01)  \frac{(q|0|p]}{[0 p]} \frac{(q|0|n|q)}{(0+n)^2}J(2\dots n)
    = \frac{(01)^4}{(01)(12)(23)\dots(n-1~n)(n0)}.
\end{align}
Here we have used $q\to 1$ to write the last expression. This is the famous Park-Taylor formula for the MHV amplitude. We have thus reproduced the derivation in \cite{Berends:1987me}, via essentially the same method. The only difference with this reference is that we have kept the reference spinor $p$ of the negative helicity particle general, and thus have the non-trivial first term in our formula \eqref{YM-single-minus-explicitly}. This is to prepare ourselves for what happens for gravity. 

\section{Gravity}
\label{sec:GR}

\subsection{Action}

In any Feynman diagram calculation, it is essential to start with as simple action for the theory as possible, in order to minimize the number of diagrams that need to be calculated. The conventional wisdom of the community is that Feynman diagram calculations with gravity are too complicated to be feasible. This, in particular, justifies the $Gravity=(YM)^2$ philosophy that relates amplitudes in "simple" YM theory to "hard" and "impossible to calculate in any other way" amplitudes in gravity. However, we shall see that with the right starting point the scattering amplitudes in gravity are not more complicated than those in YM theory, in the sense that both can be derived by solving an appropriate BG recursion relation, and the YM and gravity recursion relations exhibit complete parallel.  

Our starting point action for gravity will be what can be called the chiral Einstein-Cartan action. We follow the terminology of the book \cite{Krasnov:2020lku}. We will write this action as follows
\begin{align}\label{chiral-EC}
    S[e,A] = \frac{1}{16\pi G} \int \Sigma^i \wedge (dA^i + \frac{1}{2} \epsilon^{ijk} A^j \wedge A^k).
\end{align}
Here the indices $i,j,\ldots = 1,2,3$ and the object $A^i$ is an $\mathbb{R}^3$-valued 1-form. The $\mathbb{R}^3$-valued 2-form $\Sigma^i$ is not considered an independent object in the above action. Rather, it is constructed from the co-frame field $e^I, I=1,2,3,4$, which is a $\mathbb{R}^4$-valued 1-form, as the self-dual part of the $\Lambda^2 \mathbb{R}^4$-value 2-form $e^I\wedge e^J$
\begin{align}
    \Sigma^i = P_+^i{}_{IJ} \, e^I\wedge e^J.
\end{align}
Here $P_+^i{}_{IJ}$ is the projector from $\Lambda^2 \mathbb{R}^4$ to $\Lambda_+ \mathbb{R}^4$. The variation of the above action with $A^i$ leads to an algebraic equation for this field, which determines $A^i$ to be the self-dual part of the spin connection for the frame $e^I$, see \cite{Krasnov:2020lku}, Chapter 5. When substituted into the Lagrangian, the Lagrangian becomes the scalar curvature of the metric encoded by the frame $e^I$ plus a total derivative term. 

For the purposes of this paper, we need to rewrite the above action in spinor notations. We have
\begin{align}\label{chiral-GR-spinor}
    S[e,A] = \frac{1}{16\pi G} \int e_{A}{}^{A'}\wedge e_{BA'} \wedge ( d A^{AB} + A^{AC} \wedge A^{B}{}_C). 
\end{align}
This action also appears in \cite{Capovilla:1991qb}. The benefit of the spinor formalism is that the projection on the self-dual part that appears in \eqref{chiral-EC} becomes easy to write explicitly, and is given by the primed spinor index contraction in $e_{A}{}^{A'}\wedge e_{BA'}$. 

\subsection{Gravitational perturbation theory}

We now expand \eqref{chiral-GR-spinor} around the configuration $e^{AA'} = E^{AA'}, A^{AB}=0$, where $E^{AA'}$ is the co-frame for the flat metric. The signature of the background metric is unimportant for our purposes, because, as is usual in the spinor helicity formalism, some complexification is inherent in the approach in order to render amplitudes vanishing in Minkowski space to exist. At the end of calculations physical results can be obtained by evaluating everything on the Lorentzian signature background metric. 

We will also rescale the fields describing the perturbations, by absorbing the factor $1/16\pi G= M_p^2$ into the fields. This is done by absorbing the factor of $M_p$ into each of the fields $e,a$, so that the new mass dimensions of the fields become $[e]=1, [a]=2$. Keeping only the terms of second and higher order in the perturbations, the expanded Lagrangian reads
\begin{align}\label{gravity-perturbative-L}
    S[e,a]=  \int 
    E_{A}{}^{A'}\wedge E_{BA'} \wedge a^{AC} \wedge a^{B}{}_C + 2 E_{A}{}^{A'}\wedge e_{BA'} \wedge  d a^{AB}
     \\ \nonumber
+ \frac{2}{M_p} E_{A}{}^{A'}\wedge e_{BA'} \wedge a^{AC} \wedge a^{B}{}_C + \frac{1}{M_p} e_{A}{}^{A'}\wedge e_{BA'} \wedge d a^{AB} \\ \nonumber
    + \frac{1}{M_p^2} e_{A}{}^{A'}\wedge e_{BA'} \wedge a^{AC} \wedge a^{B}{}_C. 
\end{align}
Here every line represents terms of a given order in the perturbations $e^{AA'}, a^{AB}$. The action appears complicated, but, as we will soon see, only two of the terms present in the action are actually relevant for the computation we need to perform. 

To analyse the structure of the perturbation theory further, and, in particular, to understand the gauge-fixing that is necessary to invert the kinetic term, we rewrite the action fully in terms of the spinor notations, converting also the form indices into spinor ones. 

We start with the "kinetic", derivative containing term in the quadratic part of the action. It reads
\begin{align}\label{gravity-kinetic}
    2 E_{A}{}^{A'}{}_{MM'} e_{BA' NN'} \partial_{RR'} a^{AB}{}_{SS'} \epsilon^{MM' NN' RR' SS'}.
\end{align}
We will use the following formula for the 4-dimensional $\epsilon$-tensor
\begin{align}
    \epsilon^{MM' NN' RR' SS'} = \epsilon^{MN}\epsilon^{RS}\epsilon^{M'R'}\epsilon^{N'S'} - \epsilon^{M'N'}\epsilon^{R'S'} \epsilon^{MR}\epsilon^{NS}.
\end{align}
This formula is true up to a signature dependent factor, which we ignore, in line with our general convention that overall constants are to be ignored. Using this formula, as well as the fact that the background frame is just the product of two factors of the spinor metric
\begin{align}
    E^{AA'}{}_{MM'} = \epsilon_{M}{}^A \epsilon_{M'}{}^{A'}, 
\end{align}
the kinetic term \eqref{gravity-kinetic} simplifies to
\begin{align}\label{kinetic-gravity}
    2  e_{BA' AB'} \partial_{R}{}^{A'} a^{ABRB'}
    - 2 e_{BA' N}{}^{A'} \partial_{RR'} a^{RBNR'}.
\end{align}
The general gauge-fixing needed to make the quadratic part of the Lagrangian non-degenerate is described in \cite{Krasnov:2020bqr}. A part of the story consists in adding to $a^{AB}{}_{MM'}$, which is $AB$-symmetric, an $AB$ anti-symmetric component, and so effectively a vector field, which then plays the role of the Lagrange multilier imposing the de Donder gauge on the frame perturbation. Then one also adds a new auxiliary field that imposes the Lorentz gauge condition for $\partial_{RR'} a^{RBNR'}$. After this, the kinetic term becomes the sum of two Dirac operators, whose inversion is straightforward. The metric perturbation field, together with the Lagrange multiplier for the Lorentz gauge for the connection, split into two fields that in \cite{Krasnov:2020bqr} are called $H^{(AB)A'B'}$ and $h^{AB}$. The first field does not have any symmetry property in the unprimed indices, while being symmetric in the primed ones. It carries 12 components. The second field does not have any symmetry property, and carries 4 components. The gauge-fixed kinetic term is then the sum of two terms. One contains the Dirac operator acting on $H^{(AB)A'B'}$ and coupling it to a connection field $\Omega^{(AB){}_{CC'}}$. The second contains the Dirac operator acting on $h^{AB}$ and coupling it to the connection field $\omega_{AA'}$. The details of this procedure are not important for our purposes here, with the only important fact being that it can be carried out. 

The reason why things simplify in the present case is that we will only need to invert the kinetic term on the connection fields satisfying the Lorentz gauge condition $\partial_{RR'} a^{RBNR'}=0$, in which case the second term in \eqref{kinetic-gravity} vanishes, and the first term contains the simple Dirac operator whose inversion is straightforward. 

With this discussion out of the way, we can state the only relevant for our purposes propagator. It is given by
\begin{align}
    \langle e_{AA' BB'} a^{CD RR'}\rangle = \frac{1}{k^2} \epsilon_A{}^{(C} \epsilon_B{}^{D)} k_{A'}{}^R \epsilon_{B'}{}^{R'}. 
\end{align}
Here we also ignored the factor of $1/2$ that comes from the factor of $2$ in \eqref{gravity-kinetic}. There is also a propagator of the type $\langle ee\rangle$, but this will not play any role in the comptutation below, so we do not need to state it. 

\subsection{Polarisation spinors}

We now state the polarisation spinors for the two helicity gravitons. They are given by the standard expressions
\begin{align}
    \varepsilon^+_{ABA'B'}(k) = \frac{q_Aq_B k_{A'} k_{B'}}{(qk)^2}~,~ \quad \varepsilon^-_{ABA'B'}(k) = \frac{k_Ak_B p_{A'} p_{B'}}{[pk]^2}.
\end{align}
As in the case of YM theory, the spinors $q,p$ are the reference spinors. 

We will also state the on-shell values for the connection field $a^{AB}{}_{MM'}$. These are determined by solving the $a$-field equation arising by varying the quadratic part of the Lagrangian with respect to this field. Schematically, the field one obtains is the derivative of the tetrad perturbation field. For the perturbations of two different helicities one gets
\begin{align}
    a^+_{ABCC'}(k) = 0, \qquad a^-_{ABCC'}(k) =\frac{k_Ak_B k_C p_{C'}}{[pk]}.
\end{align}
The fact that one of these is zero plays an important role in the analysis below. It means that the positive polarisation states can only be inserted into the $e$-legs of the vertices. 

\subsection{Derivation of the all-plus Berends-Giele recursion}

The purpose of this subsection is to derive the recursion relation for the all-plus BG currents. As in the YM case, we motivate it by considering the simplest process, in which two positive helicity gravitons get combined by the cubic vertex, and then the final propagator is applied. There appear to be two different types of cubic vertices in \eqref{gravity-perturbative-L}. However, since we want to insert two on-shell states and for on-shell states of positive helicity $a$ vanishes, only the second cubic vertex is relevant. We first rewrite it in the fully spinor notation with form indices converted into spinor indices 
\begin{align}\label{gravity-cubic-vertex}
    \frac{1}{M_p} e_A{}^{A'}{}_{MM'} e_{BA' NN'} \partial_{RR'} a^{AB}{}_{SS'} \epsilon^{MM'NN'RR'SS'}. 
\end{align}
In momentum space, with two positive polarisation states inserted, this takes the form
\begin{align}
    \frac{1}{M_p} \frac{q_A q_M 1^{A'} 1_{M'}}{(1q)^2} \frac{q_B q_N 2_{A'} 2_{N'}}{(2q)^2} (1+2)_{RR'} a^{AB}{}_{SS'} (\epsilon^{MN}\epsilon^{RS}\epsilon^{M'R'}\epsilon^{N'S'} - \epsilon^{M'N'}\epsilon^{R'S'} \epsilon^{MR}\epsilon^{NS}). 
\end{align}
It is clear that only the second term contributes, because the first term causes the contraction of two factors of $q$, and we get
\begin{align}
    \frac{1}{M_p}   \frac{[12]^2}{(1q)^2(2q)^2} q_A q_B q^S  (q|1+2|_{R'} a^{AB}{}_{S}{}^{R'}  . 
\end{align}
The object $a^{AB}{}_{S}{}^{R'}$ plays the role of the placeholder here. We now apply the propagator $\langle a e\rangle$ to this result to get the $J(12)$ current
\begin{align}
    J(12)_{AB A'B'} &=  
    q_A q_B (q|1+2|_{A'} (q|1+2|_{B'} J(12), \\ \nonumber 
    J(12) &= \frac{1}{M_p} \frac{1}{(1+2)^2}  \frac{[12]^2}{(1q)^2(2q)^2} = \frac{1}{M_p} \frac{1}{(q1)^2 (q2)^2} \frac{[12]}{(12)}.
\end{align}
This results suggests the pattern. Indeed, we can rewrite the positive helicity polarisation spinors in the following way
\begin{align}
    \epsilon^+_{ABA'B'}(k) = q_A q_B (q|k|_{A'} (q|k|_{B'} J(k), \qquad J(k) = \frac{1}{(qk)^4}.
\end{align}
This makes it nature to conjecture that the all-plus current for any number of gravitons is of the form
\begin{align}\label{all-plus-current-ansatz}
    J(1\dots n)_{ABA'B'} = q_Aq_B (q|1\ldots n|_{A'} (q|1\ldots n|_{B'} J(1\dots n).
\end{align}

We now make the assumption \eqref{all-plus-current-ansatz} for currents with up to $n-1$ gravitons, and show that it holds for the $J(1\ldots n)$ current. To do this we notice that the sum of Feynman diagrams that contributes to this current can be arranged into a sum over groups, with each group containing diagrams that can be understood as the pairing of $J(\mathcal{I})$ current with $J(\mathcal{J})$, where the sets $\mathcal{I,J}$ together form $\mathcal{K}=\{1,\ldots,n\}$. The pairing of such two currents in the cubic vertex \eqref{gravity-cubic-vertex} is computed as follows. We have
\begin{align}
     \frac{1}{M_p} q_A q_M (q|\mathcal{I}|^{A'} (q|\mathcal{I}|_{M'} J(\mathcal{I}) q_B q_N (q|\mathcal{J}|_{A'} (q|\mathcal{J}|_{N'} J(\mathcal{J})(\mathcal{K})_{RR'} a^{AB}{}_{SS'} \\ \nonumber (\epsilon^{MN}\epsilon^{RS}\epsilon^{M'R'}\epsilon^{N'S'} - \epsilon^{M'N'}\epsilon^{R'S'} \epsilon^{MR}\epsilon^{NS}). 
\end{align}
Again only the second term in the bracket contributes and we get
\begin{align}
     \frac{1}{M_p} q_A q_B  (q|\mathcal{I}|\mathcal{J}|q)^2   J(\mathcal{I}) J(\mathcal{J}) (q|\mathcal{K}|_{R'} q^S a^{AB}{}_{S}{}^{R'}. 
\end{align}
Finally, applying to this the $\langle ae\rangle$ propagator gives a contribution to the $n$-th current
\begin{align}
    \frac{1}{M_p} \frac{1}{\Box} q_A q_B  (q|\mathcal{I}|\mathcal{J}|q)^2   J(\mathcal{I}) J(\mathcal{J})  (q|\mathcal{K}|_{A'} (q|\mathcal{K}|_{B'}. 
\end{align}
This contribution is of the form \eqref{all-plus-current-ansatz}, which means that the full $n$-th current is also of this form. This also shows that the scalar part of the all-plus gravity BG current satisfies the following recursion relation 
\begin{align}\label{gravity recursion}
    J(\mathcal{K}) =\frac{1}{\Box}\sum_{|\mathcal{I}| < |\mathcal{J}| , \mathcal{I}\cup \mathcal{J} = \mathcal{K}} (q|\mathcal{I}|\mathcal{J}|q)^2J(\mathcal{I})J(\mathcal{J}) .
\end{align}
We have now ignored the factor of $1/M_p$, which can be easily restored if needed. The sum is over all splittings of the set of momenta $\mathcal{K}=\{1,\ldots,n\}$ into two subsets $\mathcal{I,J}$. One can also always assume that $\mathcal{I}<\mathcal{J}$ to avoid double-counting. 

This is quite similar to the Yang-Mills recursion relation \eqref{YM recursion}, with the exception of two crucial differences. The factor containing the auxiliary spinor $q$ now comes squared. This is in agreement with the general philosophy of $Gravity=(YM)^2$. The other crucial difference is that there is no longer colour ordering, and the sum is taken over all graphs. Both facts change the combinatorics considerably. 

In the derivation above we have ignored the quartic vertex. This is because at tree level this vertex can only contribute if it gave a non-vanishing answer when inserted three on-shell states. However, one of these on-shell states is necessarily the $a$-state, which vanishes for positive helicity gravitons.

\subsection{Solution for the all plus current}

The solution of the all plus gravity current is given by
\begin{align}\label{GR-all-plus-current}
    J(\mathcal{K}) = \sum_{\Gamma_\mathcal{K}} \prod_{i\in \mathcal{K}} (qi)^{2\alpha_i - 4} \prod_{\langle jk\rangle \in \mathcal{K}} \frac{[jk]}{(jk)} .
\end{align}
Here $\mathcal{K}$ is the set of momenta of the positive helicity particles being scattered. The sum is over all possible trees $\Gamma_{\mathcal{K}}$ with vertices in $\mathcal{K}$, $\alpha_i$ represents the number of edges of $\Gamma_{\mathcal{K}}$ connecting to the vertex $i$ and the last product is over edges $\langle jk\rangle \in \mathcal{K}$ of the tree. We can also interpret this as a sum over currents for the individual tree graphs $J(\Gamma_\mathcal{K})$
\begin{align}\label{tree decomposition}
    J(\mathcal{K}) = \sum_{\Gamma_\mathcal{K}} J(\Gamma_\mathcal{K}), \quad 
        J(\Gamma_\mathcal{K}) = \prod_{i\in \mathcal{K}} (qi)^{2\alpha_i - 4} \prod_{\langle jk\rangle \in \mathcal{K}} \frac{[jk]}{(jk)} .
\end{align}
The proof of this formula is already contained in \cite{Bern:1998sv}, and in \cite{Krasnov:2013wsa}, but we present it for completeness, and also as the springboard for the more involved computation of the all-but-one-plus current. 

\subsection{Derivation of the $J(123)$ solution}

The purpose of this and the following subsections is to explain why \eqref{GR-all-plus-current} provides the solution to the recursion \eqref{gravity recursion}. The proof we present is one by induction, where we assume that the formula \eqref{GR-all-plus-current} works for lower order currents, substitute it into the recursion relation and verify that it works for the current with a larger set of momenta. It is useful to first treat an example, and then explain the general principle.

Let us consider the current $J(123)$. We first write the expression for it that the recursion relation gives, and then perform some rearrangement of the terms that explains the combinatorics. We have
\begin{align}\nonumber
    J(123) &= \frac{1}{\Box} \bigg((q|12|3|q)^2J(12)J(3) + (q|13|2|q)^2J(13)J(2) + (q|23|1|q)^2J(23)J(1)\bigg)
    \\
    &= \frac{1}{\Box} \bigg((q|12|3|q) \bigg[\frac{(13)}{(q1)(3q)} (q1)^2 (q3)^2 \frac{[13]}{(13)}+ \frac{(23)}{(q2)(3q)}(q2)^2 (q3)^2 \frac{[23]}{(23)}\bigg]J(12)J(3) \nonumber
    \\
    &~~~~~+ (q|13|2|q)\bigg[\frac{(12)}{(q1)(2q)} (q1)^2 (q2)^2 \frac{[12]}{(12)}+ \frac{(32)}{(q3)(2q)}(q3)^2 (q2)^2 \frac{[32]}{(32)}\bigg]J(13)J(2) \nonumber
    \\
    &~~~~~+ (q|23|1|q)\bigg[\frac{(21)}{(q2)(1q)} (q2)^2 (q1)^2 \frac{[21]}{(21)}+ \frac{(31)}{(q3)(1q)}(q3)^2 (q1)^2 \frac{[31]}{(31)}\bigg]J(23)J(1)\bigg).
\end{align}
Here, to get the second expression we took one of the $(q|\mathcal{I}|\mathcal{J}|q)$ factors and expanded the sum over momenta present in $\mathcal{I,J}$. We have also multiplied and divided the resulting expression by factors $(ij)/(qi)(jq)$. The reason for doing this will become clear below. 

We now note that each expression of the type 
\begin{align}
    (q1)^2 (q3)^2 \frac{[13]}{(13)} J(12)J(3) 
\end{align}
is just the weight $J(\Gamma)$ of a certain graph $\Gamma$ on 3 points. Indeed, the current $J(12)$ consists of a single term, as there is only one connected graph that can be drawn with two vertices. So, we can write suggestively $J(12)=J(1-2)$, where the last notation $J(1-2)$ stands for the weight $J(\Gamma)$ of the specific graph, in this case the graph with two vertices $1,2$ connected by the single edge. It is then not difficult to see that
\begin{align}
    (q1)^2 (q3)^2 \frac{[13]}{(13)} J(12)J(3) = (q1)^2 (q3)^2 \frac{[13]}{(13)} 
    \frac{1}{(q1)^2 (q2)^2} \frac{[12]}{(12)} \frac{1}{(q3)^4} \\ \nonumber 
    = \frac{1}{(q2)^2 (q3)^2} \frac{[21][13]}{(21)(13)} = J(2-1-3)
\end{align}
because the prefactor in front of $J(12) J(3)$ provides just the correct missing weight to give $J(2-1-3)$. This is the general pattern and we have
\begin{align}
    J(123) = \frac{1}{\Box}\bigg( &J(1-2-3) \bigg[(q|1|23|q)\frac{(12)}{(q1)(2q)} + (q|12|3|q)\frac{(23)}{(q2)(3q)}\bigg]\nonumber
    \\
    +&J(1-3-2) \bigg[(q|1|32|q)\frac{(13)}{(q1)(3q)} + (q|13|2|q)\frac{(32)}{(q3)(2q)}\bigg]\nonumber
    \\
    +&J(2-1-3) \bigg[(q|2|13|q)\frac{(21)}{(q2)(1q)} + (q|21|3|q)\frac{(13)}{(q1)(3q)}\bigg]
    \bigg)
\end{align}
We can now make use of the key identity \eqref{YM Identity} we already encountered in the case of YM. It shows that each square bracket in the above expression is equal to $(1+2+3)^2=\Box$. This cancels the $1/\Box$ in front of the current and gives
\begin{align}
    J(123) = J(1-2-3) + J(1-3-2) + J(2-1-3),
\end{align}
which proves \eqref{tree decomposition} for 3 momenta.

\subsection{Extension of \eqref{YM Identity} to arbitrary graphs}

This proof can be extended to an arbitrary number of particles. However, to do this we need resort to a more general version of the identity \eqref{YM Identity}, which would be valid for any graph, not necessarily ordered graphs of the type that are relevant in YM theory. To this end, we first represent \eqref{YM Identity} in a graphical notation. For the example of the graph $1-2-3$ this takes the already familiar from the YM case form
\begin{center}
    
\scalebox{0.5}{
\begin{tikzpicture}[
Vertex/.style={circle, draw=black, fill=white, very thick, minimum size=7mm},
Equation/.style={rectangle, draw=white, fill=white, very thick, minimum size=40},]
\node[Equation] (01) [left = of 11]{\huge $(q|1|23|q)\frac{(12)}{(q1)(2q)} + (q|12|3|q)\frac{(23)}{(q2)(3q)}~~=$};

\node[Vertex] (11) {1};
\node[Vertex] (12) [above right=of 11]{2};
\node[Vertex] (13) [below right=of 12]{3};
\node[Equation] (02) [right = of 13]{\huge $\oplus$};
\node[Vertex] (21) [right = of 02]{1};
\node[Vertex] (22) [above right=of 21]{2};
\node[Vertex] (23) [below right=of 22]{3};

\node[Equation] (03) [right = of 23]{\huge $=~~(1+2+3)^2$};

\draw[snake,very thick](11.north east) to (12.south west);
\draw[-,very thick](12.south east) to (13.north west);

\draw[-,very thick](21.north east) to (22.south west);
\draw[snake,very thick](22.south east) to (23.north west);

\end{tikzpicture}
}
\end{center}

However, it is not difficult to check that the same relation holds for a general tree. We can write this as
\begin{align}\label{Gravity relation}
    \sum_{\langle ij\rangle \in \Gamma}(q|\mathcal{I}|\mathcal{J}|q) \frac{(ij)}{(qi)(jq)} = (\sum_{k\in\mathcal{K}} k)^2.
\end{align}
The sum on the left is one over all edges $\langle ij\rangle$ in a graph $\Gamma$. Every such edge divides $\Gamma$ into two subgroups $\mathcal{I,J}$. 
This relation is a generalization of the Yang-Mills relation \eqref{YM Identity}, and can be shown to hold by using the Schouten identity to exchange the spinor contraction factor
\begin{align}\label{Schouten}
    \frac{(ik)}{(qi)(kq)} = \frac{(ij)}{(qi)(jq)} + \frac{(jk)}{(qj)(kq)}
\end{align}
The relation \eqref{Gravity relation} holds for any tree $\Gamma$ on $n$ points forming the set $\mathcal{K}$. Graphically, it is represented by the sum of graphs with one of the edges of $\Gamma$ replaced with the wiggly line as in \eqref{Wiggly}. 

Let us illustrate the identity \eqref{Gravity relation} for a 5-vertex tree graph $\Gamma_{\mathcal{K}}, \mathcal{K}=\{12345\}$. We have
\begin{center}
\scalebox{0.5}{
\begin{tikzpicture}[
Vertex/.style={circle, draw=black, fill=white, very thick, minimum size=7mm},
Equation/.style={rectangle, draw=white, fill=white, very thick, minimum size=40},]
\node[Vertex] (11) {1};
\node[Vertex] (12) [below=of 11]{2};
\node[Vertex] (13) [below left=of 12]{3};
\node[Vertex] (14) [below right=of 12]{4};
\node[Vertex] (15) [above right=of 14]{5};
\node[Equation] (e1) [below = of 14]{\LARGE$(q|1|2345|q)\frac{(12)}{(q1)(2q)}$};

\node[Vertex] (23) [below right=of 15]{3};
\node[Vertex] (22) [above right=of 23]{2};
\node[Vertex] (21) [above=of 22]{1};
\node[Vertex] (24) [below right=of 22]{4};
\node[Vertex] (25) [above right=of 24]{5};
\node[Equation] (e2) [below = of 24]{\LARGE$(q|3|1245|q)\frac{(32)}{(q3)(2q)}$};

\node[Vertex] (33) [below right=of 25]{3};
\node[Vertex] (32) [above right=of 33]{2};
\node[Vertex] (31) [above=of 32]{1};
\node[Vertex] (34) [below right=of 32]{4};
\node[Vertex] (35) [above right=of 34]{5};
\node[Equation] (e3) [below = of 34]{\LARGE$(q|123|45|q)\frac{(24)}{(q2)(4q)}$};

\node[Vertex] (43) [below right=of 35]{3};
\node[Vertex] (42) [above right=of 43]{2};
\node[Vertex] (41) [above=of 42]{1};
\node[Vertex] (44) [below right=of 42]{4};
\node[Vertex] (45) [above right=of 44]{5};
\node[Equation] (e4) [below = of 44]{\LARGE$(q|1234|5|q)\frac{(45)}{(q4)(5q)}$};

\node[Equation] (equal) [right= of 45]{\huge $= (1+2+3+4+5)^2$};

\draw[snake,very thick](11.south) to (12.north);
\draw[-,very thick](12.south west) to (13.north east);
\draw[-,very thick](12.south east) to (14.north west);
\draw[-,very thick](14.north east) to (15.south west);

\draw[-,very thick](21.south) to (22.north);
\draw[snake,very thick](22.south west) to (23.north east);
\draw[-,very thick](22.south east) to (24.north west);
\draw[-,very thick](24.north east) to (25.south west);

\draw[-,very thick](31.south) to (32.north);
\draw[-,very thick](32.south west) to (33.north east);
\draw[snake,very thick](32.south east) to (34.north west);
\draw[-,very thick](34.north east) to (35.south west);

\draw[-,very thick](41.south) to (42.north);
\draw[-,very thick](42.south west) to (43.north east);
\draw[-,very thick](42.south east) to (44.north west);
\draw[snake,very thick](44.north east) to (45.south west);
\end{tikzpicture}
}
\end{center}
This relation and graphical notation will be very useful for calculations in the next section.

\subsection{Alternate form of the relation}
We note that the relation \eqref{Gravity relation} can also be written as
\begin{align}\label{Gravity relation 2.0}
    \sum_{\langle ij\rangle \in \Gamma}(q|\mathcal{I}|\mathcal{J}|q) \frac{(ij)}{(qi)(jq)} = \sum_{k\in \mathcal{K}}(q|k|\mathcal{K}|q) \frac{(kK)}{(qk)(Kq)}
\end{align}
where we choose a single momenta $K\in\mathcal{K}$ as an anchor instead. The factor $(kK)/(qk)(Kq)$ can be decomposed as above by tracing the edges along the graph so that we can see that the above equation is correct.
For example, for the graph above, we choose the momenta $2$ to be $K$. Then, the relation is given by
\begin{align}
    \sum_{k\in \mathcal{K}}(q|k|\mathcal{K}|q) \frac{(kK)}{(qk)(Kq)} = &(q|1|12345|q) \frac{(12)}{(q1)(2q)}  + (q|3|12345|q) \frac{(32)}{(q3)(2q)} \nonumber
    \\
    + &(q|4|12345|q) \frac{(42)}{(q4)(2q)} + (q|5|12345|q) \frac{(52)}{(q5)(2q)}  \nonumber
    \\
    = &(q|1|12345|q) \frac{(12)}{(q1)(2q)}  + (q|3|12345|q) \frac{(32)}{(q3)(2q)} \nonumber
    \\
    + &(q|45|12345|q) \frac{(42)}{(q4)(2q)} + (q|5|12345|q) \frac{(54)}{(q5)(2q)}  
\end{align}
where we use this decomposition
\begin{align}
    \frac{(52)}{(q5)(2q)} = \frac{(54)}{(q5)(4q)} +\frac{(42)}{(q4)(2q)} 
\end{align}
This result is identical to the formula used in the previous section.
Thus, it is clear that these two ways of writing the relation are equivalent.

\subsection{Proof of the solution for an arbitrary number of particles}

We now spell out the details of the proof of the solution for $J(\mathcal{K})$ for a general set of momenta $\mathcal{K}$. As before, we assume that the solution \eqref{GR-all-plus-current} holds for smaller sets of momenta, and plug it into the recursion to derive the expression for a larger set. 

We start with the recursion \eqref{gravity recursion} and write the sum over momenta in sets $\mathcal{I},\mathcal{J}$ explicitly, similar to what was done in the case of $J(123)$. We have
\begin{align}
    J(\mathcal{K}) &=\frac{1}{\Box}\left(\sum_{|\mathcal{I}| < |\mathcal{J}| , \mathcal{I}\cup \mathcal{J}= \mathcal{K}} \sum_{i\in \mathcal{I},j\in \mathcal{J}} (q|\mathcal{I}|\mathcal{J}|q) (q|i|j|q)J(\mathcal{I})J(\mathcal{J}) \right)\nonumber
    \\
    &= \frac{1}{\Box}\left(\sum_{|\mathcal{I}| < |\mathcal{J}| , \mathcal{I}\cup \mathcal{J} = \mathcal{K}} \sum_{i\in \mathcal{I},j\in \mathcal{J}}(q|\mathcal{I}|\mathcal{J}|q) \frac{(ij)}{(qi)(jq)}\cdot(qi)^2(qj)^2\frac{[ij]}{(ij)}J(\mathcal{I})J(\mathcal{J}) \right)\label{J-gravity}.
\end{align}
As befoe, we multiplied and divided every term by $(ij)/(qi)(jq)$ to get the second expression. The expression 
\begin{align}
    (q|\mathcal{I}|\mathcal{J}|q) \frac{(ij)}{(qi)(jq)}
\end{align}
can be recognised as the summand in \eqref{Gravity relation}. On the other hand, the remaining quantities
\begin{align}
    (qi)^2(qj)^2\frac{[ij]}{(ij)}J(\mathcal{I})J(\mathcal{J}) = \sum_{\Gamma_\mathcal{I}}\sum_{\Gamma_\mathcal{J}} (qi)^2(qj)^2\frac{[ij]}{(ij)}J(\Gamma_\mathcal{I})J(\Gamma_\mathcal{J})
\end{align}
are all weights for some tree graphs $\Gamma_\mathcal{K}$. Indeed, each quantity appearing in the second sum is the weight for some tree graph $\Gamma_\mathcal{K}$ obtained by connecting the trees
 $\Gamma_\mathcal{I},\Gamma_\mathcal{J}$, where the vertex $i\in \mathcal{I}$ is connected to vertex $j\in \mathcal{J}$. Thus, the sum can be reorganized in terms of tree graphs $\Gamma_\mathcal{K}$ instead. 

Thus, we can see that the combinatorics of the sum \eqref{J-gravity} is
\begin{align}
    J(\mathcal{K}) = \frac{1}{\Box}\left(\sum_{\Gamma_\mathcal{K}} \sum_{\langle ij\rangle \in \mathcal{K}}(q|\mathcal{I}|\mathcal{J}|q) \frac{(ij)}{(qi)(jq)}J(\Gamma_\mathcal{K}) \right) = \sum_{\Gamma_\mathcal{K}} J(\Gamma_\mathcal{K}).
\end{align}
We have used \eqref{Gravity relation} to cancel the overall propagator and obtain the final expression. This proves the all plus current formula. 

\section{All plus one minus current}
\label{sec:one-minus-current}

We now tackle the more complicated problem of finding the all-but-one-plus gravity current. We first derive the recursion relation for this current.

\subsection{Derivation of the all-but-one-plus gravity recursion}

As in the YM case, we take the negative graviton to be the particle numbered as $1$, and the positive helicity gravitons to be $2\ldots n$. We also take the momentum spinor of the negative helicity graviton to be the reference spinor $q$ of the positive helicity gravitons. This means that the negative helicity graviton polarisation spinor can be written as
\begin{align}
    \epsilon^-_{ABA'B'}(1) = \frac{q_A q_B p_{A'}p_{B'}}{[1p]^2}. 
\end{align}

There are now two potential cubic vertices that can contribute, namely the first term in the second line in \eqref{gravity-perturbative-L} of the schematic type $e aa$ and the second term of the schematic type $eeda$. The on-shell value of the field $a$ is non-zero for a negative helicity graviton, and so potentially one has a contribution in which the negative helicity graviton gets inserted into the $a$-leg of this vertex, and the positive helicity gets inserted into the $e$ vertex. However, it is not difficult to check that this contribution actually vanishes, because it produces a contraction of two $q$ reference spinors. To see this, we first rewrite this cubic vertex with in the fully spinorial notation
\begin{align}\label{cubic-irrelevant}
    \epsilon_{MA} \epsilon_{M'}{}^{A'}  e_{BA'NN'}  a^{AC}{}_{RR'}  a^{B}{}_{CSS'} (\epsilon^{MN}\epsilon^{RS}\epsilon^{M'R'}\epsilon^{N'S'} - \epsilon^{M'N'}\epsilon^{R'S'} \epsilon^{MR}\epsilon^{NS})\\ \nonumber
    =    e_{BA'AN'}  a^{AC}{}_{R}{}^{A'}  a^{B}{}_{C}{}^{RN'} 
    -   e_{BA'N}{}^{A'}  a^{AC}{}_{AR'}  a^{B}{}_{C}{}^{NR'} .
\end{align}
We now use the fact that the $a$ field for the negative helicity graviton, with the assumption that $q=1$ takes the form
\begin{align}
    a^-_{ABCC'}(1) = \frac{q_A q_B q_C p_{C'}}{[1p]}. 
\end{align}
This shows that, however the negative helicity graviton gets inserted into \eqref{cubic-irrelevant} together with a positive helicity graviton inserted into the $e$-leg, there is always a contraction of two copies of $q$. This shows that this part of the cubic vertex is irrelevant for the processes we are considering. 

We also need to deal with the quartic vertex now, as two of the positive states can be inserted into the $e$-legs, and a negative state into one of the $a$-legs. Let us rewrite the quartic vertex in the fully spinorial notation. We have
\begin{align}
    e_{A}{}^{A'}{}_{MM'} e_{BA'NN'}  a^{AC}{}_{RR'} a^{B}{}_{CSS'}(\epsilon^{MN}\epsilon^{RS}\epsilon^{M'R'}\epsilon^{N'S'} - \epsilon^{M'N'}\epsilon^{R'S'} \epsilon^{MR}\epsilon^{NS}) = \\ \nonumber
    e_{A}{}^{A'}{}_{MM'} e_{BA'}{}^{M}{}_{N'}  a^{AC}{}_{R}{}^{M'} a^{B}{}_{C}{}^{RN'}- e_{A}{}^{A'}{}_{MM'} e_{BA'N}{}^{M'}  a^{ACM}{}_{R'} a^{B}{}_{C}{}^{NR'}.
\end{align}
The first term vanishes when both $e$ factors are substituted by a positive polarisation spinor. Indeed, there is then a contraction of two $q$ factors. In the second term the contraction of two $e$'s is not in the unprimed indices, and so does not vanish. However, when evaluation on the positive polarisations it produces a factor of $q_A q_M q_B q_N$, which then gets contracted with the two $a$'s. When one of these is an on-shell state containing 3 factors of $q$, the result vanishes. So, the quartic vertex cannot contribute even in the all-but-one-plus case. 

We thus only need to insert the negative helicity state into the vertex $ee da$. Let us first show that the $a$-leg of this vertex continues to be the off-shell one even when one of the on-shell states is negative. Let us write this vertex in fully spinor notation 
\begin{align}
   e_A{}^{A'}{}_{MM'} e_{BA' NN'} \partial_{RR'} a^{AB}{}_{SS'} (\epsilon^{MN}\epsilon^{RS}\epsilon^{M'R'}\epsilon^{N'S'} - \epsilon^{M'N'}\epsilon^{R'S'} \epsilon^{MR}\epsilon^{NS}) = \\ \nonumber
   e_A{}^{A'}{}_{MM'} e_{BA'}{}^{M}{}_{N'} \partial_{R}{}^{M'} a^{ABRN'}-e_A{}^{A'}{}_{MM'} e_{BA' N}{}^{M'} \partial^M{}_{R'} a^{ABNR'}.
\end{align}
When the $a$-leg of this vertex is evaluated on a negative on-shell state, it will produce 3 factors of $q$. These will then contract with at least one of the $q$ factors in the one-shell $e$-leg. So, the only way to obtain a non-zero result from this vertex is to insert the on-shell states into the $e$-legs. 

Let us now consider the process in which a single negative helicity state gets inserted into the $ee da$ vertex together with the all-plus current $J(2\ldots n)$. We have
\begin{align}
     \frac{1}{M_p} \frac{q_A q_M p^{A'}p_{M'}}{[1p]^2}        
     q_B q_N (q|2\ldots n|_{A'} (q|2\ldots n|_{N'} J(\mathcal{J})(1+2+\ldots +n)_{RR'} a^{AB}{}_{SS'} \\ \nonumber (\epsilon^{MN}\epsilon^{RS}\epsilon^{M'R'}\epsilon^{N'S'} - \epsilon^{M'N'}\epsilon^{R'S'} \epsilon^{MR}\epsilon^{NS}). 
\end{align}
As before, only the second term is relevant and after simplifications we get
\begin{align}
     \frac{1}{M_p} q_A      
     q_B \frac{(q|2\ldots n|p]^2}{[1p]^2} J(\mathcal{J}) (q|1+2+\ldots +n|_{R'} q^S a^{AB}{}_{S}{}^{R'}. 
\end{align}
Applying the final propagator gives
\begin{align}
    \frac{1}{M_p} q_A  q_B(q|1+2+\ldots +n|_{A'} (q|1+2+\ldots +n|_{B'}  \frac{1}{\Box}  \frac{(q|2\ldots n|p]^2}{[1p]^2} J(\mathcal{J}).
\end{align}
This is of the same type as the all-positive currents \eqref{current-ansatz}, which shows that the all-but-one-plus currents follow the same pattern. This also shows that the recurrence relation for the scalar part of the all-but-one-plus currents is given by
\begin{align}\label{gravity-recursion}
    J(1|\mathcal{K}) = \frac{1}{\Box}\bigg(\frac{(q|\mathcal{K}|p]^2}{[1p]^2}J(\mathcal{K})+\sum_{|\mathcal{I}| < |\mathcal{J}| , \mathcal{I}\cup \mathcal{J} = \mathcal{K}} (q|\mathcal{I}|\mathcal{J}|q)^2 J(1|\mathcal{I})J(\mathcal{J})  \bigg) 
\end{align}

\subsection{Previous results}

It has been shown explicitly in \cite{Delfino:2014xea} that the all-but-one-plus current can be separated into $p$-dependent and $p$-independent terms as
\begin{align}\label{gravity-current}
    J(1|\mathcal{K}) = \sum_{i\in\mathcal{K}}\frac{(qi)[ip]^2}{[qi][1p]^2}J(\mathcal{K}) + S(1|\mathcal{K})
\end{align}
where the recursion relation for the $p$-independent object $S(1|\mathcal{K})$ is given by
\begin{align}\label{S-current}
    S(1|\mathcal{K}) = - \frac{1}{\Box}\bigg(\sum \frac{(qi)(qj)}{[1i][1j]}[ij]^2 J(\mathcal{K}) + \sum_{\mathcal{I}\cup \mathcal{J} = \mathcal{K}}(q|1\mathcal{I}|\mathcal{J}|q)^2 S(1|\mathcal{I})J(\mathcal{J})\bigg)
\end{align}
This recursion starts with $S(1|2) = 0$ with the first non-trivial term being the 3-current $S(1|23)$. The 3-current is then only the first term
\begin{align}\label{2-3 S-current}
    S(1|23) = - \frac{(q2)(q3)}{[12][13]}[23]^2 J(23)
\end{align}
The main objective of this paper is to provide a general solution for the S-current and show that the gravity MHV amplitude can be obtained from this solution directly. Note that the first term in \eqref{gravity-current} is analogous to the $p$-dependent first term in the YM solution \eqref{YM-single-minus-explicitly}. Our task is to find the $p$-independent terms. We will not be using the results from \cite{Delfino:2014xea} here, providing all the derivations from scratch. 

\subsection{Motivating the form of the solution}

We can rewrite the first $p$-dependent term in \eqref{gravity-current} as 
\begin{align}
    J(1|\mathcal{K}) = \sum_{\Gamma_\mathcal{K}} \sum_{i\in\mathcal{K}}\frac{(qi)[ip]^2}{[1i][1p]^2} J(\Gamma_\mathcal{K}) + S(1|\mathcal{K}).
\end{align} 
The sum over $i\in \mathcal{K}$ is over all the vertices of the graph $\Gamma_\mathcal{K}$, and since every graph $\Gamma_\mathcal{K}$ is a tree connecting all the vertices of $\mathcal{K}$, this sum is independent of $\Gamma_\mathcal{K}$. Nevertheless, we can rewrite the above expression in the following suggestive form
\begin{align}
     J(1|\mathcal{K})= \sum_{\Gamma_\mathcal{K}}\phi_1(\Gamma_\mathcal{K})J(\Gamma_\mathcal{K})+S(1|\mathcal{K}), \quad 
     \phi_1(\Gamma_\mathcal{K}) = \sum_{i\in\Gamma_\mathcal{K}}\frac{(qi)[ip]^2}{[1i][1p]^2}.
\end{align}
This is similar to the first term in the YM formula \eqref{YM-single-minus-explicitly}, with the exception that in the YM case we have only one possible graph, as dictated by the color ordering. There is no ordering in the gravity case, and a sum over all trees built from the set $\mathcal{K}$ arises. 

The above discussion suggests that we look for the solution for the full current $J(1|\mathcal{K})$ in a similar form
\begin{align}\label{solution-ansatz}
    J(1|\mathcal{K}) = \sum_{\Gamma_\mathcal{K}} \Phi(\Gamma_\mathcal{K})J(\Gamma_\mathcal{K}),
\end{align}
where $\Phi(\Gamma_\mathcal{K})$ is some scalar factor associated with the tree graph. 

\subsection{Obtaining a recursion for $\Phi(\Gamma_\mathcal{K})$}

We now substitute the solution ansatz \eqref{solution-ansatz} into the recursion \eqref{gravity-recursion}. This gives
\begin{align}
     J(1|\mathcal{K}) = \frac{1}{\Box}\bigg(\frac{(q|\mathcal{K}|p]^2}{[1p]^2}
     \sum_{\Gamma_\mathcal{K}}
     J(\Gamma_\mathcal{K})+\sum_{|\mathcal{I}| < |\mathcal{J}| , \mathcal{I}\cup \mathcal{J} = \mathcal{K}} 
     (q|\mathcal{I}|\mathcal{J}|q)^2 \sum_{\Gamma_\mathcal{I},\Gamma_\mathcal{J}} \Phi(\Gamma_\mathcal{I})J(\Gamma_\mathcal{I})J(\Gamma_\mathcal{J})  \bigg).
\end{align}
We now do the same trick we used in the all plus case and expand out one of the square factors in the second term. This provides a factor of $(ij)/(qi)(jq)$, as well as weights required to connect graphs $\Gamma_\mathcal{I},\Gamma_\mathcal{J}$ into some graph $\Gamma_\mathcal{K}$ containing all the vertices. The sum reorganises itself into a sum over $\Gamma_\mathcal{K}$
\begin{align}
    J(1|\mathcal{K}) = \frac{1}{\Box}\bigg(\frac{(q|\mathcal{K}|p]^2}{[1p]^2}
     \sum_{\Gamma_\mathcal{K}}
     J(\Gamma_\mathcal{K}) + \sum_{\Gamma_\mathcal{K}} \sum_{\langle ij\rangle\in\Gamma_\mathcal{K}}(q|1\mathcal{I}|\mathcal{J}|q) \frac{(ij)}{(qi)(jq)} \Phi(\Gamma_\mathcal{I})J(\Gamma_\mathcal{K})\bigg).
\end{align}
Here the second sum in the last term is that over all edges $\langle ij\rangle$ of the graph $\Gamma_\mathcal{K}$, and $\mathcal{I,J}$ are the two subgraphs that $\mathcal{K}$ splits into once the edge $\langle ij\rangle$ is deleted. 
With this simplification every term on both sides is a sum over $\Gamma_\mathcal{K}$. Equating the coefficients in front of $J(\Gamma_\mathcal{K})$ we get
\begin{align}
    \Phi(\Gamma_\mathcal{K}) = \frac{1}{\Box}\bigg(\frac{(q|\mathcal{K}|p]^2}{[1p]^2}
      +  \sum_{\langle ij\rangle\in\Gamma_\mathcal{K}}(q|1\mathcal{I}|\mathcal{J}|q) \frac{(ij)}{(qi)(jq)} \Phi(\Gamma_\mathcal{I})\bigg).
\end{align}
The sum in the second term here is over all edges $\langle ij\rangle$ of $\Gamma_\mathcal{K}$. They split $\Gamma_\mathcal{K}$ into two subgraphs $\Gamma_\mathcal{I},\Gamma_\mathcal{J}$. Both $\Gamma_\mathcal{I},\Gamma_\mathcal{J}$ will contribute to the $\Phi(\Gamma_\mathcal{I})$ term, so we could as well write this recursion as
\begin{align}
    \Phi(\Gamma_\mathcal{K}) = \frac{1}{\Box}\bigg(\frac{(q|\mathcal{K}|p]^2}{[1p]^2}
      +  \sum_{\langle ij\rangle\in\Gamma_\mathcal{K}}(q|1\mathcal{I}|\mathcal{J}|q) \frac{(ij)}{(qi)(jq)} \Phi(\Gamma_\mathcal{I})+(q|1\mathcal{J}|\mathcal{I}|q) \frac{(ji)}{(qj)(iq)} \Phi(\Gamma_\mathcal{J})
      \bigg).
\end{align}
Since we can drop the momentum $1$ from $(q|1\mathcal{I}|\mathcal{J}|q)$, the quantities $(q|1\mathcal{I}|\mathcal{J}|q)$ are actually anti-symmetric in $i,j$, as the factors $(ij)/(qi)(jq)$. This means we can rewrite the recursion as
\begin{align}
    \Phi(\Gamma_\mathcal{K}) = \frac{1}{\Box}\bigg(\frac{(q|\mathcal{K}|p]^2}{[1p]^2}
      +  \sum_{\langle ij\rangle\in\Gamma_\mathcal{K}}(q|1\mathcal{I}|\mathcal{J}|q) \frac{(ij)}{(qi)(jq)} (\Phi(\Gamma_\mathcal{I})+ \Phi(\Gamma_\mathcal{J}))
      \bigg).
\end{align}

\subsection{Obtaining $p$-dependent terms}

We first show that the $p$-depedent terms in $\Phi(\Gamma_\mathcal{K})$ are given by the sum over vertices of $\Gamma_\mathcal{K}$
\begin{align}
    \Phi(\Gamma_\mathcal{K}) = \sum_{i\in \Gamma_\mathcal{K}} \phi_1(i)+\ldots, \qquad \phi_1(i)= \frac{(qi)[ip]^2}{[1i][1p]^2} = \frac{(q|i|p]^2}{(1+i)^2 [1p]^2},
\end{align}
where the last equality is a suggestive rewriting of $\phi_1(i)$. We have denoted by $\phi_1$ the $p$-dependent terms, anticipating that they are a sum of contributions from individual vertices. The dots stand for the higher-order, $p$-independent terms. 

We can now do some more suggestive rewritings, expanding the square to get
\begin{align}
    \frac{(q|\mathcal{K}|p]^2}{[1p]^2} =  \left(\frac{(q|\mathcal{K}|p]}{[1p]} \right)^2=    
    \sum_{i\in \Gamma_\mathcal{K}} \frac{(q|i|p]^2}{[1p]^2} + \sum_{i\not = j} \frac{(q|i|p](q|j|p]}{[1p]^2}.
    \end{align}
    We can rewrite the first term here as 
  \begin{align}  
   \frac{(q|i|p]^2}{[1p]^2}= \phi_1(i) (1+i)^2 = \sum_{i\in \Gamma_\mathcal{K}} \phi_1(i) (q|1|i|q) \frac{(1i)}{(q1)(iq)},
\end{align}
where the last expression was obtained by using the main identity \eqref{YM Identity} to write $(1+i)^2$ as $(q|1|i|q) ((1i)/(q1)(iq))$. Note that the problematic in the limit $q\to 1$ factors cancel on both sides, and this way of rewriting is only for motivation purposes. 

We now assume that the $p$-dependent terms in $\Phi(\Gamma_\mathcal{K})$ is given by the sum of $\phi_1(i)$, for graphs $\Gamma_\mathcal{I}<\Gamma_\mathcal{K}$, and use this to prove that this continues to be true for $\Gamma_\mathcal{K}$. With this assumption, the recursion takes the form
\begin{align}
   & \Phi(\Gamma_\mathcal{K}) = \frac{1}{\Box}\bigg(\sum_{i\in \Gamma_\mathcal{K}} \phi_1(i) (q|1|i|q) \frac{(1i)}{(q1)(iq)}+ \sum_{i\not = j} \frac{(q|i|p](q|j|p]}{[1p]^2}
   \\ \nonumber
      &+  \sum_{\langle ij\rangle\in\Gamma_\mathcal{K}}(q|\mathcal{I}|\mathcal{J}|q) \frac{(ij)}{(qi)(jq)} \left(\sum_{k\in \Gamma_\mathcal{K}} \phi_1(k) + \tilde{\Phi}(\Gamma_\mathcal{I})+ \tilde{\Phi}(\Gamma_\mathcal{J})\right) 
   \bigg).
\end{align}
Here $\tilde{\Phi}(\Gamma_\mathcal{I})$ is the $p$-independent part of $\Phi(\Gamma_\mathcal{I})$. We now want to combine the $\Phi$-containing terms from the first term with such term from the first term in the second line, and use the main identity \eqref{YM Identity} to cancel the $\Box$. This should happen for the coefficient of every $\phi_1(i)$ term, for every $i\in \Gamma_\mathcal{K}$. We should massage both $\Phi$-containing terms to exhibit this. 

We rewrite 
\begin{align}
    (q|1|i|q) = (q|1| \mathcal{K}|q) - (q|1| \mathcal{K}/\{i\}|q) = (q|1| \mathcal{K}|q) - 
    \sum_{j\not= i} (q|1| j|q).
\end{align}
The first term can then be written as
\begin{align}
    \sum_{i\in \Gamma_\mathcal{K}} \phi_1(i) (q|1|\mathcal{K}|q) \frac{(1i)}{(q1)(iq)}-
    \sum_{i\in \Gamma_\mathcal{K}} \phi_1(i) \sum_{j\not= i} (q|1| j|q) \frac{(1i)}{(q1)(iq)}.
\end{align}

We can now use \eqref{Gravity relation} to sum up the coefficient of every $\phi_1(i)$. Indeed, we can rewrite this relation as
\begin{align}
    (q|1|\mathcal{K}|q) \frac{(1i)}{(q1)(iq)} + \sum_{\langle ij\rangle\in\Gamma_\mathcal{K}}(q|1\mathcal{I}|\mathcal{J}|q) \frac{(ij)}{(qi)(jq)} = (1+ \mathcal{K})^2 = \Box.
\end{align}
Note that the first term is actually $i$-independent
\begin{align}
    (q|1|\mathcal{K}|q) \frac{(1i)}{(q1)(iq)} = -[1|\mathcal{K}|q),
\end{align}
and in the second term the sum is as usual over all edges $\langle ij\rangle\in\Gamma_\mathcal{K}$. This means that we get
\begin{align}\label{manipulating-phi-rec-1}
   & \Phi(\Gamma_\mathcal{K}) = \sum_{i\in \Gamma_\mathcal{K}} \phi_1(i) + \frac{1}{\Box}\bigg(
    - \sum_{i\in \Gamma_\mathcal{K}} \phi_1(i) \sum_{j\not= i} (q|1| j|q) \frac{(1i)}{(q1)(iq)}+ \sum_{i\not = j} \frac{(q|i|p](q|j|p]}{[1p]^2}
   \\ \nonumber
      &+  \sum_{\langle ij\rangle\in\Gamma_\mathcal{K}}(q|1\mathcal{I}|\mathcal{J}|q) \frac{(ij)}{(qi)(jq)} \left( \tilde{\Phi}(\Gamma_\mathcal{I})+ \tilde{\Phi}(\Gamma_\mathcal{J})\right) 
   \bigg).
\end{align}

We now want to show that all the terms in the first line apart from the first combine into something $p$-independent. We have
\begin{align}
    - \sum_{i\in \Gamma_\mathcal{K}} \phi_1(i) \sum_{j\not= i} (q|1| j|q) \frac{(1i)}{(q1)(iq)}=
    \sum_{i\in \Gamma_\mathcal{K}} \sum_{j\not= i}
    \frac{(q|i|p]^2}{(1i)[1i][1p]^2} [1|j|q),
\end{align}
and so the terms in the brackets in the first line are
\begin{align}
    \sum_{i\in \Gamma_\mathcal{K}} \sum_{j\not= i}
    \frac{(q|i|p]^2}{(1i)[1i][1p]^2} [1|j|q)=
    -\sum_{i\in \Gamma_\mathcal{K}} \sum_{j\not= i} \frac{(qi)(qj)}{[1p]^2[1i][1j]} [ip]^2 [1j]^2.
\end{align}
We wrote the expression in a suggestive form. Now, both terms in the brackets in the first line of \eqref{manipulating-phi-rec-1} can be written as
\begin{align}
    -\sum_{i\in \Gamma_\mathcal{K}} \sum_{j\not= i} \frac{(qi)(qj)}{[1p]^2[1i][1j]}( [ip]^2 [1j]^2 - [ip]^2[jp]^2).
\end{align}
This can be written as a sum over pairs $\langle ij\rangle$
\begin{align}
    &-\sum_{\langle ij\rangle} \frac{(qi)(qj)}{[1p]^2[1i][1j]} ( [ip]^2 [1j]^2 + [jp]^2 [1i]^2 - 2 [ip][jp]) = -\sum_{\langle ij\rangle} \frac{(qi)(qj)}{[1p]^2[1i][1j]} ( [ip] [1j] - [jp] [1i])^2 
    \\ \nonumber &=
    -\sum_{\langle ij\rangle} \frac{(qi)(qj)}{[1p]^2[1i][1j]} ([1p][ij])^2 = -\sum_{\langle ij\rangle} \frac{(qi)(qj)[ij]^2}{[1i][1j]} 
\end{align}
This shows that $[1p]^2$ cancels and we have
\begin{align}\label{manipulating-phi-rec-2}
   & \Phi(\Gamma_\mathcal{K}) = \sum_{i\in \Gamma_\mathcal{K}} \phi_1(i) + \frac{1}{\Box}\bigg(
    -\sum_{\langle ij\rangle} \frac{(qi)(qj)[ij]^2}{[1i][1j]}
   \\ \nonumber
      &+  \sum_{\langle ij\rangle\in\Gamma_\mathcal{K}}(q|\mathcal{I}|\mathcal{J}|q) \frac{(ij)}{(qi)(jq)} \left( \tilde{\Phi}(\Gamma_\mathcal{I})+ \tilde{\Phi}(\Gamma_\mathcal{J})\right) 
   \bigg).
\end{align}
This proves that the $p$-dependent part of $\Phi$ is given by the sum of contributions $\phi_1(i)$, and also gives us a recursion for the $p$-independent part, which we write as
\begin{align}\label{manipulating-phi-rec-2}
    \tilde{\Phi}(\Gamma_\mathcal{K}) =\frac{1}{\Box}\bigg(
    - \sum_{\langle ij\rangle} \frac{(q|i|j|q)^2}{(qi)(qj)[1i][1j]}
   +  \sum_{\langle ij\rangle\in\Gamma_\mathcal{K}}(q|\mathcal{I}|\mathcal{J}|q) \frac{(ij)}{(qi)(jq)} \left( \tilde{\Phi}(\Gamma_\mathcal{I})+ \tilde{\Phi}(\Gamma_\mathcal{J})\right) 
   \bigg).
\end{align}

\subsection{An important identity}

The sum in the first term in \eqref{manipulating-phi-rec-2} is over all pairs $\langle ij\rangle$ belonging to the set $\mathcal{K}$. We want to show that it can be rewritten in a graph-specific way. To this end, we use the Schouten identity similar to \eqref{schouten-appendix}, but now for the square brackets instead.

We start with an example for $\mathcal{K}=\{234\}$. For the case of three momenta we have
\begin{align}
    \sum_{\langle ij\rangle\in \{234\}} \frac{(qi)(qj)[ij]^2}{[1i][1j]} = \frac{(q2)(q4)[24]^2}{[12][14]}+ \frac{(q2)(q3)[23]^2}{[12][13]}+\frac{(q3)(q4)[34]^2}{[13][14]}.
\end{align}
We can decide to associate this quantity, which is graph agnostic, to the graph $2-3-4$. In order to do this, we use Schouten identity to rewrite the term $[24]/[12][14]$ as
\begin{align}
    \frac{[24]}{[12][14]} = \frac{[23]}{[12][13]}+\frac{[34]}{[13][14]}.   
\end{align}
This gives
\begin{align}\nonumber
      \sum_{\langle ij\rangle\in \{234\}} \frac{(qi)(qj)[ij]^2}{[1i][1j]} = (q2)(q4)[24]\left( \frac{[23]}{[12][13]}+\frac{[34]}{[13][14]}\right)
      + \frac{(q2)(q3)[23]^2}{[12][13]}+\frac{(q3)(q4)[34]^2}{[13][14]}= \\ \nonumber
      \frac{[23]}{[12][13]}( (q2)(q4)[24]+ (q2)(q3)[23]) + \frac{[34]}{[13][14]}( (q2)(q4)[24]+ (q3)(q4)[34]) = \\ \nonumber
      \frac{(q|2|3|q)}{(1+2)^2(1+3)^2} (q|2|34|q) + \frac{(q|3|4|q)}{(1+3)^2(1+4)^2} (q|23|4|q) = \sum_{\langle ij\rangle \in (2-3-4)} (q|\mathcal{I}|\mathcal{J}|q) \frac{(q|i|j|q)}{(1+i)^2(1+j)^2}.
\end{align}
To write the equality in the last line we have used $q\to 1$, and wrote the result in a suggestive way. This shows that the graph-agnostic sum on the left can be written in a graph-specific way, as a sum over all edges of the graph, splitting the graph into subgraphs $\mathcal{I,J}$.

A straightforward generalization of this example gives
\begin{align}\label{identity-sum-edges}
    \sum_{\langle ij\rangle} \frac{(q|i|j|q)^2}{(qi)(qj)[1i][1j]} = \sum_{\langle ij\rangle \in \Gamma_{\mathcal{K}}} (q|\mathcal{I}|\mathcal{J}|q) \frac{(q|i|j|q)}{(1+i)^2(1+j)^2}.
\end{align}
Here the sum on the left is graph-agnostic, but it can be rewritten as a sum over edges of any graph $\Gamma_{\mathcal{K}}$. Thus, the above formula can also be interpreted as the statement that the sum over edges of some graph $\Gamma_{\mathcal{K}}$ is actually graph-independent, and is given by the sum on the left-hand side. 

Using the identity \eqref{identity-sum-edges}, we can rewrite the recursion for the $p$-indepedent part as
\begin{align}
    \tilde{\Phi}(\Gamma_\mathcal{K}) = \frac{1}{\Box}\bigg(-\sum_{\langle ij\rangle \in \Gamma_\mathcal{K}} \frac{(q|i|j|q)}{(1+i)^2(1+j)^2}(q|\mathcal{I} |  \mathcal{J}|q)  +  \sum_{\langle ij\rangle\in\Gamma_\mathcal{K}}(q|\mathcal{I}|\mathcal{J}|q) \frac{(ij)}{(qi)(jq)} \left(  \tilde{\Phi}(\Gamma_\mathcal{I})+ \tilde{\Phi}(\Gamma_\mathcal{J})\right) 
   \bigg).
\end{align}
Now, to rewrite this in a nicer way we use the identity
\begin{align}
    (q|i|j|q) = - (q|j|i|q)
\end{align}
Thus, the recursion relation above can be written as
\begin{align}\label{phi-recursion}
    \tilde{\Phi}(\Gamma_\mathcal{K}) = \frac{1}{\Box}\bigg(\sum_{\langle ij\rangle \in \Gamma_\mathcal{K}} \frac{(q|j|i|q)}{(1+i)^2(1+j)^2}(q|\mathcal{I} |  \mathcal{J}|q)  +  \sum_{\langle ij\rangle\in\Gamma_\mathcal{K}}(q|\mathcal{I}|\mathcal{J}|q) \frac{(ij)}{(qi)(jq)} \left(  \tilde{\Phi}(\Gamma_\mathcal{I})+ \tilde{\Phi}(\Gamma_\mathcal{J})\right) 
   \bigg).
\end{align}

\subsection{$\tilde{\Phi}$ on two points and conjecture for $\tilde{\Phi}$}

We can immediately get the $p$-independent current $\tilde{\Phi}$ for a graph on two points from \eqref{phi-recursion}. Indeed, $\tilde{\Phi}$ for a graph on one point is zero by construction (as this only contains the $p$-dependent part). This means that according to \eqref{2-3 S-current}
\begin{align}\label{phi-23}
    \tilde{\Phi}(23) \equiv \phi_2(23) = \frac{(q|2|3|q)(q|3|2|q)}{(1+2)^2(1+3)^2(1+2+3)^2}.
\end{align}
The minus sign appearing in \eqref{phi-23} is absorbed into the spinor contraction above.
We have introduced a new notation, and denoted the contribution to $\Phi$ that depends on a pair of momenta as $\phi_2(23)$. This is in anticipation that $\Phi$ is given by a sum of contributions from individual vertices of $\mathcal{K}$ (this is the $p$-dependent part), then from pairs of vertices, and then progressively by contributions involving more and more vertices. In general, for an edge $\langle ij \rangle$, the formula of $\phi_2(ij)$ is given by
\begin{align}\label{phi-2-point-sol}
    \phi_2(ij) = \frac{(q|i|j|q)(q|j|i|q)}{(1+i)^2(1+j)^2(1+i+j)^2}.
\end{align}

Thus, we conjecture that $\tilde{\Phi}$ for an arbitrary graph is given by a sum of $\phi_2(ij)$ for all edges of the graph, plus higher order contributions
\begin{align}
    \tilde{\Phi}(\Gamma_\mathcal{K}) = \sum_{\langle ij\rangle \in \Gamma_\mathcal{K}} {\phi}_2(ij) + \ldots
\end{align}
We will prove this after we explicitly treat the example of a 3-point graph. Interestingly, we can already see that the first term in \eqref{phi-recursion} is almost identical to the solution of $\phi_2(ij)$ \eqref{phi-2-point-sol} with a modified spinor contraction and without the final propagator. This motivates us to massage the first term to look like the missing contribution for $\phi_2(ij)$ to complete the momentum square.

\subsection{Recursive relation for $\phi_2$ and higher}

Motivated by the previous considerations, we can conjecture that the $p$-independent object $\tilde{\Phi}(\Gamma_\mathcal{K})$ is given by the sum of quantities $\phi_i(\Gamma_\mathcal{I})$, for each subgraph $\Gamma_\mathcal{I}\subset \Gamma_\mathcal{K}$
\begin{align}\label{phi-structure}
    \tilde{\Phi}(\Gamma_\mathcal{K}) = \sum_{\langle ij\rangle\in \Gamma_\mathcal{K}} {\phi}_2(ij) + \sum_{\langle ijk\rangle\in \Gamma_\mathcal{K}} {\phi}_3(ijk) +\ldots =
    \sum_{\Gamma_\mathcal{I}\subset \Gamma_\mathcal{K}, |\mathcal{I}|=i} \phi_i(\Gamma_\mathcal{I}).
\end{align}

Substituting this into the recursion \eqref{phi-recursion}, and collecting terms in front of each ${\phi}_i(\Gamma_\mathcal{I})$, we get
\begin{align}\label{phi-recursion-rewrite}
    \tilde{\Phi}(\Gamma_\mathcal{K}) = \frac{1}{\Box}\bigg(\sum_{\braket{ij}\in \Gamma_\mathcal{K}} (q|\mathcal{I}|\mathcal{J}|q) \frac{(q|j|i|q)}{(1+i)^2(1+j)^2} + \sum_{\Gamma_\mathcal{I}\subset \Gamma_\mathcal{K},i = |\mathcal{I}|\geq2} {\phi}_i(\Gamma_\mathcal{I}) \sum_{\braket{mn}\in \Gamma_\mathcal{K}/\Gamma_\mathcal{I}} (q|\mathcal{M}|\mathcal{N}|q)\frac{(mn)}{(qm)(nq)}\bigg).
\end{align}
Here we collect the terms in the second sum based on the factors $\phi_i(\Gamma_\mathcal{I})$. The set of momenta $\mathcal{M}$ and $\mathcal{N}$ are the vertices of the graph $\Gamma_\mathcal{K}$ that are split by edge $\braket{mn}\in \Gamma_\mathcal{K}/\Gamma_\mathcal{I}$. Note that the sets of momenta that appear in $\mathcal{M}$ and $\mathcal{N}$ are those of 
$\Gamma_\mathcal{K}$, not of $\Gamma_\mathcal{K}/\Gamma_\mathcal{I}$.
We can see that the last sum is over the edges contained in $\Gamma_\mathcal{K}$ modulo edges in $\Gamma_\mathcal{I}$. This structure is identical for the Yang-Mills all plus one minus current calculation in the sense that the factor $\Phi(\Gamma_\mathcal{I})$ is multiplied with a factor that is almost the total momentum square $\Box$. As the sum goes over the bigger graph $\Gamma_\mathcal{I}$, the number of terms in the last sum decreases accordingly, similar to \eqref{YM-p-independent-1}. 

Similar to the YM case, we can make the ansatz \eqref{phi-structure} for $\tilde{\Phi}(\Gamma_\mathcal{K})$. Then the left hand side is given by the sum of $\phi_i$ for all subgraphs of $\Gamma_\mathcal{K}$, times the total momentum squared. On the left hand side, the second term is also given by the sum of $\phi_i$ for subgraphs, but the factor that multiplies this sum contains only some of the terms needed to get the total momentum squared. This means that the missing factors come from the first term. Which means that we must have
\begin{align}\label{relation-phi}
    \sum_{\braket{ij}\in \Gamma_\mathcal{K}} (q|\mathcal{I}|\mathcal{J}|q) \frac{(q|j|i|q)}{(1+i)^2(1+j)^2} = \sum_{\Gamma_\mathcal{I}\subset \Gamma_\mathcal{K},i = |\mathcal{I}|\geq2} {\phi}_i(\Gamma_\mathcal{I}) \sum_{\braket{mn}\in \Gamma_\mathcal{I}+1} (q|\mathcal{M}|\mathcal{N}|q)\frac{(mn)}{(qm)(nq)}.
\end{align}
Here the second sum is taken over the edges of the graph that we denoted as $\Gamma_\mathcal{I}+1$, which is the graph $\Gamma_\mathcal{I}$ with the additional vertex $1$ appended. It is important to remember that in this formula the sets $\mathcal{M}$ and $\mathcal{N}$ are those of 
$\Gamma_\mathcal{K}$, not of $\Gamma_\mathcal{I}$.
For example, applying this to $\mathcal{K}$ consisting of just momenta $2,3$ we immediately recover $\phi_2$ as given by \eqref{phi-2-point-sol}. The relation \eqref{relation-phi} is best illustrated by considering an example of a graph with 3 vertices. 

\subsection{Solution for the graph on three points}

We now compute the quantity $\phi_3$. This is done by considering a graph with 3 vertices. 
There are three possible tree graphs for 3 vertices

\begin{center}
    
\scalebox{0.5}{
\begin{tikzpicture}[
Vertex/.style={circle, draw=black, fill=white, very thick, minimum size=7mm},
Equation/.style={rectangle, draw=white, fill=white, very thick, minimum size=40},]

\node[Vertex] (11) {2};
\node[Vertex] (12) [above right=of 11]{3};
\node[Vertex] (13) [below right=of 12]{4};
\node[Equation] (01) [left = of 11]{\huge $\Gamma_\mathcal{K} ~~=~$};
\node[Equation] (02) [right = of 13]{\huge or};
\node[Vertex] (21) [right = of 02]{2};
\node[Vertex] (22) [above right=of 21]{3};
\node[Vertex] (23) [below right=of 22]{4};

\node[Equation] (03) [right = of 23]{\huge or};
\node[Vertex] (31) [right = of 03]{2};
\node[Vertex] (32) [above right=of 31]{3};
\node[Vertex] (33) [below right=of 32]{4};

\draw[-,very thick](11.north east) to (12.south west);
\draw[-,very thick](12.south east) to (13.north west);

\draw[-,very thick](21.north east) to (22.south west);
\draw[-,very thick](21.east) to (23.west);

\draw[-,very thick](31.east) to (33.west);
\draw[-,very thick](32.south east) to (33.north west);

\end{tikzpicture}
}
\end{center}

We now use relation \eqref{relation-phi} for the first tree, namely $2-3-4$. We have
\begin{align}\label{2-3}
   & \frac{(q|3|2|q)(q|2|34|q)}{(1+2)^2(1+3)^2}  +  \frac{(q|4|3|q)(q|23|4|q)}{(1+3)^2(1+4)^2} 
   = 
   \\ \nonumber
   & {\phi}_2(23) \Big((q|1|234|q) \frac{(12)}{(q1)(2q)} 
    + (q|12|34|q) \frac{(23)}{(q2)(3q)} \Big)  
    \\ \nonumber
     + & {\phi}_2(34) \Big((q|23|41|q) \frac{(34)}{(q3)(4q)} + (q|234|1|q) \frac{(41)}{(q4)(1q)} \Big)
     \\ \nonumber
     + & \phi_3(234) (1+2+3+4)^2.
\end{align}
There is a choice of how to interpret the $\Gamma_{\mathcal{I}}+1$ graph prescription. The ambiguity is where the node $\{1\}$ is placed with respect to the graph $\Gamma_{\mathcal{I}}$. All such choices are equivalent, because one can decompose the momentum squared for a set of nodes in many different ways, the ambuguity being which graph one decides to associate with this set of nodes. We have made a convenient choice in the third line of \eqref{2-3}. In that case the graph $\Gamma_{\mathcal{I}}$ is $3-4$, and we have placed the node $\{1\}$ next to the node $\{4\}$ so that 
$\Gamma_{\mathcal{I}}+1 = 3-4-1$. 

The formula \eqref{2-3} gives the answer for $\phi_3$ since we know $\phi_2$. We just need to simplify the arising expression. To proceed, we note that the left-hand side of \eqref{2-3} can be written as
\begin{align}\label{23-comp-1}
\phi_2(23) (1+2+3)^2 + \phi_2(34) (1+3+4)^2 \\ \nonumber
    +\frac{(q|3|2|q)(q|2|4|q)}{(1+2)^2(1+3)^2}  
    +\frac{(q|4|3|q)(q|2|4|q)}{(1+3)^2(1+4)^2}. 
\end{align}
The right-hand side can also be rewritten in a similar way, namely
\begin{align}\label{23-comp-2}
    & {\phi}_2(23)(1+2+3)^2 + {\phi}_2(23) \Big((q|1|4|q) \frac{(12)}{(q1)(2q)} 
    + (q|12|4|q) \frac{(23)}{(q2)(3q)} \Big)  
    \\ \nonumber
     + & {\phi}_2(34)(1+3+4)^2 + {\phi}_2(34) \Big((q|2|41|q) \frac{(34)}{(q3)(4q)} + (q|2|1|q) \frac{(41)}{(q4)(1q)} \Big)
     \\ \nonumber
     + & \phi_3(234) (1+2+3+4)^2.
\end{align}
Canceling the same factors from both sides we get a formula for $\phi_3(234)$. Let us massage the arising expression to put it into the most useful form. 

Let us first consider the terms related to $\phi_2(23)$, namely the first term in the second line of \eqref{23-comp-1} and the second term in the first line of \eqref{23-comp-2}. Multiplying-dividing the term from \eqref{23-comp-1} by $(1+2+3)^2$ and expanding this, we get for these terms
\begin{align}\label{23-comp-3}
    \frac{(q|3|2|q)}{(1+2)^2(1+3)^2(1+2+3)^2}\Big[ (q|2|4|q) \left( (q|1|23|q) \frac{(12)}{(q1)(2q)} 
    + (q|12|3|q) \frac{(23)}{(q2)(3q)}\right)\\
    \nonumber
    -(q|2|3|q) \Big((q|1|4|q) \frac{(12)}{(q1)(2q)} 
    + (q|12|4|q) \frac{(23)}{(q2)(3q)} \Big)\Big].
\end{align}
We then note that the last terms in each brackets simply cancel each other, while the first term can be combined using Schouten identity to produce
\begin{align}
    (q|2|4|q) (q|1|23|q)- (q|2|23|q)(q|1|4|q)=
    (q|1|2|q) (q|23|4|q) .
\end{align}
Using $(q|1|2|q) (12)/(q1)(2q) = (1+2)^2$ shows that \eqref{23-comp-3} equals
\begin{align}
    \frac{(q|3|2|q)(q|23|4|q)}{(1+3)^2(1+2+3)^2}.
\end{align}
Analogous manipulations for the other terms give the answer
\begin{align}
     {\phi}_3(234)(1+2+3+4)^2 = \frac{(q|3|2|q)(q|23|4|q)}{(1+3)^2(1+2+3)^2}+\frac{(q|4|3|q)(q|2|34|q)}{(1+3)^2(1+3+4)^2}.
 \end{align}
Alternatively, we can write
 \begin{align}
     {\phi}_3(234) = {\phi}([2[3]]4)+{\phi}(2[[3]4]),
 \end{align}
    and
 \begin{align}\label{phi-234}
  {\phi}([2[3]]4)=  \frac{(q|2|3|q)(q|4|23|q)}{(1+3)^2(1+2+3)^2 (1+2+3+4)^2}, \\ \nonumber
  {\phi}(2[[3]4])=\frac{(q|4|3|q)(q|2|34|q)}{(1+3)^2(1+3+4)^2(1+2+3+4)^2}.
\end{align}
This is a sum of contributions of the type ${\phi}_2(ij)$ for every edge of the graph, as well as contributions that we have denoted as ${\phi}([i[j]]k)$ and ${\phi}(i[[j]k])$. The meaning of this notation is that $[i[j]]k$ is obtained by first taking the vertex $j$, then appending $i$ to it, and finally appending the momentum $k$. This corresponds to the expression
\begin{align}
   {\phi}([i[j]]k) =  \frac{(q|i|j|q)(q|ij|k|q)}{(1+j)^2(1+i+j)^2 (1+i+j+k)^2}.
\end{align}
Similarly, 
\begin{align}
{\phi}(i[[j]k])=\frac{(q|j|k|q)(q|i|jk|q)}{(1+j)^2(1+j+k)^2(1+i+j+k)^2},
\end{align}
so that
\begin{align}\label{phi-ijk}
    {\phi}_3(ijk) = {\phi}([i[j]]k) +{\phi}(i[[j]k]).
\end{align}

\subsection{Recursive relation for $\phi_3$ and higher}

We can massage each term in the left-hand side of \eqref{relation-phi} in a suggestive way, generalising what happened in the example of the graph with 3 vertices. Consider
\begin{align}
    (q|\mathcal{I}|\mathcal{J}|q)\frac{(q|j|i|q)}{(1+i)^2 (1+j)^2}.
\end{align}
The first step is to rewrite this as $\phi_2(ij)(1+i+j)^2$ plus a contribution. We have
\begin{align}\label{ij-left}
    (q|\mathcal{I}|\mathcal{J}|q)\frac{(q|j|i|q)}{(1+i)^2 (1+j)^2} = &\phi_2(ij) (1+i+j)^2 
    \\ \nonumber
    +& \frac{(q|j|i|q)}{(1+i)^2 (1+j)^2} \Big( 
(q|\mathcal{I}/i|j|q) + (q|i|\mathcal{J}/j|q) + (q|\mathcal{I}/i|\mathcal{J}/j|q)\Big).    
\end{align}
On the other hand, we can also rewrite the $\phi_2$-containing terms on the right-hand side of \eqref{relation-phi}. Such terms become
\begin{align}\label{ij-right}
    \sum_{\langle ij\rangle \in \Gamma_{\mathcal{K}}} \phi_2(ij) \left(
    (q|1| \mathcal{K}|q)\frac{(1i)}{(q1)(iq)} 
+(q|\mathcal{I}|\mathcal{J}|q)\frac{(ij)}{(qi)(jq)}\right) = \sum_{\langle ij\rangle \in \Gamma_{\mathcal{K}}} \phi_2(ij) (1+i+j)^2 
\\ \nonumber
+ \sum_{\langle ij\rangle \in \Gamma_{\mathcal{K}}} \phi_2(ij) \Big(
    (q|1| \mathcal{I}/i +\mathcal{J}/j |q)\frac{(1i)}{(q1)(iq)} 
\\ \nonumber    
+(q|\mathcal{I}/i|j|q)\frac{(ij)}{(qi)(jq)}+(q|i|\mathcal{J}/j|q)\frac{(ij)}{(qi)(jq)}+(q|\mathcal{I}/i|\mathcal{J}/j|q)\frac{(ij)}{(qi)(jq)}\Big) 
\end{align}
Note that in the terms in the second line it does not matter whether we write $(1i)/(q1)(iq)$ or $(1j)/(q1)(jq)$, because when $q\to 1$ both coincide. 

We now consider the difference of $\phi_2(ij)$-related terms on both sides. Multiplying and dividing the terms in \eqref{ij-left} by $(1+i+j)^2$, and expanding this in the numerator we get
\begin{align}
    \frac{(q|j|i|q)}{(1+i)^2 (1+j)^2(1+i+j)^2} \Big( 
(q|\mathcal{I}/i|j|q) + (q|i|\mathcal{J}/j|q) + (q|\mathcal{I}/i|\mathcal{J}/j|q)\Big)\times
\\ \nonumber
\left( (q|1|ij|q) \frac{(1i)}{(q1)(iq)} + (q|i|j|q) \frac{(ij)}{(qi)(jq)}\right)
\\ \nonumber
- \frac{(q|j|i|q)}{(1+i)^2 (1+j)^2(1+i+j)^2} (q|i|j|q) \Big(
    (q|1| \mathcal{I}/i +\mathcal{J}/j |q)\frac{(1i)}{(q1)(iq)} 
\\ \nonumber    
+(q|\mathcal{I}/i|j|q)\frac{(ij)}{(qi)(jq)}+(q|i|\mathcal{J}/j|q)\frac{(ij)}{(qi)(jq)}+(q|\mathcal{I}/i|\mathcal{J}/j|q)\frac{(ij)}{(qi)(jq)}\Big) 
\end{align}
We immediately note that all $(ij)/(qi)(jq)$ containing terms cancel. The terms containing $(1i)/(q1)(iq)$ can be simplified as follows. We have
\begin{align}\nonumber
    (q|i|j|q) (q|1| \mathcal{I}/i +\mathcal{J}/j |q)= (q|1|i|q) (q|j |\mathcal{I}/i +\mathcal{J}/j|q) - (q|1|j|q) (q|i |\mathcal{I}/i +\mathcal{J}/j|q).
\end{align}
On the other hand, we have
\begin{align}\nonumber
    (q|1|ij|q) \Big( 
(q|\mathcal{I}/i|j|q) + (q|i|\mathcal{J}/j|q) + (q|\mathcal{I}/i|\mathcal{J}/j|q)\Big)=
\\ \nonumber
(q|1|i|q) \Big( 
(q|\mathcal{I}/i|j|q) + (q|i|\mathcal{J}/j|q) + (q|\mathcal{I}/i|\mathcal{J}/j|q)\Big)
\\ \nonumber
+(q|1|j|q) \Big( 
(q|\mathcal{I}/i|j|q) + (q|i|\mathcal{J}/j|q) + (q|\mathcal{I}/i|\mathcal{J}/j|q)\Big).
\end{align}
This gives, after some simple algebra
\begin{align}\nonumber
    (q|1|ij|q) \Big( 
(q|\mathcal{I}/i|j|q) + (q|i|\mathcal{J}/j|q) + (q|\mathcal{I}/i|\mathcal{J}/j|q)\Big)-(q|i|j|q) (q|1| \mathcal{I}/i +\mathcal{J}/j |q)= 
\\ \nonumber
(q|1|i|q)  (q|\mathcal{I}+j|\mathcal{J}/j|q)
+(q|1|j|q)  (q|\mathcal{I}/i|\mathcal{J}+j|q).
\end{align}
This multiplied by $(1i)/(q1)(iq)$ (or by $(1j)/(q1)(jq)$) produces
\begin{align}
    (1+i)^2 (q|\mathcal{I}+j|\mathcal{J}/j|q)
+(1+j)^2 (q|\mathcal{I}/i|\mathcal{J}+j|q).
\end{align}

Substituting this into our recursion eliminates all the $\phi_2$-containing terms, with the resulting 
relation for functions $\phi_3$ and higher being
\begin{align}\label{relation-phi-3}
    \sum_{\braket{ij}\in \Gamma_\mathcal{K}} \frac{(q|i|j|q)(q|\mathcal{J}/\{j\}|\mathcal{I}+j|q)}{(1+j)^2 (1+i+j)^2}
    + \frac{(q|j|i|q)(q|\mathcal{I}/\{i\}|\mathcal{J}+i|q)}{(1+i)^2 (1+i+j)^2} = 
    \\ \nonumber
    \sum_{\Gamma_\mathcal{I}\subset \Gamma_\mathcal{K},i = |\mathcal{I}|\geq3} {\phi}_i(\Gamma_\mathcal{I}) \sum_{\braket{mn}\in \Gamma_\mathcal{I}+1} (q|\mathcal{M}|\mathcal{N}|q)\frac{(mn)}{(qm)(nq)}.
\end{align}
Applying it to a graph with 3 vertices immediately recovers our solution \eqref{phi-ijk} for $\phi_3$. 
And we can use \eqref{relation-phi-3} to compute $\phi_4$. 

\subsection{Graphical notation}

All our considerations can be illustrated by a graphical notation. Our ansatz for the current $\tilde{\Phi}(\Gamma_\mathcal{K})$ for an arbitrary graph $\Gamma_\mathcal{K}$ is that it is given by a sum over all possible subgraphs $\Gamma_\mathcal{I}\subset \Gamma_\mathcal{K}$. If we denote the irreducible part of $\tilde{\Phi}(\Gamma_\mathcal{K})$ by a graph with diamond vertices then graphically, for a graph on 3 vertices, we would draw
\begin{center}
    
\scalebox{0.5}{
\begin{tikzpicture}[
Diam/.style={diamond, draw=black, fill=white, very thick, minimum size=5mm},
Vertex/.style={rectangle, draw=black, fill=white, very thick, minimum size=5mm},
Equation/.style={rectangle, draw=white, fill=white, very thick, minimum size=20},]

\node[Equation] (01) {\huge $\tilde{\Phi}(2-3-4) ~~=~$};

\node[Diam] (21) [right=of 01]{2};
\node[Diam] (22) [right=of 21]{3};

\node[Equation] (E1) at (7.3,0) {\Large $\oplus$};
\node[Equation] (E2) at (11.1,0) {\Large $\oplus$};

\node[Diam] (31) [right=of 22]{3};
\node[Diam] (32) [right=of 31]{4};

\node[Diam] (41) [right=of 32]{2};
\node[Diam] (42) [right=of 41]{3};
\node[Diam] (43) [right=of 42]{4};

\draw[-,very thick](21.east) to (22.west);

\draw[-,very thick](31.east) to (32.west);

\draw[-,very thick](41.east) to (42.west);
\draw[-,very thick](42.east) to (43.west);

\end{tikzpicture}
}
\end{center}
where each term in the sum is given by
\begin{center}
    
\scalebox{0.5}{
\begin{tikzpicture}[
Diam/.style={diamond, draw=black, fill=white, very thick, minimum size=5mm},
Vertex/.style={rectangle, draw=black, fill=white, very thick, minimum size=5mm},
Equation/.style={rectangle, draw=white, fill=white, very thick, minimum size=20},]

\node[Diam] (21) {2};
\node[Diam] (22) [right=of 21]{3};
\node[Equation] (E1) at (6,0) {\huge $=~~ \frac{(q|2|3|q)(q|3|2|q)}{(1+2)^2(1+3)^2(1+2+3)^2}$};

\node[Diam] (31) at (11,0){3};
\node[Diam] (32) [right=of 31]{4};
\node[Equation] (E2) at (18,0) {\huge $=~~ \frac{(q|3|4|q)(q|4|3|q)}{(1+3)^2(1+4)^2(1+3+4)^2}$};

\draw[-,very thick](21.east) to (22.west);
\draw[-,very thick](31.east) to (32.west);

\end{tikzpicture}
}
\end{center}
and 
\begin{center}
    
\scalebox{0.5}{
\begin{tikzpicture}[
Diam/.style={diamond, draw=black, fill=white, very thick, minimum size=5mm},
Vertex/.style={rectangle, draw=black, fill=white, very thick, minimum size=5mm},
Equation/.style={rectangle, draw=white, fill=white, very thick, minimum size=20},]

\node[Diam] (21) {2};
\node[Diam] (22) [right=of 21]{3};
\node[Diam] (23)[right = of 22]{4};
\node[Equation] (E1) at (13,0) {\huge $=~~ \frac{(q|2|3|q)(q|4|23|q)}{(1+3)^2(1+2+3)^2 +(1+2+3+4)^2} + 
    \frac{(q|4|3|q)(q|2|34|q)}{(1+3)^2(1+3+4)^2(1+2+3+4)^2}$};

\draw[-,very thick](21.east) to (22.west);
\draw[-,very thick](22.east) to (23.west);

\end{tikzpicture}
}
\end{center}

Similarly, for $\tilde{\Phi}(2-3-4-5)$, the answer would be given by 
\begin{center}
    
\scalebox{0.5}{
\begin{tikzpicture}[
Diam/.style={diamond, draw=black, fill=white, very thick, minimum size=5mm},
Vertex/.style={rectangle, draw=black, fill=white, very thick, minimum size=5mm},
Equation/.style={rectangle, draw=white, fill=white, very thick, minimum size=20},]
\node[Diam] (11) {2};
\node[Diam] (12) [right = of 11]{3};
\draw[-,very thick](12.west) to (11.east);
\node[Equation] (E1) [right = of 12] {\Large$\oplus$};

\node[Diam] (21) [right=of E1]{3};
\node[Diam] (22) [right = of 21]{4};
\draw[-,very thick](22.west) to (21.east);
\node[Equation] (E2) [right = of 22] {\Large$\oplus$};

\node[Diam] (31) [right = of E2]{4};
\node[Diam] (32) [right = of 31]{5};
\draw[-,very thick](32.west) to (31.east);

\node[Diam] (41) [below right=of 11]{2};
\node[Diam] (42) [right = of 41]{3};
\node[Diam] (43) [right = of 42]{4};
\draw[-,very thick](42.west) to (41.east);
\draw[-,very thick](43.west) to (42.east);
\node[Equation] (E3) [right = of 43] {\Large$\oplus$};

\node[Diam] (51) [right=of E3]{3};
\node[Diam] (52) [right = of 51]{4};
\node[Diam] (53) [right = of 52]{5};
\draw[-,very thick](52.west) to (51.east);
\draw[-,very thick](53.west) to (52.east);

\node[Diam] (61) [below right=of 42]{2};
\node[Diam] (62) [right = of 61]{3};
\node[Diam] (63) [right = of 62]{4};
\node[Diam] (64) [right = of 63]{5};
\draw[-,very thick](62.west) to (61.east);
\draw[-,very thick](63.west) to (62.east);
\draw[-,very thick](64.west) to (63.east);

\end{tikzpicture}
}
\end{center}

\subsection{Computation of $\phi_4$ - Linear graph}

We now use \eqref{relation-phi-3} to compute $\phi_4$. We do it for two different possible graph topologies. We first consider a linear graph that contains no vertices of valency more than two. 

The left-hand side of \eqref{relation-phi-3}, for the graph $2-3-4-5$, becomes a sum of 4 terms
\begin{align}\label{4pt-lhs}
    \frac{(q|2|3|q)(q|45|23|q)}{(1+3)^2(1+2+3)^2} 
    +\frac{(q|4|3|q)(q|2|345|q)}{(1+3)^2(1+3+4)^2} 
    + \frac{(q|3|4|q)(q|5|234|q)}{(1+4)^2(1+3+4)^2} 
    +\frac{(q|5|4|q)(q|23|45|q)}{(1+4)^2(1+4+5)^2}. 
\end{align}
On the other hand, the right-hand side is
\begin{align}\label{4pt-rhs}
    &\phi_3(234) \Big( (q|1|2345|q) \frac{(12)}{(q1)(2q)}+(q|12|345|q) \frac{(23)}{(q2)(3q)} +(q|123|45|q) \frac{(34)}{(q3)(4q)}\Big) \\ \nonumber
    +& \phi_3(345) \Big((q|23|451|q) \frac{(34)}{(q3)(4q)}+(q|234|51|q) \frac{(45)}{(q4)(5q)}+(q|2345|1|q) \frac{(51)}{(q5)(1q)} \Big)
    \\ \nonumber
    +   &\phi_4(2345) (1+2+3+4+5)^2.
\end{align}
Again, there is a choice of how to interpret the $\Gamma_{\mathcal{I}}+1$ graph in that the node $\{1\}$ can be be connected to any node of $\Gamma_{\mathcal{I}}$. We have chosen to connect the node $\{1\}$ to the node $\{5\}$ in the bracket multiplying $\phi_3(345)$, as this gives the most convenient set of terms for later manipulations. 

The expressions for $\phi_3(234)$ and $\phi_3(345)$ are
\begin{align}
    \phi_3(234) = \frac{(q|2|3|q)(q|4|23|q)}{(1+3)^2(1+2+3)^2 (1+2+3+4)^2}+ \frac{(q|4|3|q)(q|2|34|q)}{(1+3)^2(1+3+4)^2(1+2+3+4)^2}, \\ \nonumber
    \phi_3(345) = \frac{(q|3|4|q)(q|5|34|q)}{(1+4)^2(1+3+4)^2 (1+3+4+5)^2}+ \frac{(q|5|4|q)(q|3|45|q)}{(1+4)^2(1+4+5)^2(1+3+4+5)^2}.
\end{align}
We can note the the terms in \eqref{4pt-lhs} are very closely related to $\phi_3(234)$ and $\phi_3(345)$, with the first two terms in \eqref{4pt-lhs} being basically $\phi_3(234) (1+2+3+4)^2$, with the only difference that the $q$-factors in the numerators contains an additional dependence on the momentum $5$. Similarly, the last two terms in \eqref{4pt-lhs} are basically $\phi_3(345) (1+3+4+5)^2$, again with some additional dependence on the momentum $2$ in the numerators. The terms in \eqref{4pt-rhs} are similarly multiples of $\phi_3(234)$ and $\phi_3(345)$ by terms that want to complete to $(1+2+3+4)^2$ and $(1+3+4+5)^2$ apart from the fact that there is dependence on an additional momentum.
This means that we can rewrite \eqref{4pt-lhs} as
\begin{align}\nonumber
    &\phi_3(234) (1+2+3+4)^2 + \phi_3(345) (1+3+4+5)^2 \\ \nonumber 
    +&\frac{(q|2|3|q)(q|5|23|q)}{(1+3)^2(1+2+3)^2} 
    +\frac{(q|4|3|q)(q|2|5|q)}{(1+3)^2(1+3+4)^2} 
    + \frac{(q|3|4|q)(q|5|2|q)}{(1+4)^2(1+3+4)^2} 
    +\frac{(q|5|4|q)(q|2|45|q)}{(1+4)^2(1+4+5)^2}.
\end{align}
The right-hand side can be rewritten as
\begin{align}\label{4pt-rhs}
&\phi_3(234) (1+2+3+4)^2 + \phi_3(345) (1+3+4+5)^2 \\ \nonumber 
    + &\phi_3(234) \Big( (q|1|5|q) \frac{(12)}{(q1)(2q)}+(q|12|5|q) \frac{(23)}{(q2)(3q)} +(q|123|5|q) \frac{(34)}{(q3)(4q)}\Big) \\ \nonumber
    +& \phi_3(345) \Big((q|2|451|q) \frac{(34)}{(q3)(4q)}+(q|2|51|q) \frac{(45)}{(q4)(5q)}+(q|2|1|q) \frac{(51)}{(q5)(1q)} \Big)
    \\ \nonumber
    +   &\phi_4(2345) (1+2+3+4+5)^2.
\end{align}

Substituting the formula for $\phi_3$ we get
\begin{align}\nonumber
    \phi_4(2345) (1+2+3+4+5)^2 =
    \\ \nonumber
    \frac{(q|2|3|q)(q|5|23|q)}{(1+3)^2(1+2+3)^2} 
    +\frac{(q|4|3|q)(q|2|5|q)}{(1+3)^2(1+3+4)^2}
    + \frac{(q|3|4|q)(q|5|2|q)}{(1+4)^2(1+3+4)^2} 
    +\frac{(q|5|4|q)(q|2|45|q)}{(1+4)^2(1+4+5)^2}
    \\ \nonumber
    - \left(\frac{(q|2|3|q)(q|4|23|q)}{(1+3)^2(1+2+3)^2 (1+2+3+4)^2}+ \frac{(q|4|3|q)(q|2|34|q)}{(1+3)^2(1+3+4)^2(1+2+3+4)^2}\right) \times
    \\ \nonumber
    \Big( (q|1|5|q) \frac{(12)}{(q1)(2q)}+(q|12|5|q) \frac{(23)}{(q2)(3q)} +(q|123|5|q) \frac{(34)}{(q3)(4q)}\Big) \\ \nonumber
    - \left(\frac{(q|3|4|q)(q|5|34|q)}{(1+4)^2(1+3+4)^2 (1+3+4+5)^2}+ \frac{(q|5|4|q)(q|3|45|q)}{(1+4)^2(1+4+5)^2(1+3+4+5)^2}\right)\times
    \\ \nonumber
    \Big((q|2|451|q) \frac{(34)}{(q3)(4q)}+(q|2|51|q) \frac{(45)}{(q4)(5q)}+(q|2|1|q) \frac{(51)}{(q5)(1q)} \Big).
\end{align}
It should also be kept in mind that the terms in the third and the fifth line could be written differently. 

Collecting the terms with same propagators we get
\begin{align}\label{4pt-calc}
    &\phi_4(2345) (1+2+3+4+5)^2 =
    \\ \nonumber
    &\frac{(q|2|3|q)}{(1+3)^2(1+2+3)^2(1+2+3+4)^2} \Big( (q|5|23|q)(1+2+3+4)^2 \\ \nonumber
    - &(q|4|23|q)\Big( (q|1|5|q) \frac{(12)}{(q1)(2q)}+(q|12|5|q) \frac{(23)}{(q2)(3q)} +(q|123|5|q) \frac{(34)}{(q3)(4q)}\Big)
    \Big)
    \\ \nonumber
    +&\frac{(q|4|3|q)}{(1+3)^2(1+3+4)^2(1+2+3+4)^2} 
    \Big( (q|2|5|q) (1+2+3+4)^2 \\ \nonumber
    -&(q|2|34|q)
    \Big( (q|1|5|q) \frac{(12)}{(q1)(2q)}+(q|12|5|q) \frac{(23)}{(q2)(3q)} +(q|123|5|q) \frac{(34)}{(q3)(4q)}\Big)
    \Big) 
    \\ \nonumber
    + 
    &\frac{(q|3|4|q)}{(1+4)^2(1+3+4)^2(1+3+4+5)^2} 
    \Big( (q|5|2|q)(1+3+4+5)^2 \\ \nonumber
    -&(q|5|34|q)\Big((q|2|451|q) \frac{(34)}{(q3)(4q)}+(q|2|51|q) \frac{(45)}{(q4)(5q)}+(q|2|1|q) \frac{(51)}{(q5)(1q)} \Big)
    \Big) 
    \\ \nonumber  
    +&\frac{(q|5|4|q)}{(1+4)^2(1+4+5)^2(1+3+4+5)^2}
    \Big( (q|2|45|q) (1+3+4+5)^2 \\ \nonumber
    -&(q|3|45|q)
    \Big((q|2|451|q) \frac{(34)}{(q3)(4q)}+(q|2|51|q) \frac{(45)}{(q4)(5q)}+(q|2|1|q) \frac{(51)}{(q5)(1q)} \Big).
    \Big) 
\end{align}

We now analyse the brackets. For the first bracket, replacing the factor of $(1+2+3+4)^2$ by its expression from \eqref{YM Identity}, we have
\begin{align}
    (q|5|23|q) \Big( (q|1|234|q) \frac{(12)}{(q1)(2q)}+(q|12|34|q) \frac{(23)}{(q2)(3q)} +(q|123|4|q) \frac{(34)}{(q3)(4q)} \Big) 
    \\ \nonumber
    - (q|4|23|q)\Big( (q|1|5|q) \frac{(12)}{(q1)(2q)}+(q|12|5|q) \frac{(23)}{(q2)(3q)} +(q|123|5|q) \frac{(34)}{(q3)(4q)}\Big).
\end{align}
The last terms in each line cancel each other. The terms proportional to $(23)$ are
\begin{align}\nonumber
    (q|5|23|q)(q|2|34|q)- (q|4|23|q)(q|2|5|q)=
    \\ \nonumber
(q|5|23|q)(q|2|34|q) - (q|5|23|q)(q|2|4|q) + 
      (q|5|4|q)(q|2|23|q)= 
      \\ \nonumber
      (q|5|234|q)(q|2|3|q).
\end{align}
where we have used the Schouten identity. Similarly, the terms proportional to $(12)$ are
\begin{align}\nonumber
    (q|5|23|q) (q|1|234|q)- (q|4|23|q)(q|1|5|q) = 
    \\ \nonumber
(q|5|23|q)(q|1|234|q) - (q|5|23|q)(q|1|4|q) + 
      (q|5|4|q)(q|1|23|q)= 
      \\ \nonumber
    (q|5|234|q) (q|1|23|q). 
\end{align}
This means that the terms in the first bracket simplify to
\begin{align}
    (q|5|234|q) \left( (q|1|23|q)\frac{(12)}{(q1)(2q)}+ (q|2|23|q)\frac{(23)}{(q2)(3q)}\right) =(q|5|234|q) (1+2+3)^2. 
\end{align}

For the second bracket we have
\begin{align}
(q|2|5|q) \Big( (q|1|234|q) \frac{(12)}{(q1)(2q)}+(q|12|34|q) \frac{(23)}{(q2)(3q)} +(q|123|4|q) \frac{(34)}{(q3)(4q)} \Big) 
\\ \nonumber
    -(q|2|34|q)
    \Big( (q|1|5|q) \frac{(12)}{(q1)(2q)}+(q|12|5|q) \frac{(23)}{(q2)(3q)} +(q|123|5|q) \frac{(34)}{(q3)(4q)}\Big)
    \end{align}
Now the second terms in each line cancel, and the rest gives
\begin{align}
   (q|234|5|q)\left( (q|1|2|q) \frac{(13)}{(q1)(3q)}
    + (q|3|2|q)  \frac{(34)}{(q3)(4q)}\right). 
\end{align}
We have replaced $(12)/(q1)(2q)$ with $(13)/(q1)(3q)$, which is legitimate given $q\to 1$.
This means that the sum of the first two terms in \eqref{4pt-calc} can be written as
\begin{align}\label{4pt-calc-1}
    \frac{(q|2|3|q)(q|5|234|q)}{(1+3)^2(1+3+4)^2(1+2+3+4)^2} \left( (q|1|34|q) \frac{(13)}{(q1)(3q)}
    + (q|3|4|q)  \frac{(34)}{(q3)(4q)}\right) 
    \\ \nonumber
    + \frac{(q|3|4|q)(q|5|234|q)}{(1+3)^2(1+3+4)^2(1+2+3+4)^2}\left( (q|1|2|q) \frac{(13)}{(q1)(3q)}
    + (q|3|2|q)  \frac{(34)}{(q3)(4q)}\right).
\end{align}
The last terms cancel, while the $(13)$ terms combine
\begin{align}
(q|2|3|q)(q|1|34|q) + (q|3|34|q)(q|1|2|q)= (q|1|3|q) (q|2|34|q).    
\end{align}
This means that the $(1+3)^2$ from the denominator gets canceled, and the terms in \eqref{4pt-calc-1} give
\begin{align}
    \frac{(q|2|34|q)(q|5|234|q)}{(1+3+4)^2(1+2+3+4)^2}.
\end{align}
The other remaining terms from \eqref{4pt-calc} are analyzed similarly, and the final result for $\phi_4$ is 
\begin{align}
  \phi_4(2345)(1+2+3+4+5)^2 =  \frac{(q|2|34|q)(q|5|234|q)}{(1+3+4)^2(1+2+3+4)^2} + \frac{(q|5|34|q)(q|2|345|q)}{(1+3+4)^2(1+3+4+5)^2}
\end{align}
This is a sum of two contributions, which can be understood as
\begin{align}
    {\phi}( [2[34]]5) + {\phi}( 2[[34]5]) ,
\end{align}
where the grouping in the brackets denotes the order in which the end nodes $2,5$ are added to the "core" $34$. 

\subsection{Computation of $\phi_4$ - Graph with a cubic vertex}
\label{sec:t-graph}

For a graph on 4 vertices we, for the first time, get the possibility of topology as shown

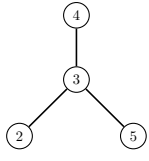
\begin{center}
\scalebox{0.5}{
\begin{tikzpicture}[
Circ/.style={circle, draw=black, fill=white, thick, minimum size=3mm},]
\node[Circ] (13) {3};
\node[Circ] (12) [below left=of 13]{2};
\node[Circ] (15) [below right=of 13]{5};
\node[Circ] (14) [above=of 13]{4};

\draw[-,very thick](12.north east) to (13.south west);
\draw[-,very thick](15.north west) to (13.south east);
\draw[-,very thick](14.south) to (13.north);
\end{tikzpicture}}
\end{center}
For this graph the sum of its subgraphs is, using the graphical notation 
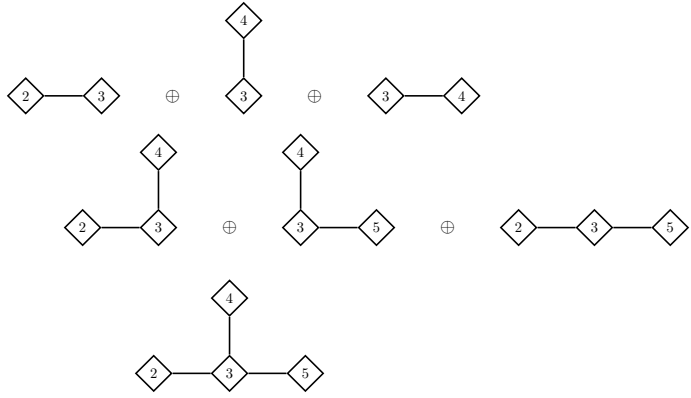
\begin{center}
\scalebox{0.5}{
\begin{tikzpicture}[
Diam/.style={diamond, draw=black, fill=white, very thick, minimum size=5mm},
Vertex/.style={rectangle, draw=black, fill=white, very thick, minimum size=5mm},
Equation/.style={rectangle, draw=white, fill=white, very thick, minimum size=20},]
\node[Diam] (11) {2};
\node[Diam] (12) [right = of 11]{3};
\draw[-,very thick](12.west) to (11.east);
\node[Equation] (E1) [right = of 12] {\Large$\oplus$};

\node[Diam] (21) [right=of E1]{3};
\node[Diam] (22) [above = of 21]{4};
\draw[-,very thick](22.south) to (21.north);
\node[Equation] (E2) [right = of 21] {\Large$\oplus$};

\node[Diam] (31) [right = of E2]{3};
\node[Diam] (32) [right = of 31]{4};
\draw[-,very thick](32.west) to (31.east);

\node[Diam] (43) [below right=of 12]{4};
\node[Diam] (42) [below = of 43]{3};
\node[Diam] (41) [left = of 42]{2};
\draw[-,very thick](42.west) to (41.east);
\draw[-,very thick](43.south) to (42.north);
\node[Equation] (E3) [right = of 42] {\Large$\oplus$};

\node[Diam] (51) [right=of E3]{3};
\node[Diam] (52) [above = of 51]{4};
\node[Diam] (53) [right = of 51]{5};
\draw[-,very thick](52.south) to (51.north);
\draw[-,very thick](53.west) to (51.east);
\node[Equation] (E4) [right = of 53] {\Large$\oplus$};

\node[Diam] (61) [right=of E4]{2};
\node[Diam] (62) [right = of 61]{3};
\node[Diam] (63) [right = of 62]{5};
\draw[-,very thick](61.east) to (62.west);
\draw[-,very thick](62.east) to (63.west);

\node[Diam] (63) [below =of E3]{4};
\node[Diam] (62) [below = of 63]{3};
\node[Diam] (64) [right = of 62]{5};
\node[Diam] (61) [left = of 62]{2};
\draw[-,very thick](62.west) to (61.east);
\draw[-,very thick](64.west) to (62.east);
\draw[-,very thick](63.south) to (62.north);

\end{tikzpicture}
}
\end{center}

The left-hand side of \eqref{relation-phi-3} for this graph is a sum of just 3 terms
\begin{align}\label{4pt-t-lhs}
     \frac{(q|2|3|q)(q|45|23|q)}{(1+3)^2(1+2+3)^2} 
    +\frac{(q|4|3|q)(q|25|34|q)}{(1+3)^2(1+3+4)^2} 
    + \frac{(q|5|3|q)(q|24|35|q)}{(1+3)^2(1+3+5)^2}.
\end{align}
On the other hand we have
\begin{align}\nonumber
    \phi_3(234) (1+2+3+4)^2 + \phi_3(435)(1+3+4+5)^2+\phi_3(235)(1+2+3+5)^2 =
    \\ \nonumber
    \frac{(q|2|3|q)(q|4|23|q)}{(1+3)^2(1+2+3)^2}+ \frac{(q|4|3|q)(q|2|34|q)}{(1+3)^2(1+3+4)^2}+ \frac{(q|4|3|q)(q|5|34|q)}{(1+3)^2(1+3+4)^2}+ \frac{(q|5|3|q)(q|4|35|q)}{(1+3)^2(1+3+5)^2}\\ \nonumber
    + \frac{(q|2|3|q)(q|5|23|q)}{(1+3)^2(1+2+3)^2}+ \frac{(q|5|3|q)(q|2|35|q)}{(1+3)^2(1+3+5)^2}=
    \\ \nonumber
    \frac{(q|2|3|q)(q|45|23|q)}{(1+3)^2(1+2+3)^2}+ \frac{(q|4|3|q)(q|25|34|q)}{(1+3)^2(1+3+4)^2}+ \frac{(q|5|3|q)(q|24|35|q)}{(1+3)^2(1+3+5)^2},
\end{align}
which equals \eqref{4pt-t-lhs}.

The right-hand side is
\begin{align}
    &\phi_3(234) \Big( (q|1|2345|q) \frac{(12)}{(q1)(2q)}+(q|12|345|q) \frac{(23)}{(q2)(3q)} +(q|1235|4|q) \frac{(34)}{(q3)(4q)}\Big) \\ \nonumber
    +&\phi_3(435) \Big( (q|4|235|q) \frac{(43)}{(q4)(3q)} +(q|243|5|q) \frac{(35)}{(q3)(5q)}+(q|2435|1|q) \frac{(51)}{(q5)(1q)}\Big) \\ \nonumber
    +&\phi_3(235) \Big( (q|1|2354|q) \frac{(12)}{(q1)(2q)}+(q|12|354|q) \frac{(23)}{(q2)(3q)} +(q|1234|5|q) \frac{(35)}{(q3)(5q)}\Big) \\ \nonumber
    +&\phi_4(\Gamma) (1+2+3+4+5)^2.
\end{align}
This can be rewritten as
\begin{align}\nonumber
&\phi_3(234) (1+2+3+4)^2 +
    \phi_3(234) \Big( (q|1|5|q) \frac{(12)}{(q1)(2q)}+(q|12|5|q) \frac{(23)}{(q2)(3q)} +(q|5|4|q) \frac{(34)}{(q3)(4q)}\Big) \\ \nonumber
    +&\phi_3(435)(1+3+4+5)^2 + \phi_3(435)\Big( (q|4|2|q) \frac{(43)}{(q4)(3q)} +(q|2|5|q) \frac{(35)}{(q3)(5q)}+(q|2|1|q) \frac{(51)}{(q5)(1q)}\Big) \\ \nonumber
    +&\phi_3(235)(1+2+3+5)^2 + \phi_3(235) \Big( (q|1|4|q) \frac{(12)}{(q1)(2q)}+(q|12|4|q) \frac{(23)}{(q2)(3q)} +(q|4|5|q) \frac{(35)}{(q3)(5q)}\Big) \\ \nonumber
    +&\phi_4(\Gamma) (1+2+3+4+5)^2,
\end{align}
which means that
\begin{align}\nonumber
    &\phi_4(\Gamma) (1+2+3+4+5)^2 = \\ \nonumber
     -&\Big(\frac{(q|2|3|q)(q|4|23|q)}{(1+3)^2(1+2+3)^2(1+2+3+4)^2}+ \frac{(q|4|3|q)(q|2|34|q)}{(1+3)^2(1+3+4)^2(1+2+3+4)^2}\Big)\times
     \\ \nonumber
     &\Big( (q|1|5|q) \frac{(12)}{(q1)(2q)}+(q|12|5|q) \frac{(23)}{(q2)(3q)} +(q|5|4|q) \frac{(34)}{(q3)(4q)}\Big) \\ \nonumber   
     -& \Big( \frac{(q|4|3|q)(q|5|34|q)}{(1+3)^2(1+3+4)^2(1+3+4+5)^2}+ \frac{(q|5|3|q)(q|4|35|q)}{(1+3)^2(1+3+5)^2(1+3+4+5)^2}\Big) \times \\ \nonumber
& \Big( (q|4|2|q) \frac{(43)}{(q4)(3q)} +(q|2|5|q) \frac{(35)}{(q3)(5q)}+(q|2|1|q) \frac{(51)}{(q5)(1q)}\Big) \\ \nonumber  
    - & \Big( \frac{(q|2|3|q)(q|5|23|q)}{(1+3)^2(1+2+3)^2(1+2+3+5)^2}+ \frac{(q|5|3|q)(q|2|35|q)}{(1+3)^2(1+3+5)^2(1+2+3+5)^2}\Big) \times \\ \nonumber
    &\Big( (q|1|4|q) \frac{(12)}{(q1)(2q)}+(q|12|4|q) \frac{(23)}{(q2)(3q)} +(q|4|5|q) \frac{(35)}{(q3)(5q)}\Big)
\end{align}

We now proceed to combine terms with related denominators. Consider the first term in the second line. Using Schouten identity we have
\begin{align}
    (q|4|23|q)\Big( (q|1|5|q) \frac{(12)}{(q1)(2q)}+(q|12|5|q) \frac{(23)}{(q2)(3q)} -(q|4|5|q) \frac{(34)}{(q3)(4q)}\Big) = \\ \nonumber
    (q|5|23|q) \Big( (q|1|4|q) \frac{(12)}{(q1)(2q)}+(q|12|4|q) \frac{(23)}{(q2)(3q)} -(q|4|4|q) \frac{(34)}{(q3)(4q)}\Big) 
    \\ \nonumber
    +(q|4|5|q) \Big( (q|1|23|q) \frac{(12)}{(q1)(2q)}+(q|12|23|q) \frac{(23)}{(q2)(3q)} -(q|4|23|q) \frac{(34)}{(q3)(4q)}\Big) =
    \\ \nonumber
    (q|5|234|q) \left( (q|1|4|q) \frac{(12)}{(q1)(2q)}+(q|2|4|q) \frac{(23)}{(q2)(3q)} \right)  + (q|4|5|q) (1+2+3+4)^2=
    \\ \nonumber
    - (q|5|234|q) \frac{(q|4|12|3)}{(q3)} + (q|4|5|q) (1+2+3+4)^2.
\end{align}
We can similarly rewrite
\begin{align}
    (q|5|23|q)\Big( (q|1|4|q) \frac{(12)}{(q1)(2q)}+(q|12|4|q) \frac{(23)}{(q2)(3q)} -(q|5|4|q) \frac{(35)}{(q3)(5q)}\Big) = 
    \\ \nonumber
    (q|4|23|q)\Big( (q|1|5|q) \frac{(12)}{(q1)(2q)}+(q|12|5|q) \frac{(23)}{(q2)(3q)} -(q|5|5|q) \frac{(35)}{(q3)(5q)}\Big)\\
    \nonumber
    -(q|4|5|q)\Big( (q|1|23|q) \frac{(12)}{(q1)(2q)}+(q|12|23|q) \frac{(23)}{(q2)(3q)} -(q|5|23|q) \frac{(35)}{(q3)(5q)}\Big)=
\\ \nonumber
(q|4|235|q) \left( (q|1|5|q) \frac{(12)}{(q1)(2q)}+(q|2|5|q) \frac{(23)}{(q2)(3q)} \right) - (q|4|5|q) (1+2+3+5)^2=
\\ \nonumber
- (q|4|235|q) \frac{(q|5|12|3)}{(q3)} - (q|4|5|q) (1+2+3+5)^2.
\end{align}
For the two analysed terms, the parts where the propagators cancel kill each other, with the remaining result being
\begin{align}\nonumber
   &\frac{(q|2|3|q)}{(1+3)^2(1+2+3)^2(1+2+3+4)^2} \frac{(q|4|12|3)(q|5|234|q)}{(q3)(1+2+3+4+5)^2} 
   \\ \nonumber
    +&\frac{(q|2|3|q)}{(1+3)^2(1+2+3)^2(1+2+3+5)^2} \frac{(q|5|12|3)(q|4|235|q)}{(q3)(1+2+3+4+5)^2}.
\end{align}
The other terms are analysed similarly, with the complete result being
\begin{align}\label{answer-t}
\phi_4(2345) = 
    &\frac{(q|2|3|q)}{(1+3)^2(1+2+3)^2(1+2+3+4)^2} \frac{(q|4|12|3)(q|5|234|q)}{(q3)(1+2+3+4+5)^2} 
    \\
    +  &\frac{(q|2|3|q)}{(1+3)^2(1+2+3)^2(1+2+3+5)^2} \frac{(q|5|12|3)(q|4|235|q)}{(q3)(1+2+3+4+5)^2}
    \\
    +  &\frac{(q|4|3|q)}{(1+3)^2(1+3+4)^2(1+3+4+5)^2} \frac{(q|5|14|3)(q|2|345|q)}{(q3)(1+2+3+4+5)^2}
    \\
    +  &\frac{(q|4|3|q)}{(1+3)^2(1+3+4)^2(1+2+3+4)^2} \frac{(q|2|14|3)(q|5|234|q)}{(q3)(1+2+3+4+5)^2}
    \\
    +  &\frac{(q|5|3|q)}{(1+3)^2(1+3+5)^2(1+2+3+5)^2} \frac{(q|2|15|3)(q|4|235|q)}{(q3)(1+2+3+4+5)^2}
    \\
    +  &\frac{(q|5|3|q)}{(1+3)^2(1+3+5)^2(1+3+4+5)^2} \frac{(q|4|15|3)(q|2|345|q)}{(q3)(1+2+3+4+5)^2}.
\end{align}
All the terms here can also be understood as arising from the procedure of starting with the "core", which in this case is given by the vertex $3$, and then adding the "ends". There are $3!$ different orders in which this can be done, which results in the above terms. 

\subsection{Solution}
\label{sec:solution}

The considered above examples are sufficient to guess the general pattern. In this subsection we describe the arising solution for the objects $\phi_i$. We will then proof this ansatz for the solution by recursion in the following section.  

The all plus one minus gravity current for a tree $\Gamma_\mathcal{K}$ is given by by a sum over all subgraphs $\Gamma_\mathcal{I}\subseteq \Gamma_\mathcal{K}$, starting with subgraphs consisting of just a single vertex. This single vertex contribution is gauge $p$-dependent. All higher subgraphs contributions are gauge-independent. We have
\begin{align}
    &J(1|\Gamma_\mathcal{K}) = \bigg(\sum_{i\in\mathcal{K}}\frac{(qi)[ip]^2}{[qi][qp]^2} + \sum_{\Gamma_\mathcal{I}\subseteq \Gamma_\mathcal{K}, i = |\mathcal{I}|\geq 2} {\phi}_i(\Gamma_\mathcal{I})\bigg)J(\Gamma_\mathcal{K}),
 \end{align}
 which can also be written as
\begin{align}    
    &J(1|\Gamma_\mathcal{K}) =  \sum_{\Gamma_\mathcal{I}\subseteq \Gamma_\mathcal{K}, i = |\mathcal{I}|\geq1} {\phi}_i(\Gamma_\mathcal{I})J(\Gamma_\mathcal{K})
\end{align}
with the $p$-dependent part ${\phi}_1(i)$ given by  as 
\begin{align}
    {\phi}_1 = \frac{(qi)[ip]^2}{[qi][qp]^2}.
\end{align}
The $p$-independent contribution ${\phi}_i(\Gamma_\mathcal{I})$ from each subgraph  is determined by the following rules
\begin{enumerate}
    \item Determine a stem of the tree graph $\Psi\subset \Gamma_\mathcal{K}$ and the leaves which is final leg vertices $\eta_i$ for $i\in \{1,\dots, E\}$, with $E$ being the number of ends of the graph. A leaf is defined as a vertex with only one edge connecting to it, and the stem of the tree graph is the part of the graph obtained after all final leg vertices are cut off. For example, below, the stem is the vertices $\Psi = \{3,4\}$ while the vertices $\eta_{1,2,3} = \{2,5,6\}$ are the leaves of this tree graph.
    \begin{center}
         \scalebox{0.5}
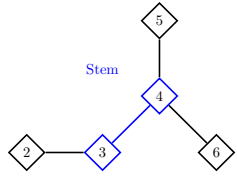
{
\begin{tikzpicture}[
Diamond/.style={diamond, draw=blue, fill=white, very thick, minimum size=5mm},
Vertex/.style={diamond, draw=black, fill=white, very thick, minimum size=5mm},
Equation/.style={rectangle, draw=white, fill=white, very thick, minimum size=40},]

\node[Vertex] (41) {2};
\node[Diamond] (42) [right=of 41]{3};
\node[Diamond] (43) [above right=of 42]{4};
\node[Vertex] (44) [above =of 43]{5};
\node[Vertex] (45) [below right=of 43]{6};
\node[Equation] (E2) [above = of 42]{\textcolor{blue}{Stem}};

\draw[-,very thick](41.east) to (42.west);
\draw[blue,-,very thick](42.north east) to (43.south west);
\draw[-,very thick](43.north) to (44.south);
\draw[-,very thick](43.south east) to (45.north west);

\end{tikzpicture}
}
    \end{center}
    \item  Form a set $\Delta$ of vertices of $\Gamma_\mathcal{K}$ (or of the stem $\Psi$) that have more than two edges connected to them. If the multiplicity of vertex $j\in \Gamma_\mathcal{K}$ is $\alpha_j$ then we denote $k_j = \alpha_j - 2$. This is an integer greater than zero for all vertices that should be included in $\Delta$. We will include a vertex $j$ into $\Delta$ the $k_j$ number of times
    \begin{align}
        \Delta = \{\dots\underbrace{j,\dots,j}_{k_j},\underbrace{l,\dots,l}_{k_l}\dots  \} 
    \end{align}
    It is not difficult to see that the number of elements in $\Delta$ is $E-2$. Let us denote the element of the set $\Delta$ by $\delta_j \in \Delta$ for $j \in {1,\dots,E-2}$.
    For example, for the graph above, the set $\Delta$ has only one element, which is given by $\{4\}$.

    \item Pick a leaf (end) $\eta_1$ of the graph, and start with the term
    \begin{align}
        \frac{(q|\eta_1|\Psi|q)}{(1+\Psi)^2}
    \end{align}
    for some end $\eta_1$. This term contains information about the core $\Psi$ and the vertex $\eta_1$ attached to it. The denominator is the momentum squared of the momenta of the core as well as the negative helicity graviton.
    
    \item  Keep adding other leaves (ends). In the case when the number of ends of the graph is bigger than two $E>2$, all ends apart from the final are added as follows. Adding an end $\eta_2$ produces the factor
    \begin{align}
        \times \frac{(q|\eta_2|1+\Psi + \eta_1|\delta_1)}{(1+\Psi+\eta_1)^2 (q \delta_1)}
    \end{align}
    This term tells us that the tree currently is bigger by one vertex $\Psi+\eta_1$, and the vertex $\eta_2$ will be attaching to the tree. However, instead of contracting with the reference spinor in the last bracket, we have $\delta_1$ contraction instead. The spinor $\delta_1$ could be taken to be the momentum spinor of the vertex from set $\Delta$ to which the end $\eta_2$ is connected in the graph. However, any other element of the set $\Delta$ could be used and will not affect the final solution. This will be discussed further below. 
    
    \item Repeat adding leaves (ends) as in step 4 to produce
    \begin{align}
        \times \prod^{E-2}_{j=1} \frac{(q|\eta_{j+1}|1+\Psi + \eta_1+\dots +\eta_j|\delta_j)}{(1+\Psi+\eta_1+\dots +\eta_j)^2 (q \delta_j)}
    \end{align}
    As we have already mentioned, the order in which elements of set $\Delta$ appear here does not matter for the solution. 
    
    \item Add the last leaf (end) by multiplying with 
    \begin{align}
        \times \frac{(q|\eta_n|\Psi + \eta_1 + \dots + \eta_{n-1}|q)}{(1+\Psi +  \eta_1 + \dots + \eta_{n-1})^2} \cdot \frac{1}{(1+\Psi +  \eta_1 + \dots + \eta_{n})^2}
    \end{align}
    where the final propagator is for the whole tree $\Gamma_\mathcal{I}$, which can also be written as $1/\Box$ for simplicity.
    
    \item The final solution is the sum over all possible $E!$ permutations of $\eta_i$ in the above formula.
 \end{enumerate}   
 
    Thus, the solution for $\phi(\Gamma_\mathcal{I})$ is given by
    \begin{align}\label{SolutionPhi}
        \sum_{\eta ~\text{perms}} \frac{1}{\Box}  \frac{(q|\eta_1|\Psi|q)}{(1+\Psi)^2} \cdot \prod^{E-2}_{j=1} \frac{(q|\eta_{j+1}|1+\Psi + \eta_1+\dots +\eta_j|\delta_j)}{(1+\Psi+\eta_1+\dots +\eta_j)^2 (q \delta_j)}\cdot \frac{(q|\eta_E|\Psi + \eta_1 + \dots + \eta_{E-1}|q)}{(1+\Psi +  \eta_1 + \dots + \eta_{E-1})^2} 
    \end{align}
    The sum is over $E!$ terms as we consider all permutations of $\eta_i$.
    The solution exhibits two symmetries. First, the order in which ends $\eta_i$ is added does not matter because we sum over all such possible orders. The second property of the solution is not manifest, and is the alredy mentioned property that it does not matter in which order the elements of the set $\Delta$ are drawn when producing the solution. This effect comes from summing over permutations of $\eta$ and could be demonstrated using the Schouten identity.

The described solution \eqref{SolutionPhi} has $E$ factors in the numerator and $E+1$ momentum squared factors in the denominator.

As a concrete example of application of the above rules, let us state the solution ${\phi}_5(23456)$ for the 5-vertex tree 
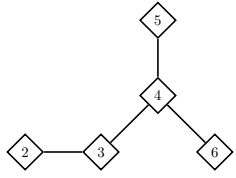
\begin{center}\label{5-point-graph}
    \scalebox{0.5}{
\begin{tikzpicture}[
Diamond/.style={diamond, draw=black, fill=white, very thick, minimum size=5mm},
Vertex/.style={diamond, draw=black, fill=white, very thick, minimum size=5mm},
Equation/.style={rectangle, draw=white, fill=white, very thick, minimum size=40},]

\node[Vertex] (41) {2};
\node[Vertex] (42) [right=of 41]{3};
\node[Vertex] (43) [above right=of 42]{4};
\node[Vertex] (44) [above =of 43]{5};
\node[Vertex] (45) [below right=of 43]{6};

\draw[-,very thick](41.east) to (42.west);
\draw[-,very thick](42.north east) to (43.south west);
\draw[-,very thick](43.north) to (44.south);
\draw[-,very thick](43.south east) to (45.north west);
\end{tikzpicture}
}
\end{center}
It is given by
\begin{align}\label{phi-23456}
    &\frac{(q|2|34|q)(q|5|123|4)(q|6|2345|q)}{\Box(1+3+4)^2(1+2+3+4)^2(q4)(1+2+3+4+5)^2}
    \\
    +&\frac{(q|2|34|q)(q|6|123|4)(q|5|2346|q)}{\Box(1+3+4)^2(1+2+3+4)^2(q4)(1+2+3+4+6)^2}
    \\
    +&\frac{(q|5|34|q)(q|2|135|4)(q|6|2345|q)}{\Box(1+3+4)^2(1+3+4+5)^2(q4)(1+2+3+4+5)^2}
    \\
    +&\frac{(q|5|34|q)(q|6|135|4)(q|2|3456|q)}{\Box(1+3+4)^2(1+3+4+5)^2(q4)(1+3+4+5+6)^2}
    \\
    +&\frac{(q|6|34|q)(q|2|136|4)(q|5|2346|q)}{\Box(1+3+4)^2(1+3+4+6)^2(q4)(1+2+3+4+6)^2}
    \\
    +&\frac{(q|6|34|q)(q|5|136|4)(q|2|3456|q)}{\Box(1+3+4)^2(1+3+4+6)^2(q4)(1+3+4+5+6)^2}.
\end{align}
The first term is obtained by starting with $\eta_1=\{2\}$, then adding $\eta_2=\{5\}$. This node connects to the node $\{4\}$, which in this case is all of the set $\Delta$. In this case, the core consists of the vertices $\{3,4\}$, the vertices $\{2,5,6\}$ are ends to be added. There are $3!$ terms. The vertex $\{4\}$ has three edges connecting to it and forms the set $\Delta$ for this graph.

\subsection{MHV Amplitudes}\label{GravityAmplitude}

We now apply the result obtained to extract the MHV graviton amplitude. This amplitude is obtained by inserting the negative helicity state into the off-shell leg, and amputating the final propagator
\begin{align}
    A_{MHV} = A(01|\mathcal{K}) = \epsilon_{-}^{ABA'B'}(0) ~\Box~ J(1|2\dots n)_{ABA'B'}.
\end{align}
This gives
\begin{align}
    A(01|\mathcal{K}) = \frac{0^A 0^B p^{A'}p^{B'}}{[0p]^2} q_A q_B (q|1+\mathcal{K}|_{A'}(q|1+\mathcal{K}|_{B'}  (1+\mathcal{K})^2 J(1|\mathcal{K}).
\end{align}
Applying the momentum conservation $0=-(1+\mathcal{K})$ gives
\begin{align}
    A(01|\mathcal{K}) = (01)^4 \sum_{\Gamma_{\mathcal{K}}}  (1+\mathcal{K})^2 \Phi(\Gamma_\mathcal{K})J(\Gamma_\mathcal{K}).
\end{align}
As in the YM case, it is clear that only the highest order term in the sum \eqref{SolutionPhi} gives a non-zero contribution, since it has a pole canceling the propagator. This gives
\begin{align}
    A(01|\mathcal{K}) = (01)^4 \sum_{\Gamma_{\mathcal{K}}}  (1+\mathcal{K})^2 \phi_{|\mathcal{K}|}(\Gamma_\mathcal{K})J(\Gamma_\mathcal{K}).
\end{align}
The MHV amplitude is thus given by a sum over trees. Let us focus on a single tree amplitude. 

Before we describe a calculation evaluating $(1+\mathcal{K})^2 \phi_{|\mathcal{K}|}$ on-shell, let us consider an illustrative example. Let us consider the same graph as in \eqref{5-point-graph}. The contribution to the MHV amplitude from this graph is given by
\begin{align}
    A(01|\Gamma) = (01)^4 J(\Gamma) \Big( &\frac{(q|2|34|q)(q|5|123|4)(q|6|2345|q)}{(1+3+4)^2(1+2+3+4)^2(q4)(1+2+3+4+5)^2}
    \\
    +&\frac{(q|2|34|q)(q|6|123|4)(q|5|2346|q)}{(1+3+4)^2(1+2+3+4)^2(q4)(1+2+3+4+6)^2}
    \\
    +&\frac{(q|5|34|q)(q|2|135|4)(q|6|2345|q)}{(1+3+4)^2(1+3+4+5)^2(q4)(1+2+3+4+5)^2}
    \\
    +&\frac{(q|5|34|q)(q|6|135|4)(q|2|3456|q)}{(1+3+4)^2(1+3+4+5)^2(q4)(1+3+4+5+6)^2}
    \\
    +&\frac{(q|6|34|q)(q|2|136|4)(q|5|2346|q)}{(1+3+4)^2(1+3+4+6)^2(q4)(1+2+3+4+6)^2}
    \\
    +&\frac{(q|6|34|q)(q|5|136|4)(q|2|3456|q)}{(1+3+4)^2(1+3+4+6)^2(q4)(1+3+4+5+6)^2}\Big).
\end{align}
We now use the momentum conservation in the very last $q$-numerator, as well as in the very last propagator. For example, we have
\begin{align}
    \frac{(q|6|2345|q)}{(1+2+3+4+5)^2}= - \frac{(q|6|0|q)}{(0+6)^2} = - \frac{(q6)[60](0q)}{[60](60)} = (01) \frac{(q6)}{(06)},
\end{align}
where we used $q\to 1$. Applying this to all the terms gives
\begin{align}
    \frac{A(01|\Gamma)}{J(\Gamma)} = (01)^5  \Big( &\frac{(q|2|34|q)(q|5|123|4)(q6)}{(1+3+4)^2(1+2+3+4)^2(q4)(06)} +\frac{(q|2|34|q)(q|6|123|4)(q5)}{(1+3+4)^2(1+2+3+4)^2(q4)(05)}
    \\
    +&\frac{(q|5|34|q)(q|2|135|4)(q6)}{(1+3+4)^2(1+3+4+5)^2(q4)(06)}
    +\frac{(q|5|34|q)(q|6|135|4)(q2)}{(1+3+4)^2(1+3+4+5)^2(q4)(02)}
    \\
    +&\frac{(q|6|34|q)(q|2|136|4)(q5)}{(1+3+4)^2(1+3+4+6)^2(q4)(05)}
   +\frac{(q|6|34|q)(q|5|136|4)(q2)}{(1+3+4)^2(1+3+4+6)^2(q4)(02)}\Big).
\end{align}
We grouped the terms according to their propagators. The terms in each line can be further combined as follows
\begin{align}
    \frac{A(01|\Gamma)}{(01)^5J(\Gamma)} = \frac{(q5)(q6)}{(05)(06)} \frac{(q|2|34|q)(0|56|123|4)}{(1+3+4)^2(1+2+3+4)^2(q4)} 
    \\
    + \frac{(q2)(q6)}{(02)(06)} \frac{(q|5|34|q)(0|26|135|4)}{(1+3+4)^2(1+3+4+5)^2(q4)}
    \\
    + \frac{(q2)(q5)}{(02)(05)} \frac{(q|6|34|q)(0|25|136|4)}{(1+3+4)^2(1+3+4+6)^2(q4)}.
\end{align}
We can then again use the momentum conservation in each numerator, for example the first numerator gives
\begin{align}
    (0|56|123|4) = (0|056|1234|4) = - (0|1234|1234|4) = (1+2+3+4)^2 (04).
\end{align}
Applying the same procedure to all 3 terms we get
\begin{align}\nonumber
    &\frac{A(01|\Gamma)}{(01)^5J(\Gamma)} = 
    \\ \nonumber
    & \frac{(q5)(q6)(04)}{(05)(06)(q4)} \frac{(q|2|34|q)}{(1+3+4)^2} 
    + \frac{(q2)(q6)(04)}{(02)(06)(q4)} \frac{(q|5|34|q)}{(1+3+4)^2} 
    + \frac{(q2)(q5)(04)}{(02)(05)(q4)} \frac{(q|6|34|q)}{(1+3+4)^2} = \\ \nonumber
    & \frac{(q2)(q5)(q6)(04)}{(02)(05)(06)(q4)}
    \frac{(0|256|34|q)}{(1+3+4)^2} = (0q) \frac{(q2)(q5)(q6)(04)}{(02)(05)(06)(q4)}.
\end{align}
This shows that 
\begin{align}
    A(01|\Gamma) = (01)^6 \frac{(q2)(q5)(q6)(04)}{(02)(05)(06)(q4)} J(\Gamma) = 
    (01)^6 \frac{(q4)(04)}{(q2)(02)(q5)(05)(q6)(06)} \prod_{\langle ij\rangle\in \Gamma} \frac{[ij]}{(ij)},
\end{align}
which is the correct contribution of this graph to the MHV amplitude. 

Let us now give the general argument. It follows exactly the same pattern as the already considered example. We first apply the momentum conservation in the very last set of factors to get
\begin{align}
    (1+\mathcal{K})^2 \phi(\Gamma_\mathcal{K})
    \\
    = &  \sum_{\eta ~\text{perms}} \frac{(q|\eta_1|\Psi|q)}{(1+\Psi)^2} \cdot \prod^{E-2}_{j=1} \frac{(q|\eta_{j+1}|1+\Psi + \eta_1+\dots +\eta_j|\delta_j)}{(1+\Psi+\eta_1+\dots +\eta_j)^2 (q \delta_j)}\cdot (-1) \frac{(q|\eta_E|0|q)}{(0 + \eta_{E})^2} 
\end{align}
We can now multiply and divide this expression by $\prod_{i=1}^E[ (0 \eta_i)/(1 \eta_i)]$, converting it to
\begin{align}
     (01) \prod_{i=1}^E \frac{(1 \eta_i)}{(0 \eta_i)}\sum_{\eta ~\text{perms}} \frac{(0|\eta_1|\Psi|q)}{(1+\Psi)^2} \cdot \prod^{E-2}_{j=1} \frac{(0|\eta_{j+1}|1+\Psi + \eta_1+\dots +\eta_j|\delta_j)}{(1+\Psi+\eta_1+\dots +\eta_j)^2 (q \delta_j)} 
\end{align}
We can now collect terms in this sum according to their denominators. It is clear that there are precisely two terms with the same denominators obtained by permuting $\eta_E$ and $\eta_{E-1}$. Adding these gives
\begin{align}
     (01) \prod_{i=1}^E \frac{(1 \eta_i)}{(0 \eta_i)}\sum_{(E-1)! \,\, \eta ~\text{perms}} \frac{(0|\eta_1|\Psi|q)}{(1+\Psi)^2} \cdot \prod^{E-3}_{j=1} \frac{(0|\eta_{j+1}|1+\Psi + \eta_1+\dots +\eta_j|\delta_j)}{(1+\Psi+\eta_1+\dots +\eta_j)^2 (q \delta_j)} \times \\ \nonumber
     \frac{(0|\eta_{E-1} + \eta_E| \Psi + \eta_1+\dots +\eta_{E-2}|\delta_{E-2})}{(1+\Psi+\eta_1+\dots +\eta_{E-2})^2(q\delta_{E-2})}.
\end{align}
Using the momentum conservation 
\begin{align}
    1+\Psi + \eta_1+ \dots +\eta_{E-2} = -(0+\eta_{E-1}+\eta_E)
\end{align}
cancels the propagator in the last factor and gives
\begin{align}
     (01) \prod_{i=1}^E\bigg[ \frac{(1 \eta_i)}{(0 \eta_i)}\bigg] \sum_{(E-1)! \,\, \eta ~\text{perms}} \frac{(0|\eta_1|\Psi|q)}{(1+\Psi)^2} \cdot \prod^{E-3}_{j=1} \frac{(0|\eta_{j+1}|1+\Psi + \eta_1+\dots +\eta_j|\delta_j)}{(1+\Psi+\eta_1+\dots +\eta_j)^2 (q \delta_j)} \frac{(0\delta_{E-2})}{(q\delta_{E-2})}.
\end{align}

The general pattern is now clear. One keeps grouping the terms according to their propagators. At each step more and more terms are grouped in this way, and at each step one uses the momentum conservation to eliminate one of the propagators, producing a factor of $(0\delta)/(q\delta)$. This process is repeated to give the following extremely simple final answer
\begin{align}\label{answer-phi}
    (1+\mathcal{K})^2 \phi(\Gamma_\mathcal{K})\Big|_{on\,\, shell}=
     (01)^2  \prod_{i=1}^E\frac{(1 \eta_i)}{(0 \eta_i)} \prod_{j = 1}^{E-2} \frac{(0 \delta_j)}{(1 \delta_j)}.
\end{align}
We now multiplying this by $(01)^4 J(\Gamma_\mathcal{K})$. Recall that $J(\Gamma_\mathcal{K})$ is given by 
\begin{align}
    J(\Gamma_\mathcal{K}) = \prod_{i\in \mathcal{K}} (qi)^{2(\alpha_i-2)} \prod_{\langle ij\rangle \in \Gamma_\mathcal{K}} \frac{[ij]}{(ij)}.
\end{align}
This means that, apart from the edge contributions, the vertices come with a factor of $(qi)^{-2}$ if they are the end vertices and with a factor of $(qi)^{2k_i}, k_i = \alpha_i -2$ if they are vertices of the core with multiplicity more than two. We can now see that the first prefactor in \eqref{answer-phi} converts $(1i)^{-2}$ for each end vertex into $(0i)^{-1}(1i)^{-1}$, while the second factor converts $(qi)^{2k_i}$ for each vertex of multiplicty more than two in the core of the graph to $(0i)^{k_i} (1i)^{k_i}$. This produces the known correct MHV amplitude formula
\begin{align}
    A_{MHV}(01|\mathcal{K}) = (01)^6 \sum _{\Gamma_\mathcal{K}}\prod_{i\in \mathcal{K}} (0i)^{\alpha_i-2}(1i)^{\alpha_i -2} \prod_{\langle ij\rangle \in \Gamma_\mathcal{K}} \frac{[jk]}{(jk)}.
\end{align}

\section{Proof of the formula for $\phi_i$}
\label{sec:proof}

The purpose of this section is to spell out the argument for why the ansatz for the solution \eqref{SolutionPhi} solves the recursion relation.
The idea of the proof is very similar to the one used in the Appendix to prove the formula in the case of YM theory. Complications arise because there is no longer linear ordering. 

\subsection{General argument}

Noticing the pattern that led to \eqref{relation-phi} and then \eqref{relation-phi-3}, we can see that both of these recurrence relations for $\phi_i$ can be written as
\begin{align}\label{phi-star-relation}
    \sum_{\Gamma_\mathcal{I}\subset \Gamma_\mathcal{K},i = |\mathcal{I}|} {\phi}^*_i(\Gamma_\mathcal{I}) (1+\mathcal{I})^2 = &\sum_{\Gamma_\mathcal{I}\subset \Gamma_\mathcal{K},i = |\mathcal{I}|} {\phi}_i(\Gamma_\mathcal{I}) \sum_{\braket{mn}\in 1+\Gamma_\mathcal{I}} (q|\mathcal{M}|\mathcal{N}|q)\frac{(mn)}{(qm)(nq)}\nonumber
    \\ 
    + &\sum_{\Gamma_\mathcal{J}\subset \Gamma_\mathcal{K},j = |\mathcal{J}| = |\mathcal{I}|+1} {\phi}^*_j(\Gamma_\mathcal{J}) (1+\mathcal{J})^2.
\end{align}
Here the object $\phi^*_i(\Gamma_\mathcal{I})$ is built from the same ingredients as $\phi_i(\Gamma_\mathcal{I})$, with the difference being that it depends on the momenta of all of the graph $\Gamma_\mathcal{K}$. For example, in these notations the expression \eqref{4pt-lhs} is written as 
\begin{align}
    \phi_3^*(234)(1+2+3+4)^2 + \phi_3^*(345) (1+3+4+5)^2 .
\end{align}
Analysing the examples one can see that the way $\phi^*_i$ is produced from $\phi_i$ is to change only the final spinor numerator. This is done according to the following rule
\begin{align}
     (q|\eta_E|\mathcal{I}|q) \rightarrow  (q| \sum \alpha + \eta_E|\mathcal{K}|q)
\end{align} 
The difference between $\mathcal{I}$ and $\mathcal{K}$ is the momenta $\sum \alpha + \sum \beta$ defined as a group of vertices connecting to $\eta_E$ and $\mathcal{I}/\eta_E$ respectively, as in the following drawing

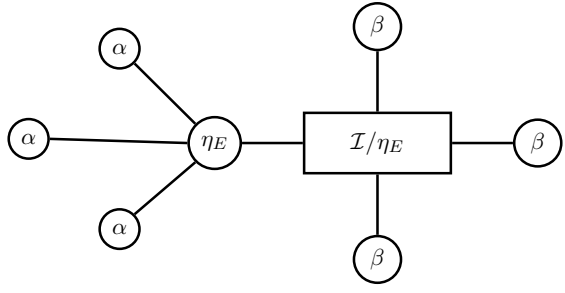
\begin{center}
\scalebox{0.8}{
\begin{tikzpicture}[
Diam/.style={diamond, draw=black, fill=white, very thick, minimum size=5mm},
Alpha/.style={circle, draw=black, fill=white, very thick, minimum size=5mm},
Beta/.style={circle, draw=black, fill=white, very thick, minimum size=5mm},
Eta/.style={circle, draw=black, fill=white, very thick, minimum size=5mm},
Psi/.style={rectangle, draw=black, fill=white, very thick, minimum size=10mm},
Equation/.style={rectangle, draw=white, fill=white, very thick, minimum size=20},]
\node[Alpha] (a1) {$\alpha$};
\node[Alpha] (a2) [above right = of 11]{$\alpha$};
\node[Alpha] (a3) [below right = of 11]{$\alpha$};
\node[Eta] (e) [below right = of a2]{$\eta_E$};
\draw[-,very thick](a3.north east) to (e.south west);
\draw[-,very thick](a2.south east) to (e.north west);
\draw[-,very thick](a1.east) to (e.west);

\node[Psi] (p) [right = of e]{$~~~~~\mathcal{I}/\eta_E~~~~~$};
\node[Beta] (b1) [above = of p]{$\beta$};
\node[Beta] (b2) [right = of p]{$\beta$};
\node[Beta] (b3) [below = of p]{$\beta$};
\draw[-,very thick](e.east) to (p.west);

\draw[-,very thick](p.north) to (b1.south);
\draw[-,very thick](p.east) to (b2.west);
\draw[-,very thick](p.south) to (b3.north);

\end{tikzpicture}
}
\end{center}

To spell out explicitly, the object $\phi^*_i$ is given by
\begin{align}\label{SolutionPhi-star-F}
        \phi_i^*(\Gamma_\mathcal{I})(1+\mathcal{I})^2 = \sum_{\eta ~\text{perms}}   \frac{(q|\eta_1|\Psi|q)}{(1+\Psi)^2} \cdot \prod^{E-2}_{j=1} \frac{(q|\eta_{j+1}|1+\Psi + \eta_1+\dots +\eta_j|\delta_j)}{(1+\Psi+\eta_1+\dots +\eta_j)^2 (q \delta_j)}\times \\ \nonumber
        \frac{(q| \sum \alpha + \eta_E|\mathcal{K}|q)}{(1+\Psi +  \eta_1 + \dots + \eta_{E-1})^2} .
\end{align}
We now prove that \eqref{phi-star-relation} holds with $\phi_i$ given by \eqref{SolutionPhi}, thus establishing that \eqref{SolutionPhi} is solution to the recursion relation \eqref{relation-phi-3}.

\subsection{Important identity}

To make the proof less complicated, we can focus on the final spinor numerator in $\phi^*_i(\Gamma_\mathcal{I})$ only, since both $\phi(\Gamma_\mathcal{I})$ and $\phi^*(\Gamma_\mathcal{I})$ share the same factors apart from the final spinor numerator.

Thus, we want to show that
\begin{align}\label{F-1-recursion}
    (q| \sum \alpha + \eta_n|\mathcal{K}|q) (1+\mathcal{I})^2 = (q|\eta_E|\mathcal{I}|q)\sum_{\braket{mn}\in \Gamma_\mathcal{I}\cup 1} (q|\mathcal{M}|\mathcal{N}|q)\frac{(mn)}{(qm)(nq)}  + \dots,
\end{align}
where the dots stand for the terms that contribute to higher $\phi_i$. We start the calculation using the identity \eqref{conventions-minus}
\begin{align}\label{F-relation}
    (q| \sum \alpha + \eta_E|\mathcal{K}|q) (1+\mathcal{I})^2 = -(q|1 +\mathcal{I}|1+\mathcal{I}|\eta_E|\mathcal{K}|q) + (q| \sum \alpha |\mathcal{K}|q) (1+\mathcal{I})^2 
\end{align}
The first term can be decomposed as
\begin{align}
    -(q|1 +\mathcal{I}|1+\mathcal{I}|\eta_E|\mathcal{K}|q) = &\sum_{i \in 1+\mathcal{I}} -(q|1 +\mathcal{I}|i](i|\eta_E|\mathcal{K}|q) \times\frac{(iq)}{(iq)}\nonumber
    \\
    = &\sum_{i \in 1+\mathcal{I}} -(q|1 +\mathcal{I}|i|q)(i|\eta_E|\mathcal{K}|q) \times\frac{1}{(iq)}
\end{align}
Now we use Schouten identity between these two spinor contractions as
\begin{align}
    = & -\sum_{i \in 1+\mathcal{I}}\bigg[(q|\mathcal{I}|\eta_E|i)(q|i|\mathcal{K}|q) \times\frac{1}{(iq)} + (q|\mathcal{I}|\mathcal{K}|q)(i|\eta_E|i|q) \times\frac{1}{(iq)}\bigg] \nonumber
    \\
    = & -\sum_{i \in 1+\mathcal{I}}\bigg[(q|\mathcal{I}|\eta_E] (\eta_E i)(q|i|\mathcal{K}|q) \times\frac{1}{(iq)} \times \frac{(\eta_E q)}{(\eta_E q)} + (q|\mathcal{I}|\mathcal{K}|q)(i|\eta_E|i]\bigg] \nonumber
    \\
    = & \sum_{i \in 1+\mathcal{I}}(q|\eta_E|\mathcal{I}|q) (q|i|\mathcal{K}|q) \times\frac{(i \eta_E)}{(qi)(\eta_E q)} - (q|\mathcal{I}|\mathcal{K}|q)(\eta_E|1+\mathcal{I}|\eta_E],
\end{align}
where we took the sum in the last term. We can now add and subtract to this the following quantity
\begin{align}
    (q|\eta_E|\mathcal{I}|q) \sum (q|\beta|\mathcal{K}|q) \times \frac{(\delta \eta_E)}{(q\delta)(\eta_E q)}
\end{align}
where the vertices $\beta$ connect to the stem $\mathcal{I}/\eta_E$ through the vertices $\delta$, the above equation can be written as
\begin{align}
    = (q|\eta_E|\mathcal{I}|q) \sum_{\braket{mn}\in \Gamma_\mathcal{I}\cup 1} (q|\mathcal{M}|\mathcal{N}|q)\frac{(mn)}{(qm)(nq)} - &(q|\eta_E|\mathcal{I}|q) \sum (q|\beta|\mathcal{K}|q) \times \frac{(\delta \eta_E)}{(q\delta)(\eta_E q)} \nonumber
    \\
    - &(q|\mathcal{I}|\mathcal{K}|q)(\eta_E|1+\mathcal{I}|\eta_E]
\end{align}
Then, \eqref{F-relation} can be written as
\begin{align}
    (q| \sum \alpha + \eta_E|\mathcal{K}|q) (1+\mathcal{I})^2 = &(q|\eta_E|\mathcal{I}|q) \sum_{\braket{mn}\in \Gamma_\mathcal{I}
    \cup1} (q|\mathcal{M}|\mathcal{N}|q)\frac{(mn)}{(qm)(nq)} \nonumber
    \\
    - &(q|\eta_E|\mathcal{I}|q) \sum (q|\beta|\mathcal{K}|q) \times \frac{(\delta \eta_E)}{(q\delta)(\eta_E q)} \nonumber
    \\
    - &(q|\mathcal{I}|\mathcal{K}|q)(\eta_E|1+\mathcal{I}|\eta_E]
    + (q|\sum\alpha|\mathcal{K}|q) (1+ \mathcal{I})^2\label{F-first-equation}
\end{align}
It is clear that the first term is what we want from the recursion relation \eqref{F-1-recursion}.
Next, we notice that $\mathcal{K} = \mathcal{I} + \sum \alpha + \sum \beta$ and 
\begin{align}
    (q|\mathcal{I}|\mathcal{K}|q) = -(q|\sum \alpha + \sum\beta|\mathcal{K}|q)
\end{align}
Then, we use this to simplify the remaining terms in \eqref{F-first-equation} 
\begin{align}
    = - &(q|\eta_E|\mathcal{I}|q) \sum (q|\beta|\mathcal{K}|q) \times \frac{(\delta \eta_E)}{(q\delta)(\eta_E q)} + (q|\sum\beta|\mathcal{K}|q)(\eta_E|1+\mathcal{I}|\eta_E]\nonumber
    \\
    + &(q|\sum \alpha |\mathcal{K}|q)(\eta_E|1+\mathcal{I}|\eta_E]
    + (q|\sum\alpha|\mathcal{K}|q) (1+ \mathcal{I})^2
    \nonumber
    \\
    = &\sum (q|\beta|\mathcal{K}|q) (q|\eta_E|1+\mathcal{I}|\delta)\frac{1}{(q\delta)}
     +(q|\sum\alpha|\mathcal{K}|q) (1+ \mathcal{I} - \eta_E)^2,
\end{align}
where we used the relation
\begin{align}
    &\sum (q|\beta|\mathcal{K}|q) \bigg[ - (q|\eta_E|\mathcal{I}|q)\frac{(\delta \eta_E)}{(q\delta)(\eta_E q)} + (\eta_E|1+\mathcal{I}|\eta_E]\frac{(q\delta)(\eta_E q)}{(q\delta)(\eta_E q)}\bigg]\nonumber
    \\
    = &\sum (q|\beta|\mathcal{K}|q) \bigg[ - (q|\eta_E|1+\mathcal{I}|q)\frac{(\delta \eta_E)}{(q\delta)(\eta_E q)} + (q|\eta_E|1+\mathcal{I}|\eta_E)\frac{(\delta q)}{(q\delta)(\eta_E q)}\bigg]\nonumber
    \\
    = & \sum (q|\beta|\mathcal{K}|q) (q|\eta_E|1+\mathcal{I}|\delta)\frac{1}{(q\delta)}\label{beta-F-simplify}
\end{align}
for the first term and $(\eta_E|1+\mathcal{I}|\eta_E] = - \eta_E\cdot(1+\mathcal{I})$ for the second term respectively. The Schouten identity is used from second to third line of \eqref{beta-F-simplify}.

Now, \eqref{F-first-equation} can be written nicely as
\begin{align}\label{identity-phi-proof}
    (q| \sum \alpha + \eta_E|\mathcal{K}|q) (1+\mathcal{I})^2 = &(q|\eta_E|\mathcal{I}|q) \sum_{\braket{mn}\in 1+\Gamma_\mathcal{I}} (q|\mathcal{M}|\mathcal{N}|q)\frac{(mn)}{(qm)(nq)} \nonumber
    \\
    + &(q|\sum \beta|\mathcal{K}|q) (q|\eta_E|1+\mathcal{I}|\delta)\frac{1}{(q\delta)}
     +(q|\sum\alpha|\mathcal{K}|q) (1+ \mathcal{I} - \eta_E)^2.
\end{align}

\subsection{Proof continued}

Using the identity \eqref{identity-phi-proof} and  reinstating the prefactors for $\phi_i$ and $\phi^*_i$, we have
\begin{align}\label{phi-star-relation-1}
    {\phi}^*_i(\Gamma_\mathcal{I}) &(1+\mathcal{I})^2 = {\phi}_i(\Gamma_\mathcal{I}) \sum_{\braket{mn}\in 1+\Gamma_\mathcal{I}} (q|\mathcal{M}|\mathcal{N}|q)\frac{(mn)}{(qm)(nq)}
    \\ 
    + &\sum_{\eta ~\text{perms}}  \frac{(q|\eta_1|\Psi|q)}{(1+\Psi)^2} \cdot \prod^{E-2}_{j=1} \frac{(q|\eta_{j+1}|1+\Psi + \eta_1+\dots +\eta_j|\delta_j)}{(1+\Psi+\eta_1+\dots +\eta_j)^2 (q \delta_j)}\cdot \frac{ (1+ \mathcal{I} - \eta_E)^2(q|\sum\alpha|\mathcal{K}|q)}{(1+ \mathcal{I} - \eta_E)^2(1+\mathcal{I})^2} .\nonumber
    \\
    + &\sum_{\eta ~\text{perms}} \frac{(q|\eta_1|\Psi|q)}{(1+\Psi)^2} \cdot \prod^{E-2}_{j=1} \frac{(q|\eta_{j+1}|1+\Psi + \eta_1+\dots +\eta_j|\delta_j)}{(1+\Psi+\eta_1+\dots +\eta_j)^2 (q \delta_j)}\cdot \frac{ (q|\eta_E|1+\mathcal{I}|\delta)(q|\sum  \beta|\mathcal{K}|q) }{(q\delta)(1+ \mathcal{I} - \eta_E)^2(1+\mathcal{I})^2}\nonumber
\end{align}
It is important to note that the object $\alpha$ and $\beta$ relate to the final momenta $\eta_E$ as we sum over the permutations of $\eta_i$. 
We have used $(1+\Psi +  \eta_1 + \dots + \eta_{E-1})^2 = (1+ \mathcal{I} - \eta_E)^2$. Also in both of the terms $\Box = (1+\mathcal{I})^2$.

What is left to show is that the sum of the remaining terms in the above equation \eqref{phi-star-relation-1} over all subgraphs $\Gamma_\mathcal{I}$ can be written as a sum over subgraphs $\Gamma_\mathcal{J}$ of $\phi^*_{j}(\Gamma_\mathcal{J})$ with $|\mathcal{J}| = j = i+1$. The second line in \eqref{phi-star-relation-1} contains spinor contractions $(q|\beta|\mathcal{K}|q)$. We would like to interpret these terms as corresponding to continuing the graph $\mathcal{I}$ "through the stem" $\mathcal{I}$, because these vertices are attached to the main body of the graph $\mathcal{I}$. There are also terms containing $(q|\alpha|\mathcal{K}|q)$. These terms will correspond to enlarging the graph $\mathcal{I}$ "through the leaf" $\eta_E$.

We will proceed by considering a set of examples that show how \eqref{phi-star-relation-1} establishes \eqref{phi-star-relation}. We then give a general argument. 

\subsection{Example for an extension of a linear graph}

We start with an example in which $\Gamma_\mathcal{I}$ is a linear graph with two ends $\eta, \eta'$. It is then enlarged to a larger graph $\Gamma_\mathcal{K}$, which contains an additional node $\gamma$. The formula for $\phi_i(\Gamma_\mathcal{I})$ contains two terms, one in which $\eta$ is added to the core $\mathcal{I}/\eta,\eta'$, and then $\eta'$ is added, and vice versa. So, the process of adding $\gamma$ can be illustrated by the following figure

\begin{center}
\scalebox{0.5}{
\begin{tikzpicture}[
Diam/.style={diamond, draw=black, fill=white, very thick, minimum size=5mm},
Alpha/.style={circle, draw=black, fill=white, very thick, minimum size=5mm},
Beta/.style={circle, draw=black, fill=white, very thick, minimum size=5mm},
Eta/.style={circle, draw=blue, fill=white, very thick, minimum size=5mm},
Psi/.style={rectangle, draw=black, fill=white, very thick, minimum size=10mm},
Equation/.style={rectangle, draw=white, fill=white, very thick, minimum size=20},]
\node[Alpha] (a1) {$\gamma$};

\node[Eta] (e) [right = of a1]{$\eta$};

\draw[dashed,very thick](a1.east) to (e.west);

\node[Psi] (p) [right = of e]{$~~~~~\mathcal{I}/\eta,\eta'~~~~~$};

\node[Beta] (b2) [right = of p]{$\eta'$};

\draw[blue,-,very thick](e.east) to (p.west);

\draw[-,very thick](p.east) to (b2.west);

\node[Alpha] (aa1) [right = of b2]{$\gamma$};

\node[Beta] (ee) [right = of aa1]{$\eta$};

\draw[dashed,very thick](aa1.east) to (ee.west);

\node[Psi] (pp) [right = of ee]{$~~~~~\mathcal{I}/\eta,\eta'~~~~~$};

\node[Eta] (bb2) [right = of pp]{$\eta'$};

\draw[-,very thick](ee.east) to (pp.west);

\draw[blue, -,very thick](pp.east) to (bb2.west);

\end{tikzpicture}
}
\end{center}

In the first of this, to use \eqref{phi-star-relation-1} we need to interpret $\eta'$ as $\eta_E$, , while $\beta=\gamma$, and $\eta$ is the vertex $\delta$ to which $\gamma$ is attached. This is described by the term in the last line of \eqref{phi-star-relation-1}. The second part of the above figure represents a process in which $\eta=\eta_E$ and $\alpha=\gamma$. This is described by the second line in \eqref{phi-star-relation-1}. This gives the following two terms 
\begin{align}\label{linear-graph-example}
       \frac{(q|\eta|\mathcal{I} - \eta - \eta'|q)}{(1+\mathcal{I} - \eta - \eta')^2} \cdot \frac{(q|\eta'|1+\mathcal{I}|\eta) }{(q\eta)(1+ \mathcal{I} - \eta')^2} \frac{(q|\gamma|\mathcal{K}|q) }{(1+\mathcal{I})^2} 
       + 
       \frac{(q|\eta'|\mathcal{I} - \eta - \eta'|q)}{(1+\mathcal{I} - \eta - \eta')^2} \cdot \frac{(1+ \mathcal{I} - \eta')^2 }{(1+ \mathcal{I} - \eta')^2} \frac{(q|\gamma|\mathcal{K}|q) }{(1+\mathcal{I})^2}
\end{align}

The arising expression is simplified using the following identity. We have
\begin{align}\label{identity-simplify-phi-star}
    (q|\eta |1+ \mathcal{J}|\delta) \frac{(q|\eta'|1+ \eta + \mathcal{J}|\eta)}{(q\eta)} + (q|\eta'|1+\mathcal{J}|\delta)(1+\eta +\mathcal{J} )^2 = (q|\eta'|1 + \eta + \mathcal{J}|\delta) ( 1 + \mathcal{J})^2
\end{align}
Here $\eta$ and $\eta'$ are a single momenta while $\mathcal{J}$ is a sum of some momentum. This identity is a consequence of the Schouten identity.
The key point here is the repeat of the momenta $\eta$ as there is $(q|\eta |$ and $|\eta)$ term present in the first term. This relation can be interpreted as "shifting" the momentum squared by one momentum factor. 

Using the relation \eqref{identity-simplify-phi-star} to simplify \eqref{linear-graph-example} we obtain
\begin{align}
    \frac{(q|\eta'|\mathcal{I}-\eta'|q)}{(1+ \mathcal{I} - \eta')^2} \frac{(q|\gamma|\mathcal{K}|q) }{(1+\mathcal{I})^2}
\end{align}
One can recognize that this is one out of two terms corresponding to the object $\phi^*_{i+1}(\Gamma_\mathcal{I}\cup \{\gamma\})$. 
The other term is obtained analogously, by considering a version of the above picture in which $\gamma$ is attached to $\eta'$ instead
\begin{align}
    \frac{(q|\gamma|\mathcal{I} - \eta'|q)}{(1+ \mathcal{I} - \eta')^2} \frac{(q|\eta'|\mathcal{K}|q) }{(1+\mathcal{I}-\eta' + \gamma)^2}
\end{align}
This establishes \eqref{phi-star-relation} in the form 
\begin{align}
    \sum_{\Gamma_\mathcal{I}} \phi^*_i(\Gamma_\mathcal{I}) (1+\mathcal{I})^2 = \sum_{\Gamma_\mathcal{I}}\phi_i(\Gamma_\mathcal{I}) \sum_{\braket{mn}\in \Gamma_{\mathcal{I}\cup 1}} (q|\mathcal{M}|\mathcal{N}|q)\frac{(mn)}{(qm)(nq)}
    + \phi^*_{i+1}(\Gamma_{\mathcal{I}\cup \gamma})
\end{align}
for the case of a tree $\Gamma_\mathcal{I}$ with two ends $\eta, \eta'$ and  $\mathcal{K} = \mathcal{I}+ \gamma$. 

\subsection{Extending through stem}

Another simpler example is to consider the situation when a graph $\mathcal{I}$, whose final end is $\eta_E$, is extended by a single vertex $\beta$ which is attached to the core $\mathcal{I}/\eta_E$. There are no simplifications among the terms in \eqref{phi-star-relation-1}, and  the terms in the second line of \eqref{phi-star-relation-1} can be recognized as a part of $\phi^*$ for the graph with one more node. Indeed, we have
\begin{align}
    &\sum_{\eta ~\text{perms}}   \frac{(q|\eta_1|\Psi|q)}{(1+\Psi)^2} \cdot \prod^{E-2}_{j=1} \frac{(q|\eta_{j+1}|1+\Psi + \eta_1+\dots +\eta_j|\delta_j)}{(1+\Psi+\eta_1+\dots +\eta_j)^2 (q \delta_j)}\cdot \frac{ (q|\eta_E|1+\mathcal{I}|\delta)(q|\sum \beta|\mathcal{K}|q) }{(q\delta)(1+ \mathcal{I} - \eta_n)^2(1+\mathcal{I})^2} \nonumber
    \\
    =& \sum_{\eta ~\text{perms}}   \frac{(q|\eta_1|\Psi|q)}{(1+\Psi)^2} \cdot \prod^{E-1}_{j=1} \frac{(q|\eta_{j+1}|1+\Psi + \eta_1+\dots +\eta_j|\delta_j)}{(1+\Psi+\eta_1+\dots +\eta_j)^2 (q \delta_j)}\cdot \frac{ (q|\sum \beta|\mathcal{K}|q) }{(1+ \mathcal{I})^2} 
\end{align}
where we relabel $\delta \rightarrow \delta_{E-1}$. This is indeed just the object $\phi_{|\mathcal{I}|+1}^*(\beta+ \mathcal{I})$.

The general situation is a combination of the two patterns already considered. It is best illustrated by considering a non-trivial example. 

\subsection{A non-trivial graph example}

The purpose of this subsection is to consider 
a non-trivial example of the graph $\Gamma_\mathcal{K}$ containing 5 vertices with 3 ends $\{2,5,6\}$
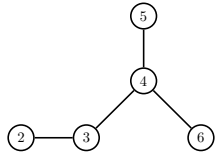
\begin{center}
 \scalebox{0.5}{
 \begin{tikzpicture}[
 Vertex/.style={circle, draw=black, fill=white, very thick, minimum size=5mm},
 Equation/.style={rectangle, draw=black, fill=white, very thick, minimum size=40},]
 \node[Vertex] (11) {2};
 \node[Vertex] (12) [right=of 11]{3};
 \node[Vertex] (13) [above right=of 12]{4};
 \node[Vertex] (14) [above =of 13]{5};
 \node[Vertex] (15) [below right=of 13]{6};

 \draw[-,very thick](11.east) to (12.west);
 \draw[-,very thick](12.north east) to (13.south west);
 \draw[-,very thick](13.north) to (14.south);
 \draw[-,very thick](13.south east) to (15.north west);
 \end{tikzpicture}
 }
 \end{center}

There are 3 possible subgraphs with 4 vertices, giving rise to quantities $\phi_4(2345),\phi_4(2346),\phi_4(3456)$.
We apply the formula \eqref{phi-star-relation-1} to obtain the quantity $\phi_4^*$ for each of these graphs. We have 
\begin{align}
    \phi_4^*(2345) (1+2+3+4+5)^2 = &\phi_4(2345) \sum_{\braket{mn}\in \Gamma_{\mathcal{I}\cup 1}} (q|\mathcal{M}|\mathcal{N}|q)\frac{(mn)}{(qm)(nq)}\nonumber
    \\
    + & \frac{(q|2|34|q)(q|5|123|4)}{(1+3+4)^2(1+2+3+4)^2(q4)}\frac{(q|6|2345|q)}{(1+2+3+4+5)^2}\nonumber
    \\
    + & \frac{(q|5|34|q)(q|2|135|4)}{(1+3+4)^2(1+3+4+5)^2(q4)}\frac{(q|6|2345|q)}{(1+2+3+4+5)^2}
\end{align}
for the first tree graph $\phi^*_4(2345)$. There are two additional terms here corresponding to two ways to build the graph $2-3-4-5$, by first adding the end $2$ and then $5$ or vice versa. The vertex $6$ is connected to the graph $\mathcal{I}$ through the vertex $4$ from the stem $\{3,4\}$. Thus, there is no further simplification here.

Another graph is similar
\begin{align}
    \phi_4^*(2346) (1+2+3+4+6)^2 = &\phi_4(2346) \sum_{\braket{mn}\in \Gamma_{\mathcal{I}\cup 1}} (q|\mathcal{M}|\mathcal{N}|q)\frac{(mn)}{(qm)(nq)}\nonumber
    \\
    + & \frac{(q|2|34|q)(q|6|123|4)}{(1+3+4)^2(1+2+3+4)^2(q4)}\frac{(q|5|2346|q)}{(1+2+3+4+6)^2}\nonumber
    \\
    + & \frac{(q|6|34|q)(q|2|136|4)}{(1+3+4)^2(1+3+4+6)^2(q4)}\frac{(q|5|2346|q)}{(1+2+3+4+6)^2}
\end{align}
The expression for $\phi_4^*(2346)$ is identical to $\phi_4^*(2345)$ with the only difference being  $5\leftrightarrow6$. 

Lastly, we obtain the equation for $\phi_4^*(3456)$
\begin{align}
    \phi^*(3456)& (1+3+4+5+6)^2 = \phi_4(3456) \sum_{\braket{mn}\in \Gamma_{\mathcal{I}\cup 1}} (q|\mathcal{M}|\mathcal{N}|q)\frac{(mn)}{(qm)(nq)}\nonumber
    \\
    +&\textcolor{blue}{\frac{(q|5|4|q)(q|6|15|4)(q|2|3456|q)}{(1+4)^2(1+4+5)^2(q4)(1+3+4+5+6)^2}}\nonumber
    \\
    +&\textcolor{red}{\frac{(q|6|4|q)(q|5|16|4)(q|2|3456|q)}{(1+4)^2(1+4+6)^2(q4)(1+3+4+5+6)^2}}\nonumber
    \\
    +&\textcolor{blue}{\frac{(q|5|4|q)(q|3|15|4)(q|6|145|3)(q|2|3456|q)}{(1+4)^2(1+4+5)^2(1+3+4+5)^2(q4)(q3)(1+3+4+5+6)^2}}\nonumber
    \\
    +&\textcolor{blue}{\frac{(q|3|4|q)(q|5|13|4)(q|6|145|3)(q|2|3456|q)}{(1+4)^2(1+3+4)^2(1+3+4+5)^2(q4)(q3)(1+3+4+5+6)^2}}\nonumber
    \\
    +&\textcolor{red}{\frac{(q|6|4|q)(q|3|16|4)(q|5|146|3)(q|2|3456|q)}{(1+4)^2(1+4+6)^2(1+3+4+6)^2(q4)(q3)(1+3+4+5+6)^2}}\nonumber
    \\
    +&\textcolor{red}{\frac{(q|3|4|q)(q|6|13|4)(q|5|146|3)(q|2|3456|q)}{(1+4)^2(1+3+4)^2(1+3+4+6)^2(q4)(q3)(1+3+4+5+6)^2}}\label{4-point}
\end{align}
There are six terms here which are the $3!$ terms in the formula for a graph with 3 ends. In this case, there are additional simplifications. The first two terms come from choosing the last momentum $\eta_E$ to be momentum 3 in \eqref{phi-star-relation-1}. In this case the momentum $2$ not in $\mathcal{I}$ is connected to $\mathcal{I}$ via the last end $\eta_E$, and so in this case we are dealing with $2=\alpha$ terms in \eqref{phi-star-relation-1}. The next two terms come from choosing the final leaf $\eta_E$ to be momenta $6$. The momenta $2$  is attached to the vertex $3$ which corresponds to $\delta \rightarrow 3$ and $\beta \rightarrow 2$. Similarly, the last two terms are when we choose $\eta_E$ to be momentum 5. This example illustrates that the momentum that is being added to the set $\mathcal{I}$ may correspond both to $\alpha$ and $\beta$ terms in \eqref{phi-star-relation-1}. 

We now start massaging the arising terms to bring them into the expected expression for $\phi^*_5(23456)(1+2+3+4+5+6)^2$. First, we use 
using the Schouten identity to shift the spinor contraction $|3)$ to be close to term with $(q|3|$ as
\begin{align}
    \phi^*(3456)& (1+3+4+5+6)^2 = \phi_4(3456) \sum_{\braket{mn}\in \Gamma_{\mathcal{I}\cup 1}} (q|\mathcal{M}|\mathcal{N}|q)\frac{(mn)}{(qm)(nq)}\nonumber
    \\
    +&\textcolor{blue}{\frac{(q|5|4|q)(q|6|15|4)(q|2|3456|q)}{(1+4)^2(1+4+5)^2(q4)(1+3+4+5+6)^2}}\nonumber
    \\
    +&\textcolor{red}{\frac{(q|6|4|q)(q|5|16|4)(q|2|3456|q)}{(1+4)^2(1+4+6)^2(q4)(1+3+4+5+6)^2}}\nonumber
    \\
    +&\textcolor{blue}{\frac{(q|5|4|q)(q|3|15|4)(q|6|145|3)(q|2|3456|q)}{(1+4)^2(1+4+5)^2(1+3+4+5)^2(q4)(q3)(1+3+4+5+6)^2}}\nonumber
    \\
    +&\textcolor{blue}{\frac{(q|3|4|q)(q|5|14|3)(q|6|135|4)(q|2|3456|q)}{(1+4)^2(1+3+4)^2(1+3+4+5)^2(q4)(q3)(1+3+4+5+6)^2}}\nonumber
    \\
    +&\textcolor{red}{\frac{(q|6|4|q)(q|3|16|4)(q|5|146|3)(q|2|3456|q)}{(1+4)^2(1+4+6)^2(1+3+4+6)^2(q4)(q3)(1+3+4+5+6)^2}}\nonumber
    \\
    +&\textcolor{red}{\frac{(q|3|4|q)(q|6|14|3)(q|5|136|4)(q|2|3456|q)}{(1+4)^2(1+3+4)^2(1+3+4+6)^2(q4)(q3)(1+3+4+5+6)^2}}\label{4-point-proof-1}
\end{align}
where the forth and the sixth line shift simultaneously.
Now, we are ready to use \eqref{identity-simplify-phi-star} starting with the combination of the first and the third terms
\begin{align}
    &\textcolor{blue}{\frac{(q|5|4|q)(q|6|15|4)(1+3+4+5)^2(q|2|3456|q)}{(1+4)^2(1+4+5)^2(1+3+4+5)^2(q4)(1+3+4+5+6)^2}}\nonumber
    \\
    +&\textcolor{blue}{\frac{(q|5|4|q)(q|3|15|4)(q|6|145|3)(q|2|3456|q)}{(1+4)^2(1+4+5)^2(1+3+4+5)^2(q4)(q3)(1+3+4+5+6)^2}}\nonumber
    \\
    = &\textcolor{blue}{\frac{(q|5|4|q)(q|2|3456|q)}{(1+4)^2(1+4+5)^2(1+3+4+5)^2(q4)(1+3+4+5+6)^2}}\nonumber
    \\
    &~~\times\bigg(\frac{(q|6|15|4)(1+3+4+5)^2 + (q|3|15|4)(q|6|145|3)}{(q3)}\bigg)
\end{align}
Using \eqref{identity-simplify-phi-star}, we obtain
\begin{align}
    \textcolor{blue}{\frac{(q|5|4|q)(1+4+5)^2(q|6|135|4)(q|2|3456|q)}{(1+4)^2(1+4+5)^2(1+3+4+5)^2(q4)(1+3+4+5+6)^2}}
\end{align}
Adding the second line from \eqref{4-point-proof-1} to the above expression gives
\begin{align}
    &\textcolor{blue}{\frac{(q|5|4|q)(1+4+5)^2(q|6|135|4)(q|2|3456|q)}{(1+4)^2(1+4+5)^2(1+3+4+5)^2(q4)(1+3+4+5+6)^2}}\nonumber
    \\
    +&\textcolor{blue}{\frac{(q|3|4|q)(q|5|14|3)(q|6|135|4)(q|2|3456|q)}{(1+4)^2(1+3+4)^2(1+3+4+5)^2(q4)(q3)(1+3+4+5+6)^2}}\nonumber
    \\
    =&\textcolor{blue}{\frac{(q|6|135|4)(q|2|3456|q)}{(1+4)^2(1+3+4)^2(1+3+4+5)^2(q4)(1+3+4+5+6)^2}}
    \nonumber
    \\
    &~~\times \bigg((q|5|4|q)(1+3+4)^2 + (q|3|4|q)(q|5|14|3)\bigg)
\end{align}
which is equal to 
\begin{align}
    &\textcolor{blue}{\frac{(q|5|34|q)(1+4)^2(q|6|135|4)(q|2|3456|q)}{(1+4)^2(1+3+4)^2(1+3+4+5)^2(q4)(1+3+4+5+6)^2}}\nonumber
    \\
    =&\textcolor{blue}{\frac{(q|5|34|q)(q|6|135|4)(q|2|3456|q)}{(1+3+4)^2(1+3+4+5)^2(q4)(1+3+4+5+6)^2}}
\end{align}
We can see the pattern of simplification as it shifts the momentum square term until it can cancel with the smallest momentum square in the denominator. 
The final result here can be interpreted as the stem grows larger from $\{3\} \rightarrow \{3,4\}$. 
The remaining terms from \eqref{4-point-proof-1} can be combined similarly, yielding
\begin{align}
    \textcolor{red}{\frac{(q|6|34|q)(q|5|135|4)(q|2|3456|q)}{(1+3+4)^2(1+3+4+6)^2(q4)(1+3+4+5+6)^2}}
\end{align}
Thus, we can rewrite \eqref{4-point-proof-1} as
\begin{align}
    \phi^*(3456) (1+3+4+5+6)^2 = &\phi_4(3456) \sum_{\braket{mn}\in \Gamma_{\mathcal{I}\cup 1}} (q|\mathcal{M}|\mathcal{N}|q)\frac{(mn)}{(qm)(nq)}\nonumber
    \\
    + &\frac{(q|5|34|q)(q|6|135|4)(q|2|3456|q)}{(1+3+4)^2(1+3+4+5)^2(q4)(1+3+4+5+6)^2}\nonumber
    \\
    +&\frac{(q|6|34|q)(q|5|135|4)(q|2|3456|q)}{(1+3+4)^2(1+3+4+6)^2(q4)(1+3+4+5+6)^2}\nonumber
\end{align}
The rearranged equation \eqref{4-point-proof-1}, such that the term with $(q|3|$ spinor contraction is next to the term with $|3)$ spinor contraction, helps us see the simplification clearly. This rearrangement will be important for the general case as well.

We have thus shown that the sum of $\phi^*_4(\Gamma_\mathcal{I})(1+\mathcal{I})^2$ over all possible trees $\Gamma_\mathcal{I}$ is given by
\begin{align}    \sum_{\Gamma_\mathcal{I}}\phi^*_4(\Gamma_\mathcal{I})(1+\mathcal{I})^2 = &\sum_{\Gamma_\mathcal{I}}\phi_4(\Gamma_\mathcal{I})\sum_{\braket{mn}\in \Gamma_{\mathcal{I}\cup 1}} (q|\mathcal{M}|\mathcal{N}|q)\frac{(mn)}{(qm)(nq)}\nonumber
    \\
    + & \frac{(q|2|34|q)(q|5|123|4)(q|6|2345|q)}{(1+3+4)^2(1+2+3+4)^2(q4)(1+2+3+4+5)^2}\nonumber
    \\
    + & \frac{(q|5|34|q)(q|2|135|4)(q|6|2345|q)}{(1+3+4)^2(1+3+4+5)^2(q4)(1+2+3+4+5)^2}\nonumber
    \\
    + & \frac{(q|2|34|q)(q|6|123|4)(q|5|2346|q)}{(1+3+4)^2(1+2+3+4)^2(q4)(1+2+3+4+6)^2}\nonumber
    \\
    + & \frac{(q|6|34|q)(q|2|136|4)(q|5|2346|q)}{(1+3+4)^2(1+3+4+6)^2(q4)(1+2+3+4+6)^2}\nonumber
    \\
    + &\frac{(q|5|34|q)(q|6|135|4)(q|2|3456|q)}{(1+3+4)^2(1+3+4+5)^2(q4)(1+3+4+5+6)^2}\nonumber
    \\
    +&\frac{(q|6|34|q)(q|5|135|4)(q|2|3456|q)}{(1+3+4)^2(1+3+4+6)^2(q4)(1+3+4+5+6)^2}
\end{align}
We can easily recognize the last six terms as $\phi^*_5(23456) = \phi_5(23456)$. Thus, in this example, we have shown that
\begin{align}  \sum_{\Gamma_\mathcal{I}}\phi^*_4(\Gamma_\mathcal{I})(1+\mathcal{I})^2 = &\sum_{\Gamma_\mathcal{I}}\phi_4(\Gamma_\mathcal{I})\sum_{\braket{mn}\in \Gamma_{\mathcal{I}\cup 1}} (q|\mathcal{M}|\mathcal{N}|q)\frac{(mn)}{(qm)(nq)}\nonumber
    \\
    + &\phi^*_5(23456)(1+2+3+4+5+6)^2.
\end{align}

\subsection{General argument}
For a general simplification, we return to \eqref{phi-star-relation-1}. The general proof will follow what we have done in the previous section, which is to rearrange the terms of $\phi^*_i(\Gamma_\mathcal{I})(1+\mathcal{I})^2$ via Schouten identity, then use the identity \eqref{identity-simplify-phi-star} $E-1$ times for a tree graph $\Gamma_\mathcal{I}$ with $E$ ends.

Considering the graph $\Gamma_\mathcal{I}$ below, which is being extended to a graph $\mathcal{K}$ that has an additional vertex $\gamma$
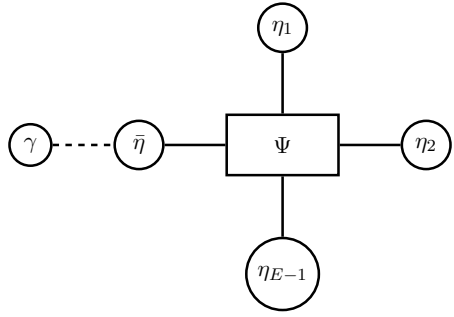
\begin{center}
\scalebox{0.8}{
\begin{tikzpicture}[
Diam/.style={diamond, draw=black, fill=white, very thick, minimum size=5mm},
Alpha/.style={circle, draw=black, fill=white, very thick, minimum size=5mm},
Beta/.style={circle, draw=black, fill=white, very thick, minimum size=5mm},
Eta/.style={circle, draw=black, fill=white, very thick, minimum size=8mm},
Psi/.style={rectangle, draw=black, fill=white, very thick, minimum size=10mm},
Equation/.style={rectangle, draw=white, fill=white, very thick, minimum size=20},]
\node[Alpha] (a1) {$\gamma$};
\node[Eta] (e) [right = of a1]{$\Bar{\eta}$};
\draw[dashed,very thick](a1.east) to (e.west);

\node[Psi] (p) [right = of e]{$~~~~~\Psi~~~~~$};
\node[Eta] (b1) [above = of p]{$\eta_1$};
\node[Eta] (b2) [right = of p]{$\eta_2$};
\node[Eta] (b3) [below = of p]{$\eta_{E-1}$};
\draw[-,very thick](e.east) to (p.west);

\draw[-,very thick](p.north) to (b1.south);
\draw[-,very thick](p.east) to (b2.west);
\draw[-,very thick](p.south) to (b3.north);

\end{tikzpicture}
}
\end{center}
where $\Psi$ is the stem of the tree and $\mathcal{I} = \Psi + \sum_E \eta$. We denote the end $\Bar{\eta}$ as a vertex connecting to $\gamma$. The idea of the proof is to show that, after simplifications, the stem of the tree will grow larger to $\Psi + \Bar{\eta}$, and a new end $\gamma$ appears. 

The solution from \eqref{phi-star-relation-1} for this graph $\Gamma_\mathcal{I}$ is given by
\begin{align}\label{proof-A}
    {\phi}^*_i(\Gamma_\mathcal{I}) &(1+\mathcal{I})^2 = {\phi}_i(\Gamma_\mathcal{I}) \sum_{\braket{mn}\in 1+\Gamma_\mathcal{I}} (q|\mathcal{M}|\mathcal{N}|q)\frac{(mn)}{(qm)(nq)}
    \\ 
    + &\sum_{\eta ~\text{perms}}  \frac{(q|\eta_1|\Psi|q)}{(1+\Psi)^2} \cdot \prod^{E-2}_{j=1} \frac{(q|\eta_{j+1}|1+\Psi + \sum_j \eta|\delta_j)}{(1+\Psi+\sum_j \eta)^2 (q \delta_j)}\cdot \frac{ (1+ \mathcal{I} - \Bar{\eta})^2(q|\gamma|\mathcal{K}|q)}{(1+ \mathcal{I} - \Bar{\eta})^2(1+\mathcal{I})^2} \nonumber
    \\
    + &\sum_{\eta ~\text{perms}} \frac{(q|\eta_1|\Psi|q)}{(1+\Psi)^2} \cdot \prod^{E-2}_{j=1} \frac{(q|\eta_{j+1}|1+\Psi +\sum_j \eta|\delta_j)}{(1+\Psi+\sum_j \eta)^2 (q \delta_j)}\bigg|_{\eta_{E-1} \rightarrow \Bar{\eta}}\cdot \frac{ (q|\eta_{E-1}|1+\mathcal{I}|\Bar{\eta})(q|\gamma|\mathcal{K}|q) }{(q\Bar{\eta})(1+ \mathcal{I} - \eta_{E-1})^2(1+\mathcal{I})^2}\nonumber
\end{align}
where we choose $\eta_E$ to be $\Bar{\eta}$ in the second line and $\eta_E \neq \Bar{\eta}$ in the final line. This corresponds to choosing $\delta$ to be $\Bar{\eta}$ in the last line. The sum in the second term is over permutations of $\eta_j$ for $j=1,\dots,E-1$ for every end of the tree $\Gamma_\mathcal{I}$. Thus, the second line consists of $(E-1)!$ terms.

The last line is much more complicated in the sum over permutation however. The momenta $\bar{\eta}$ can be in any spinor contraction with $(q|\eta|$ except the last spinor contraction $(q|\eta_{E-1}|1+\mathcal{I}|\Bar{\eta})$ while the other $\eta_j$, $j=1,\dots,E-1$ can be permuted in any order resulting in $(E-1)\times(E-1)!$ terms which, we want to show, sums up with $(E-1)!$ terms from the second line to the expected total of $E!$ terms.

Ignoring the summation over $\eta$ permutation symbol, we can write the second and last line explicitly as
\begin{align}
    & \frac{(q|\eta_1|\Psi|q)}{(1+\Psi)^2} \cdot \prod^{E-3}_{j=1} \frac{(q|\eta_{j+1}|1+\Psi + \sum_j \eta|\delta_j)}{(1+\Psi+\sum_j \eta)^2 (q \delta_j)}\cdot \frac{(q|\eta_{E-1}|1+\mathcal{I} - \Bar{\eta}|\delta_{E-2})}{(1+\mathcal{I} - \Bar{\eta} - \eta_{E-1})^2 (q \delta_{E-2})}\cdot \frac{ (1+ \mathcal{I} - \Bar{\eta})^2(q|\gamma|\mathcal{K}|q)}{(1+ \mathcal{I} - \Bar{\eta})^2(1+\mathcal{I})^2} \nonumber
    \\
    + &\frac{(q|\eta_1|\Psi|q)}{(1+\Psi)^2} \cdot \prod^{E-3}_{j=1} \frac{(q|\eta_{j+1}|1+\Psi +\sum_j \eta|\delta_j)}{(1+\Psi+\sum_j \eta)^2 (q \delta_j)}\cdot \frac{(q|\Bar{\eta}|1+\mathcal{I}- \eta_{E-1}|\delta_{E-2})}{(1+\mathcal{I} - \Bar{\eta} - \eta_{E-1})^2 (q \delta_{E-2})}\cdot  \frac{ (q|\eta_{E-1}|1+\mathcal{I}|\Bar{\eta})(q|\gamma|\mathcal{K}|q) }{(q\Bar{\eta})(1+ \mathcal{I} - \eta_{E-1})^2(1+\mathcal{I})^2}\nonumber
    \\
    + &\frac{(q|\eta_1|\Psi|q)}{(1+\Psi)^2} \cdot \prod^{E-4}_{j=1} \frac{(q|\eta_{j+1}|1+\Psi +\sum_j \eta|\delta_j)}{(1+\Psi+\sum_j \eta)^2 (q \delta_j)}\cdot \frac{(q|\Bar{\eta}|1+\mathcal{I} - \eta_{E-2}- \eta_{E-1}|\delta_{E-3})}{(1+\mathcal{I} - \Bar{\eta} - \eta_{E-2} - \eta_{E-1})^2 (q \delta_{E-3})}\nonumber
    \\
    & \cdot \frac{(q|\eta_{E-2}|1+\mathcal{I}- \eta_{E-1}|\delta_{E-2})}{(1+\mathcal{I} - \eta_{E-1} - \eta_{E-1})^2 (q \delta_{E-2})}\cdot  \frac{ (q|\eta_{E-1}|1+\mathcal{I}|\Bar{\eta})(q|\gamma|\mathcal{K}|q) }{(q\Bar{\eta})(1+ \mathcal{I} - \eta_{E-1})^2(1+\mathcal{I})^2}\nonumber
    \\
    +&\dots
    \\
    + &\frac{(q|\Bar{\eta}|\Psi|q)}{(1+\Psi)^2} \cdot \frac{(q|\eta_1|1+\Psi+\Bar{\eta}|\delta_{1})}{(1+\Psi+\Bar{\eta})^2 (q \delta_{1})} \cdot \prod^{E-2}_{j=2} \frac{(q|\eta_{j+1}|1+\Psi +\sum_j \eta|\delta_j)}{(1+\Psi+\sum_j \eta)^2 (q \delta_j)}\cdot  \frac{ (q|\eta_{E-1}|1+\mathcal{I}|\Bar{\eta})(q|\gamma|\mathcal{K}|q) }{(q\Bar{\eta})(1+ \mathcal{I} - \eta_{E-1})^2(1+\mathcal{I})^2}\nonumber
\end{align}
where we expand the last term of \eqref{proof-A} as $E-1$ sums over $(E-1)!$ permutations of $\eta_j$. Now, it is clear that we sum over the permutation of momenta $\eta_j$ for $j = 1,\dots,E-1$ as the momentum $\Bar{\eta}$ is treated as special. 
The position of $\Bar{\eta}$ shifts by one for each line similarly to \eqref{4-point}.
Now, we employ a similar shift to \eqref{4-point-proof-1} by moving term with $|\Bar{\eta)}$ through $|\delta_j)$ to be next to term with $(q|\Bar{\eta}|$. This can be done by Schouten identity over the sum of the permutation. We can write the shifted term together as
\begin{align}
    & \frac{(q|\eta_1|\Psi|q)}{(1+\Psi)^2} \cdot \prod^{E-3}_{j=1} \frac{(q|\eta_{j+1}|1+\Psi + \sum_j \eta|\delta_j)}{(1+\Psi+\sum_j \eta)^2 (q \delta_j)}\cdot \frac{(q|\eta_{E-1}|1+\mathcal{I} - \Bar{\eta}|\delta_{E-2})}{(1+\mathcal{I} - \Bar{\eta} - \eta_{E-1})^2 (q \delta_{E-2})}\cdot \frac{ (1+ \mathcal{I} - \Bar{\eta})^2(q|\gamma|\mathcal{K}|q)}{(1+ \mathcal{I} - \Bar{\eta})^2(1+\mathcal{I})^2} \nonumber
    \\
    + &\frac{(q|\eta_1|\Psi|q)}{(1+\Psi)^2} \cdot \prod^{E-3}_{j=1} \frac{(q|\eta_{j+1}|1+\Psi +\sum_j \eta|\delta_j)}{(1+\Psi+\sum_j \eta)^2 (q \delta_j)}\cdot \frac{(q|\Bar{\eta}|1+\mathcal{I}- \eta_{E-1}|\delta_{E-2})}{(1+\mathcal{I} - \Bar{\eta} - \eta_{E-1})^2 (q \delta_{E-2})}\cdot  \frac{ (q|\eta_{E-1}|1+\mathcal{I}|\Bar{\eta})(q|\gamma|\mathcal{K}|q) }{(q\Bar{\eta})(1+ \mathcal{I} - \eta_{E-1})^2(1+\mathcal{I})^2}\nonumber
    \\
    + &\frac{(q|\eta_1|\Psi|q)}{(1+\Psi)^2} \cdot \prod^{E-4}_{j=1} \frac{(q|\eta_{j+1}|1+\Psi +\sum_j \eta|\delta_j)}{(1+\Psi+\sum_j \eta)^2 (q \delta_j)}\cdot \frac{(q|\Bar{\eta}|1+\mathcal{I} - \eta_{E-2}- \eta_{E-1}|\delta_{E-3})}{(1+\mathcal{I} - \Bar{\eta} - \eta_{E-2} - \eta_{E-1})^2 (q \delta_{E-3})}\nonumber
    \\
    & \cdot \frac{(q|\eta_{E-2}|1+\mathcal{I}- \eta_{E-1}|\Bar{\eta})}{(1+\mathcal{I} - \eta_{E-1} - \eta_{E-1})^2 (q \Bar{\eta})}\cdot  \frac{ (q|\eta_{E-1}|1+\mathcal{I}|\delta_{E-2})(q|\gamma|\mathcal{K}|q) }{(q\delta_{E-2})(1+ \mathcal{I} - \eta_{E-1})^2(1+\mathcal{I})^2}\nonumber
    \\
    +&\dots\nonumber
    \\
    + &\frac{(q|\Bar{\eta}|\Psi|q)}{(1+\Psi)^2} \cdot \frac{(q|\eta_1|1+\Psi+\Bar{\eta}|\Bar{\eta})}{(1+\Psi+\Bar{\eta})^2 (q \Bar{\eta})} \cdot \prod^{E-2}_{j=2} \frac{(q|\eta_{j+1}|1+\Psi +\sum_j \eta|\delta_j)}{(1+\Psi+\sum_j \eta)^2 (q \delta_j)}\cdot  \frac{ (q|\eta_{E-1}|1+\mathcal{I}|\delta_{E-2})(q|\gamma|\mathcal{K}|q) }{(q\delta_{E-2})(1+ \mathcal{I} - \eta_{E-1})^2(1+\mathcal{I})^2}\label{n-point-proof}
\end{align}
where the sum over permutatiton of $\eta_j$ is implied every line. Then, it is clear that the simplification \eqref{identity-simplify-phi-star} can be used in a similar fashion to what we did in the previous subsection. The first term containing a square of momentum combines with the second term results in
\begin{align}
    & \frac{(q|\eta_1|\Psi|q)}{(1+\Psi)^2} \cdot \prod^{E-3}_{j=1} \frac{(q|\eta_{j+1}|1+\Psi + \sum_j \eta|\delta_j)}{(1+\Psi+\sum_j \eta)^2 (q \delta_j)}\cdot \frac{(1+ \mathcal{I} - \Bar{\eta}- \eta_{E-1})^2}{(1+\mathcal{I} - \Bar{\eta} - \eta_{E-1})^2 }\cdot \frac{ (q|\eta_{E-1}|1+\mathcal{I} |\delta_{E-2}) (q|\gamma|\mathcal{K}|q)}{(q \delta_{E-2})(1+ \mathcal{I} - \Bar{\eta})^2(1+\mathcal{I})^2} \nonumber
\end{align}
we can see that the momentum square term is shifted from $(1+\mathcal{I} - \Bar{\eta})^2$ to $(1+\mathcal{I} - \Bar{\eta} - \eta_{E-1})^2$ in the above expression.
This can combined with the third term from \eqref{n-point-proof} moving $(1+\mathcal{I} - \Bar{\eta} - \eta_{E-1})^2$ to $(1+\mathcal{I} - \Bar{\eta}  - \eta_{E-2}- \eta_{E-1} )^2$.
The pattern goes on until the last term.
Finally, the simplification results in
\begin{align}
    \sum_{\eta ~\text{perms}} \frac{(q|\eta_1|\Psi + \Bar{\eta}|q)}{(1+\Psi + \Bar{\eta})^2} \cdot \prod^{E-2}_{j=1} \frac{(q|\eta_{j+1}|\Psi  + \Bar{\eta} + \eta_1+\dots +\eta_j|\delta_j)}{(1+\Psi \Bar{\eta} + \eta_1 + \dots +\eta_j)^2 (q \delta_j)}\cdot \frac{(q|\gamma|\mathcal{K}|q)}{(1+\mathcal{I})^2}\label{simplification-solution}
\end{align}
Thus, we can clearly see that the solution above retains the structure of a tree with $E$ ends. 
When we take the sum over all possible tree graphs $\Gamma_{\mathcal{I}} \subset \Gamma_{\mathcal{K}}$ in the equation \eqref{phi-star-relation-1}, the contributions from other graph combined will result in
\begin{align}
    \sum_{\Gamma_\mathcal{I}}{\phi}^*_i(\Gamma_\mathcal{I}) (1+\mathcal{I})^2 = &\sum_{\Gamma_\mathcal{I}}{\phi}_i(\Gamma_\mathcal{I}) \sum_{\braket{mn}\in 1+\Gamma_\mathcal{I}} (q|\mathcal{M}|\mathcal{N}|q)\frac{(mn)}{(qm)(nq)}
    \\ 
    + &\sum_{\Gamma_\mathcal{K},|\mathcal{K}|=k=|\mathcal{I}|+1}{\phi}^*_k(\Gamma_\mathcal{K}) (1+\mathcal{K})^2
\end{align}
where $\phi^*_k(\Gamma_\mathcal{K}) = \phi_k(\Gamma_\mathcal{K})$.
Here, we note that in general we can attach any vertices to the vertex $\gamma$ without modifying the analysis above. This change only the final spinor contraction from $(q|\gamma|\mathcal{K}|q)$ to $(q|\sum\rho+\gamma|\mathcal{K}|q)$. This is identical to changing $\phi$ into $\phi^*$ as we change only the final spinor contraction to include the information of the whole graph instead of just an extension vertex $\gamma$, as is represented in the following figure
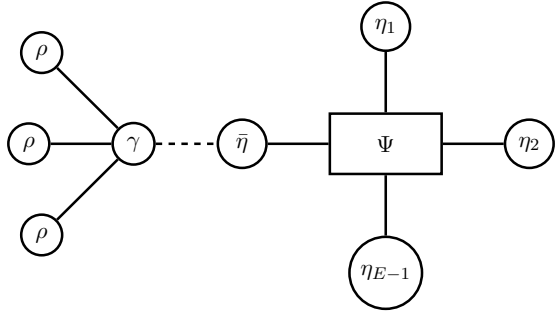
\begin{center}
\scalebox{0.8}{
\begin{tikzpicture}[
Diam/.style={diamond, draw=black, fill=white, very thick, minimum size=5mm},
Alpha/.style={circle, draw=black, fill=white, very thick, minimum size=5mm},
Beta/.style={circle, draw=black, fill=white, very thick, minimum size=5mm},
Eta/.style={circle, draw=black, fill=white, very thick, minimum size=8mm},
Psi/.style={rectangle, draw=black, fill=white, very thick, minimum size=10mm},
Equation/.style={rectangle, draw=white, fill=white, very thick, minimum size=20},]
\node[Alpha] (a1) {$\gamma$};
\node[Eta] (e) [right = of a1]{$\Bar{\eta}$};
\draw[dashed,very thick](a1.east) to (e.west);
\node[Alpha] (r1) [left = of a1]{$\rho$};
\node[Alpha] (r2) [above left= of a1]{$\rho$};
\node[Alpha] (r3) [below left= of a1]{$\rho$};
\draw[-,very thick](r1.east) to (a1.west);
\draw[-,very thick](r2.south east) to (a1.north west);
\draw[-,very thick](r3.north east) to (a1.south west);

\node[Psi] (p) [right = of e]{$~~~~~\Psi~~~~~$};
\node[Eta] (b1) [above = of p]{$\eta_1$};
\node[Eta] (b2) [right = of p]{$\eta_2$};
\node[Eta] (b3) [below = of p]{$\eta_{E-1}$};
\draw[-,very thick](e.east) to (p.west);

\draw[-,very thick](p.north) to (b1.south);
\draw[-,very thick](p.east) to (b2.west);
\draw[-,very thick](p.south) to (b3.north);

\end{tikzpicture}
}
\end{center}

Thus, we have shown that extending the tree both through the stem and through the leaf, $\phi^*$ satisfies the relation \eqref{phi-star-relation}
\begin{align}
    \sum_{\Gamma_\mathcal{I}}{\phi}^*_i(\Gamma_\mathcal{I}) (1+\mathcal{I})^2 = &\sum_{\Gamma_\mathcal{I}}{\phi}_i(\Gamma_\mathcal{I}) \sum_{\braket{mn}\in 1+\Gamma_\mathcal{I}} (q|\mathcal{M}|\mathcal{N}|q)\frac{(mn)}{(qm)(nq)}
    \\ 
    + &\sum_{\Gamma_\mathcal{J},|\mathcal{J}|=j=|\mathcal{I}|+1}{\phi}^*_j(\Gamma_\mathcal{J}) (1+\mathcal{J})^2.
\end{align}
In conclusion, we have seen that there are two different behaviors in extending the tree either through the leaf (end) or the stem. The leaf (end) extension leads to the stem becoming larger as expected, since one of the leaves (ends) becomes a part of the stem, while the stem extension simply increases the number of ends of the tree.

This discussion proves the formula \eqref{SolutionPhi} for the quantities $\phi(\Gamma_\mathcal{I})$, and thus gives a complete solution to the all-but-one-plus gravity BG current.

\section{Discussion}

The main new result of this paper is the formula for the all-but-one-plus gravitational Berends-Giele current. This current is completely described by its scalar part, which is given by the sum over spanning trees $\Gamma_\mathcal{K}$ of the complete graph on momenta $\mathcal{K}=\{2,\ldots,n\}$
\begin{align}
    J(1|\mathcal{K}) = \sum_{\Gamma_\mathcal{K}} \Phi(\Gamma_\mathcal{K}) J(\Gamma_\mathcal{K}).
\end{align}
Here 
\begin{align}
    J(\Gamma_\mathcal{K}) = \prod_{i\in \Gamma_\mathcal{K}} (qi)^{2(\alpha_i-2)} \prod_{\langle ij\rangle \in \Gamma_\mathcal{K}} \frac{[ij]}{(ij)}
\end{align}
is the contribution of the graph $\Gamma_\mathcal{K}$ to the all-plus BG current, with $\alpha_i$ being the multiplicity (number of edges coming to) of the vertex $i$ and $\langle ij\rangle \in \Gamma_\mathcal{K}$ being all edges of $\Gamma_\mathcal{K}$. At this level, the above result is not that different from what happens in Yang-Mills theory, where the all-but-one-plus current (scalar part thereof) is a multiple of the all-plus current. In the case of gravity we observe the same structure, except that now the currents are given by sums over graphs (in the case of YM only one graph plays role because of the colour ordering). But for each graph the all-but-one-plus current is a multiple of the all-plus current. 

The quantity $\Phi(\Gamma_\mathcal{K})$ which is the ratio of the all-but-one-plus to all-plus current for each graph is the most interesting new object. As in the case of YM theory, this quantity is given by the sum of contributions from all subgraphs 
\begin{align}\label{Phi-disc}
    \Phi(\Gamma_\mathcal{K}) = \sum_{\Gamma_\mathcal{I}\subset \Gamma_\mathcal{K}, i=|\mathcal{I}|} \phi_i(\Gamma_\mathcal{I}).
\end{align}
The formula for $\phi_i$ is however much more intricate as compared to the YM case, and is the main new result of this paper. It is given by 
\begin{align}\label{SolutionPhi-disc}
&\phi_i(\Gamma_\mathcal{I}) = \\ \nonumber
        &\sum_{\eta ~\text{perms}} \frac{1}{\Box}  \frac{(q|\eta_1|\Psi|q)}{(1+\Psi)^2} \cdot \prod^{E-2}_{j=1} \frac{(q|\eta_{j+1}|1+\Psi + \eta_1+\dots +\eta_j|\delta_j)}{(1+\Psi+\eta_1+\dots +\eta_j)^2 (q \delta_j)}\cdot \frac{(q|\eta_E|\Psi + \eta_1 + \dots + \eta_{E-1}|q)}{(1+\Psi +  \eta_1 + \dots + \eta_{E-1})^2}.
    \end{align}
    Here $\eta_{1,\ldots, E}$ denote the ends of the graph $\Gamma_\mathcal{I}$, and $E$ is the number of these ends. There are $E!$ terms in this formula. Details are described in Section \ref{sec:solution}.

When one wants to extract the MHV formula from the all-but-one-plus current, one is interested in the case when the sum of momenta $1+2+\ldots + n$ is null, and thus represented by the product of two spinors, in our notation $|0)\otimes [0|$. One also only wants to keep the part of the all-but-one-plus current that contains the pole when the final momentum goes null, and extract the residue of this pole (the MHV amplitude). There is only one term in the sum \eqref{Phi-disc} that contains such a pole, which is the term in which $\mathcal{I}=\mathcal{K}$. Extracting the residue of this pole, and evaluating the residue on the final momentum that is the product of two spinors brings in huge simplifications in that the sum over permutations in \eqref{SolutionPhi-disc} can be explicitly computed, as described in Section \ref{GravityAmplitude}. The answer is 
\begin{align}\label{answer-phi-disc}
    (1+\mathcal{K})^2 \phi(\Gamma_\mathcal{K})\Big|_{on\,\, shell}=
     (01)^2  \prod_{i=1}^E \frac{(1 \eta_i)}{(0 \eta_i)}  \prod_{j = 1}^{E-2} \frac{(0 \delta_j)}{(1 \delta_j)}.
\end{align}
Here $\eta_{1,\ldots, E}$ are the ends of the graph $\Gamma_\mathcal{K}$ and $\delta_j, j=1,\ldots, E-2$ are the vertices of valency more than two, taken with multiplicity equal to valency minus two. When multiplied by the all-plus current $J(\Gamma_\mathcal{K})$ one immediately obtains the known correct formula for the MHV gravity amplitudes. 

As we already discussed in the Introduction, the derivation of the gravity MHV formula that we provided is completely analogous to that which was used in the original Berends-Giele paper \cite{Berends:1987me} to obtain YM MHV amplitudes. Our starting point was the gravitational Feynman rules, to derive which we have used the chiral first-order formulation of GR. While this formulation of GR brings in some simplifications (in particular the Lagrangian is polynomial in the fields), the full set of Feynman rules is not completely straighforward even in this formalism. However, we have shown that only a single term in the kinetic term, and only a single term in the interaction terms can contribute to the answer we are after. With the arising effective set of Feynman rules the calculation that leads to the Berends-Giele recursion both for the all-plus and all-but-one-plus gravity BG current is not more involved than in the case of YM theory. Both of the arising recurrence relations exhibit the $Gravity = YM^2$ pattern. 

The most complicated part of our analysis was to solve the arising recursive relation for the all-but-one-plus gravity amplitude. This led to a rather non-trivial combinatorics. But after a suitable ansatz for the solution is guessed by considering a large number of explicit examples, one can prove that the ansatz actually satisfies the recursion using a version of proof by induction. All this is analogous to what happens in the YM case, except that in the YM case the situation is considerably simpler because only a single linearly ordered graph plays role in the combinatorics. In the case of gravity one needs to consider all possible spanning trees for a complete graph on points $2,\ldots, n$. This makes the combinatorics an order of magnitude more complex. Nevertheless, an explict answer is still obtainable, and can be seen to lead to the correct formula for the MHV amplitudes in full generality. 

We find the described here derivation of the gravity MHV formula logically very simple. Also, the amount of what is assumed in order for the construction to proceed is minimal. We start from a Lagrangian for GR, proceed to derive the parts of the Feynman rules that are needed for our computation, and then use them to establish the recursion relations, that are then solved. This is a logically self-contained derivation.  

It is clear that the reason why it is possible to establish a sufficiently simple recurrence relation for both all-plus and all-but-one-plus currents is the integrability of the self-dual sectors of both YM and GR. Indeed, it is well-known that the all-plus gravity BG current describes a self-dual gravitational background viewed as composed of a set of positive helicity gravitons. Integrability means that the "most general solution" can be described, and the possibility of computing the all-plus BG current explicitly is a testament to this integrability. Similarly, the propagation of a single negative helicity graviton on a self-dual background is an integrable problem. The existence of a closed-form solution to the all-but-one-plus BG current is similarly a testament to this integrability. So, it is the integrability of the self-dual sector of GR that makes possible the computations of this paper. 

It is interesting to compare and contrast the calculations in this paper with a recent derivation of the MHV formula \cite{Miller:2024oza} using the perturbiner method. At the level of the fully self-dual solutions, what in this paper was called the all-plus BG current is completely analogous to the perturbiner expansion of the Plebanski scalar, see formulas (2.20), (2.21) of \cite{Miller:2024oza}. Indeed, these formulas match \eqref{GR-all-plus-current} very closely. The treatment starts to differ more substantially in the way the negative helicity graviton is dealt with. In the treatment of \cite{Miller:2024oza}, this is incorporated as a linear perturbation on the self-dual perturbiner, see (3.2) of the cited paper. This perturbation is then shown to be given by a certain exponential, see (3.20) of \cite{Miller:2024oza}. This is presumably related to the fact that our all-but-one-plus current for a given graph (or rather its ratio to the all-plus current) is given by the sum over all subgraphs. The main difference in our derivation and that in \cite{Miller:2024oza} is that we can extract the MHV amplitude directly from the current. In contrast, the paper \cite{Miller:2024oza} needs to evaluate the gravity action on the obtained perturbiner solutions, and only then the MHV amplitude is extracted. There is some non-trivial combinatorics involved in this process, occupying sections 4 and 5 of the paper. Another issue with the derivation in \cite{Miller:2024oza} is that it seems to only lead to the expression (5.51). This then needs to be corrected by introducing a pair of auxiliary spinors, and further simplified by choosing these auxiliary spinors to the the momentum spinors of the two negative helicity gluons. It feels that there is some guesswork involved in this logic. In contrast, our derivation leads directly to the MHV amplitude in the expected form as soon as the formula for the all-but-one-plus current is obtained. 

It is also worth discussing what would happen if we were to add another negative helicity graviton. It is clear that in this case most of the described constructions would become impossible. First, there will be other terms in the interaction part of the Lagrangian that will start contributing. Indeed, the reason for simplifications in all-but-one-plus case was that it was possible to choose the reference spinor of the positive helicity gravitons to be the momentum spinor of the single negative helicity graviton. If there is more than one negative helicity graviton, no more analogous simplifications arise, and one has to deal with the full complexity of the perturbation theory. Some computations can probably still be done, but there is no hope for a general formula involving an arbitrary number of positive helicity gravitons. At the same time, an arbitrary graviton scattering amplitude can be computed using the BCFW recursion relation, so it is not clear if there is any value in doing the calculations of the type we described with more than one negative helicity graviton. 

In the opinion of these authors, the most interesting open question is whether any of the simplifications we have witnessed happening in flat space continues to happen for gravitons around a constant curvature background such as de Sitter. There is hope for this, because the self-dual sector of GR is still integrable even if expanded around de Sitter space. So, it is reasonable to hope that some version of the BG recursion is also possible for graviton scattering in de Sitter, and that an explicit solution is also possible. Some preliminary steps towards this problem were already done in \cite{Krasnov:2024qkh}. We hope to return to this very interesting problem in future publications. 

\section{Acknowledgments} KK is grateful to the University of Marburg, and in particular his host Ilka Agricola, for hospitality during the period when this work was done. And also to the Alexander von Humboldt foundation for financially supporting this stay. 

\appendix

\section{Yang-Mills all plus one minus current}\label{Appendix1}

\renewcommand{\theequation}{A.\arabic{equation}}

\setcounter{equation}{0}

Here we provide the proof for all plus one minus current in a way different from that in \cite{Berends:1987me}. The recursion relation is given by
\begin{align}\label{YM-recursion-minus}
    J(1|2\dots n) = \frac{1}{\Box}\bigg(\frac{(q|2\dots n|p]}{[1p]}J(2\dots n) + \sum_{m=2}^{n-1} (q|1\dots m|m+1\dots n|q) J(1|2\dots m)J(m+1\dots n)\bigg)
\end{align}
We note that we specifically chose the reference spinor for positive helicity as $q = 1$ while the reference spinor for the negative helicity gluon remains arbitrary $p$. The solution to this recursion is
\begin{align}\label{YMSol}
    J(1|2\dots n) = J(2\dots n) \bigg(\frac{[2p]}{[12][1p]} + \sum_{m=2}^{n-1} \frac{(q|1\dots m | 1\dots m+1|q)}{(1\dots m)^2(1\dots  m+1)^2}\bigg).
\end{align}
We now prove this by induction, assuming that the above formula holds for all currents $J(1|2\ldots k), k<n$, and using the recursion recursion (\ref{YM-recursion-minus}) to show that it holds for $J(1|2\ldots n)$ as well. Note that the current $J(1|2)$ contains only $p$-dependent terms, $p$-independent terms are first present in $J(1|23)$. 

\subsection{Index free notation for spinor contractions}

Here we explain our index-free notations. Given two 2-component spinors $\lambda_A,\mu_A\in S_+$, we denote by
\begin{align}
    (\lambda \mu) = \lambda^A \mu_A 
\end{align}
their spinor contraction. Note that $(\lambda \mu)=-(\mu\lambda)$. For spinors of the opposite chirality $\lambda_{A'},\mu_{A'}\in S_-$, we use square brackets to denote their contraction
\begin{align}
    [\lambda \mu] = \lambda^{A'} \mu_{A'}. 
\end{align}

Given a 4-vector $p_{AA'}$, we can view this as an endomorphism $p: S_+\to S_-$ or $p: S_-\to S_+$. To write this in index-free notations, we will always raise one of the indices of $p_{AA'}$ so that e.g. the endomorphism $p: S_+\to S_-$ is represented by
\begin{align}
    p_{A'}{}^A \mu_A.
\end{align}
We write this resulting spinor in $S_-$ as $p|\mu)$. We can then contract it with some other spinor $\lambda\in S_-$ to form
\begin{align}
    [\lambda | p | \mu) = \lambda^{A'} p_{A'}{}^{A} \mu_A. 
\end{align}
In words, our index-free notation is such that spinor indices are always contracted in such a way that it is the upper index in the preceeding object, and the lower index in the object that follows. We similarly have
\begin{align}
    (\mu | p |\lambda] = \mu^A p_{A}{}^{A'} \lambda_{A'}.
\end{align}
Note that to pass to $[\lambda | p | \mu)$ one needs to raise/lower two pairs of spinor indices, and so there is no extra minus sign in the formula
\begin{align}
    [\lambda | p | \mu) = (\mu | p |\lambda] 
\end{align}

Given a pair of 4-vectors $p,q$, we can compose the corresponding endomorphisms. For instance, when acting on an unprimed spinor $\mu$ we get the spinor
\begin{align}
    p_{A}{}^{A'} q_{A'}{}^B \mu_B \in S_+
\end{align}
We write this as $p|q|\mu)$. This can be contracted with some other spinor in $S_+$ to form the quantity
\begin{align}
    (\lambda | p|q| \mu) = \lambda^A p_{A}{}^{A'} q_{A'}{}^B \mu_B.
\end{align}
Note that 
\begin{align}
    (\lambda | p|q| \mu) = - (\mu |q| p| \lambda). 
\end{align}

When $p=q$ we have
\begin{align}\label{conventions-minus}
    (\lambda | p|p| \mu) = - p^2 (\lambda \mu),
\end{align}
with this formula defining what we mean by $p^2$. Indeed, we have
\begin{align}
    p_{A}{}^{A'} p_{A'}{}^B = - p^2 \epsilon_A{}^B,
\end{align}
which means that in our conventions
\begin{align}
    p^2 = \frac{1}{2} p^{AA'} p_{AA'}.
\end{align}
This is convenient because
\begin{align}
    (1+2)^2 = \frac{1}{2}(1+2)^{AA'}(1+2)_{AA'} =  1^{AA'} 2_{AA'} = (12) [12],
\end{align}
which is convenient for calculations. One should be careful about the minus signs coming from \eqref{conventions-minus} when doing the calculations.

\subsection{Ansatz for the solution and the corresponding recursion relation}

We look for the solution in the form
\begin{align}
    J(1|2\ldots n) = J(2\ldots n) \Phi(2\ldots n),
\end{align}
where $\Phi(2\ldots n)$ is some factor to be determined. Substituting this ansatz into the recursion \eqref{YM-recursion-minus} and taking into account that
\begin{align}
    J(2\ldots n) = \frac{1}{(q2)(23)\ldots (nq)} = J(2\ldots m) J(m+1\ldots n) \frac{(mq)(q~m+1)}{(m~ m+1)}
\end{align}
and thus
\begin{align}
   J(2\ldots m) J(m+1\ldots n)= \frac{(m~ m+1)}{(mq)(q~m+1)} J(2\ldots n),
\end{align}
we get a recursion relation for $\Phi(2\ldots n)$
\begin{align}\label{YM-recursion-phi}
    \Phi(2\dots n) = \frac{1}{\Box}\bigg(\frac{(q|2\dots n|p]}{[1p]} + \sum_{m=2}^{n-1} (q|1\dots m|m+1\dots n|q) \frac{(m~ m+1)}{(mq)(q~m+1)}\Phi(2\dots m)\bigg).
\end{align}


\subsection{A version of the basic identity \eqref{YM Identity}}

When $q\not=1$ the basic identity \eqref{YM Identity} reads
\begin{align}
    \sum_{m=1}^{n-1} (q|1\dots m|m+1\dots n|q)  \frac{(m~ m+1)}{(mq)(q~m+1)}  =  (1\ldots n)^2. 
\end{align}
The purpose of this subsection is to show that this also makes sense when $q\to 1$. Indeed, let us assume  $1\not=q$, and separate the term $m=1$. We have
\begin{align}
    (q|1|2\ldots n|q) \frac{(12)}{(1q)(q2)}  +  \sum_{m=2}^{n-1} (q|1\ldots m|m+1\ldots n) \frac{(m ~m+1)}{(mq)(q ~m+1)} = (1\ldots n)^2.
\end{align}
This can be rewritten as 
\begin{align}
   - [1|2\ldots n|q) \frac{(12)}{(q2)}  +  \sum_{m=2}^{n-1} (q|1\ldots m|m+1\ldots n|q) \frac{(m~ m+1)}{(mq)(q~m+1)}  = (1\ldots n)^2.
\end{align}
We can now send $q\to 1$ to produce
\begin{align}\label{identity-1-special}
    - [1|2\ldots n|q) +  \sum_{m=2}^{n-1} (q|1\ldots m|m+1\ldots n|q) \frac{(m~ m+1)}{(mq)(q~m+1)} =  (1\ldots n)^2.
\end{align} 
The purpose of these manipulations is to show that the basic YM identity \ref{YM Identity} also makes sense with $q\to 1$, because the problematic $0/0$ terms cancel. 

\subsection{Dealing with the $p$-dependent terms}

We first want to prove that the $p$-dependent part of $\Phi(2\ldots n)$ is given by
\begin{align}
    \phi_1 = \frac{[2p]}{[12][1p]}.
\end{align}
Concentrate on the $p$-dependent terms, and ignoring the $1/\Box$ for now, we have:
\begin{align}\label{YM1}
    \frac{(q|2\dots n|p]}{[1p]} + \frac{[2p]}{[12][1p]}\sum_{m=2}^{n-1} (q|1\dots m|m+1\dots n|q) \frac{(m~ m+1)}{(mq)(q~m+1)}.
\end{align}
We can multiply the first term by $[12]/[12]$ and use Schouten identity $[ab][cd]=[ac][bd]-[ad][bc]$ to get
\begin{align}\label{YM-shouten}
    \frac{(q|2\dots n|p][12]}{[12][1p]} = (q|2\dots n |1]\frac{[p2]}{[12][1p]} - \frac{(q|2\dots n|2|1)}{(1+2)^2},
\end{align}
where we used $(1+2)^2 = (12)[12]$ and also replaced $(q|2\ldots n|2](12)= -(q|2\ldots n|2|1)$. 

We now use the identity \ref{identity-1-special} to deal with the $p$-dependent terms. Indeed, the sum of $p$-dependent terms from \eqref{YM-shouten} and \eqref{YM1} is 
\begin{align}
    \frac{[2p]}{[12][1p]}\bigg(-(q|2\ldots n|1] + \sum_{m=2}^{n-1} (q|1\dots m|m+1\dots n|q) \frac{(m~ m+1)}{(mq)(q~m+1)} \bigg) = \frac{[2p]}{[12][1p]} (1+\dots +n)^2
\end{align}
This shows that the solution \eqref{YMSol} correctly reproduces the $p$-dependent terms. 

\subsection{Dealing with $p$-independent terms}

We now want to prove that
\begin{align}
    \Phi(2\ldots n) = \phi_1 +\sum_{j=2}^{n-1} \phi_j ,
\end{align}
    where
\begin{align} \label{phi-m}
\phi_j = 
    \frac{(q|1\dots j|1\dots j+1|q)}{(1\dots j)^2(1\dots j+1)^2}. 
\end{align}
We assume that this holds for lower values of $n$ and substitute this into the recursion relation \eqref{YM-recursion-phi}. We take $q\to 1$, and as before ignore the $1/\Box$ factor. We get
\begin{align}\label{YM-p-independent}
    -\frac{(q|2\dots n|2|q)}{(1+2)^2} + 
    \sum_{m=3}^{n-1} (q|1\dots m|m+1\dots n|q) \frac{(m~ m+1)}{(mq)(q~m+1)}  \sum_{j=2}^{m-1} \phi_j.
\end{align}
We can change the order of summation in the second term, grouping together terms with the same propagator factors 
\begin{align}\label{YM-p-independent-1}
    -\frac{(q|2\dots n|2|q)}{(1+2)^2} +
    \sum_{j=2}^{n-2} \phi_j
    \sum_{m=j+1}^{n-1} (q|1\dots m|m+1\dots n|q)  \frac{(m~ m+1)}{(mq)(q~m+1)} .
\end{align}
The second sum now is of the type we can deal with using the basic all plus recurrence relation \eqref{YM Identity}. However, in order to sum it up, we need to provide the missing terms
\begin{align}
    \sum_{m=1}^{j}(q|1\dots m |m+1\dots n|q)\frac{(m~ m+1)}{(mq)(q~m+1)} .
\end{align}
We now want to prove that the first term in \eqref{YM-p-independent} will provide these missing terms.

\begin{prop} The following identity holds
\begin{align}\label{YM-identity-1}
    \frac{(q|1\dots n|1\dots m|q)}{(1\dots m)^2} &= \frac{(q|1\dots n|1\dots m+1|q)}{(1 \dots m+1)^2} \nonumber
    \\&- \phi_m \sum_{j=1}^{m}(q|1\dots j |j+1\dots n|q)\frac{(j ~j+1)}{(j~q)(q~j+1)}.
\end{align}
Here $\phi_m$ is given by \eqref{phi-m}.
\end{prop}

$\bullet$ \textbf{Proof}
We start by multiplying the LHS by $(1+\dots + m+1)^2/(1+\dots + m+1)^2$ and write the result as
\begin{align}
    -\frac{(q|1\dots n|1\dots m|1\dots m+1|1\dots m+1|q)}{(1 \dots m)^2(1 \dots m+1)^2}.
\end{align}
We used $(a|p|p|b) = - (ab) p^2$,
which holds in our conventions. 
This can be separated into two terms
\begin{align}
    -\frac{(q|1\dots n|1\dots m|1\dots m|1\dots m+1|q)}{(1 \dots m)^2(1 \dots  m+1)^2} - \frac{(q|1\dots n|1\dots m|m+1|1\dots m+1|q)}{(1 \dots m)^2(1 \dots  m+1)^2}.
\end{align}
The first term here is equal to the first term on the right-hand side of \eqref{YM-identity-1}
\begin{align}
    -\frac{(q|1\dots n|1\dots m|1\dots m|1\dots m+1|q)}{(1\dots m)^2(1\dots  m+1)^2} = \frac{(q|1\dots n|1\dots m+1|q)}{(1\dots  m+1)^2}.
\end{align}
We now massage the second term. We have
\begin{align}\label{YM-identity-2}
    -\frac{(q|1\dots n|1\dots m|m+1|1\dots m+1|q)}{(1\dots m)^2(1\dots  m+1)^2} &=- \frac{(q~m+1)[m+1|1\dots m|q)}{(1\dots m)^2(1 \dots  m+1)^2} \frac{(q|1\dots n|1\dots m|m+1)}{(q~m+1)}\nonumber
    \\
    &=-\phi_m \sum_{i=1}^m \frac{(q|1\dots n|i](i q) (i~m+1)}{(iq)(q~m+1)}.
\end{align}
Here, to get the first equality, we multiplied the numerator and denominator by $(q~m+1)$, and to get the last equality we multiplied by $(iq)$ in the sum.
We then use the Schouten identity in the form
\begin{align}\label{schouten-appendix}
    \frac{(ab)}{(aq)(qb)} +\frac{(bc)}{(bq)(qc)} =\frac{(ac)}{(aq)(qc)}
\end{align}
The identity can be used repeatedly to give
\begin{align}
    \frac{(i~m+1)}{(iq)(q~m+1)} = \sum_{j=i}^m\frac{(j~j+1)}{(j~q)(q~j+1)}
\end{align}
This allows to rewrite \eqref{YM-identity-2} as
\begin{align}
    -\phi_m \sum_{i=1}^m (q|1\dots n|i|q) \sum_{j=i}^m\frac{(j~j+1)}{(j~q)(q~j+1)}.
    \end{align}
 We now change the order of summation, and in this way take the sum over $i$ to get
 \begin{align}\nonumber
    &-\phi_m  \sum_{j=1}^m\frac{(j~j+1)}{(j~q)(q~j+1)} (q|1\dots n|1\dots j|q) \\ \label{2nd}
    &=-\phi_m \sum_{j=1}^m (q|1\dots j|j+1\dots n|q)\frac{(j~j+1)}{(j~q)(q~j+1)}.
\end{align}
This finishes the proof of the identity \eqref{YM-identity-1}.

We can now keep applying the identity \eqref{YM-identity-1} until the numerator of the first term equals $(q|1\ldots n|1\ldots n|q)=0$. Applying this to $(q|1\ldots n|1+ 2|q)/(1+ 2)^2$ gives
\begin{prop}
    The following identity holds
    \begin{align}\label{identity-phi}
         -\frac{(q|1\ldots n|1+ 2|q)}{(1+ 2)^2} =
         \sum_{j=2}^{n-1} \phi_j  \sum_{m=1}^j (q|1\dots m|m+1\dots n|q)\frac{(m~m+1)}{(m~q)(q~m+1)}.
    \end{align}
\end{prop}

We have thus put the first term in \eqref{YM-p-independent-1} in the form of the second term. This shows that the $p$-independent terms \eqref{YM-p-independent-1} are equal to
\begin{align}
    \sum_{j=2}^{n-1} \phi_j
    \sum_{m=1}^{n-1} (q|1\dots m|m+1\dots n|q)  \frac{(m~ m+1)}{(mq)(q~m+1)} 
\end{align}
The last sum can now be computed using the basic identity \eqref{YM Identity}.

This gives the $p$-independent terms (canceling the final propagator by the ignored up to now $1/\Box$ factor), and establishes the final answer 
\begin{align}\label{phi-current}
    \Phi(2\ldots n) = \sum_{j=1}^{n-1} \phi_j.
\end{align}

We now note that we could change the logic of the above argument. Indeed, we can assume that \eqref{phi-current} holds, and substitute this relation to both sides of the recursion \eqref{YM-recursion-phi} for $\Phi(2\ldots n)$. This will imply that 
\begin{align}
    \frac{(q|2\ldots n|p]}{[1p]} = \sum_{j=1}^{n-1} \phi_j \sum_{m=1}^j (q|1\dots m|m+1\dots n|q)  \frac{(m~ m+1)}{(mq)(q~m+1)}.
\end{align}
Alternatively, taking care of the $p$-dependent terms, the identity \eqref{identity-phi} must hold. 
Thus, the formula \eqref{identity-phi} can be interpreted as a recursion relation that allows to compute the quantities $\phi_j$. 

For example, applying \eqref{identity-phi} with $n=3$, we can compute $\phi_2$ directly. Indeed, in this case the identity becomes
\begin{align}
    \frac{(q|12|123|q)}{(1+2)^2} = \phi_2 (1+2+3)^2, 
\end{align}
which implies the answer \eqref{phi-m} for $\phi_2$. One can then compute recursively, recovering all $\phi_j$. 

\section{Schouten Identity}\label{Identity}

\renewcommand{\theequation}{B.\arabic{equation}}

\setcounter{equation}{0}

There is a form of Schouten identity that is frequently used :
\begin{align}
    \sum_{a,I,j,k} (A|a|i|I)(J|j|k|K) &= \sum_{a,I,j,k}(Aa)[ai](iI)(Jj)[jk](kK)
    \\
    &= \sum_{a,I,j,k} \bigg[(A|a|j|J)(I|i|k|K) + (A|a|k|K)(J|j|i|I)\bigg]
\end{align}
For example, consider the case
\begin{align}\label{schouten 1234}
    (q|1|23|q)(q|2|4|q) = (q|1|2|q)(q|23|4|q) + (q|1|4|q)(q|2|23|q)
\end{align}
For a more complex example, 
\begin{align}
    (q|1|345|q)(q|2|45|q) &= (q|1|2|q)(q|345|45|q) + (q|1|45|q)(q|2|345|q).
\end{align}
This identity makes the calculation much faster.

\bibliographystyle{abbrv}
\bibliography{sample}
\end{document}